\documentclass[longauth]{aa}  
\usepackage{natbib}
\usepackage{graphicx}
\usepackage{txfonts}
\usepackage{hyperref}
\usepackage[normalem]{ulem}
\usepackage{xcolor}
\usepackage{epsfig}
\usepackage{xspace}
\newcommand\gaia{\textit{Gaia}}

\newcommand\gdrtwo{\gaia~DR2 }
\newcommand\gedrthree{\gaia~EDR3}
\newcommand\gdrthree{\gaia~DR3\xspace}
\newcommand\gdr[1]{\gaia~DR#1}         % used in release documentation
\newcommand\egdr[1]{\gaia~EDR#1}       % used in release documentation

\newcommand{\xp}{{XP}\xspace}

\newcommand\figref[1]{Fig.~\ref{#1}}
\newcommand\figureref[1]{Figure~\ref{#1}}

\newcommand{\orcit}[1]{\protect\href{https://orcid.org/#1}{\protect\includegraphics[width=8pt]{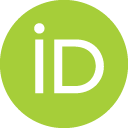}}}

\newcommand\bpminrp{\ensuremath{G_\mathrm{BP}-G_\mathrm{RP}}}

\newcommand\ebpminrp{\ensuremath{E(G_\mathrm{BP}-G_\mathrm{RP})}}

\def\bprp{\bpminrp}

\def\teff{\ensuremath{T_{\rm eff}}\xspace}
\def\a0{$A_{\rm 0}$}

\def\gmag{$G$}
\def\gbp{$G_{\rm BP }$}
\def\grp{$G_{\rm RP }$}
\def\bprp{\ensuremath{G_{\rm BP}-G_{\rm RP}}} 
\def\grvs{$G_{\rm RVS }$}
\def\kms{\,km\,s$^{-1} $}

\def\muas{\,$\mu$as}
\def\logg{$\log g$}

\providecommand{\Msun}{\ensuremath{\,{\cal M}_{\odot}}\xspace}

\makeatletter
\DeclareRobustCommand*{\fieldName}[1]{%
  \begingroup\@fieldName\scantokens{\texttt{\small {#1}}\noexpand}\endgroup}
\begingroup\lccode`\~=`\_\relax
   \lowercase{\endgroup\def\@fieldName{\catcode`\_=\active \let~\_}}
\makeatother

\makeatletter
\DeclareRobustCommand*{\tableName}[1]{%
  \begingroup\@tableName\scantokens{\texttt{#1}\noexpand}\endgroup}
\begingroup\lccode`\~=`\_\relax
   \lowercase{\endgroup\def\@tableName{\catcode`\_=\active \let~\_}}
\makeatother

\usepackage{amstext}

\begin{document} 

   \title{{\gaia} Data Release 3\\Catalogue Validation}

   \author{
C.~Babusiaux\orcit{0000-0002-7631-348X}\inst{\ref{inst:IPAG},\ref{inst:gepi}}, 
C.~Fabricius\orcit{0000-0003-2639-1372}\inst{\ref{inst:ieec}},
S.~Khanna\orcit{0000-0002-2604-4277}\inst{\ref{inst:kapteyn}},
T.~Muraveva\inst{\ref{inst:bologna}}, 
C.~Reyl\'e\orcit{0000-0003-2258-2403}\inst{\ref{inst:utinam}}, 
F.~Spoto\orcit{0000-0001-7319-5847}\inst{\ref{inst:cfa}}, 
A.~Vallenari\inst{\ref{inst:padova}}, 
X.~Luri\orcit{0000-0001-5428-9397}\inst{\ref{inst:ieec}},
%
% \TBC{Rest of authors alphabetically:}
%
F.~Arenou\inst{\ref{inst:gepi}}, 
M.A.~\'Alvarez\orcit{0000-0002-6786-2620}\inst{\ref{inst:citic}}, 
F.~Anders\inst{\ref{inst:ieec}}, 
T.~Antoja\inst{\ref{inst:ieec}}, 
E.~Balbinot\orcit{0000-0002-1322-3153}\inst{\ref{inst:kapteyn}}, 
C.~Barache\inst{\ref{inst:SYRTE}}, 
N.~Bauchet\orcit{0000-0002-2307-8973}\inst{\ref{inst:gepi}},  
D. Bossini\inst{\ref{inst:caup}}, 
D.~Busonero\inst{\ref{inst:torino}}, 
T.~Cantat-Gaudin\orcit{0000-0001-8726-2588}\inst{\ref{inst:ieec},\ref{inst:mpi}}, 
J.~M.~Carrasco\orcit{0000-0002-3029-5853} \inst{\ref{inst:ieec}}, 
C.~Dafonte\orcit{0000-0003-4693-7555}\inst{\ref{inst:citic}}, 
S.~Diakit\'e\inst{\ref{inst:mesofc}}, 
F.~Figueras\inst{\ref{inst:ieec}}, 
A.~Garcia-Gutierrez\inst{\ref{inst:ieec}}, 
A.~Garofalo\inst{\ref{inst:bologna}}, 
A.~Helmi\orcit{0000-0003-3937-7641}\inst{\ref{inst:kapteyn}}, 
\'O. Jim\'enez-Arranz\inst{\ref{inst:ieec}}, 
C.~Jordi\orcit{0000-0001-5495-9602}\inst{\ref{inst:ieec}}, 
P.~Kervella\orcit{0000-0003-0626-1749}\inst{\ref{inst:lesia}}, 
Z. Kostrzewa-Rutkowska\inst{\ref{inst:leiden},\ref{inst:leiden2}}, 
N.~Leclerc\inst{\ref{inst:gepi}},  
E.~Licata\inst{\ref{inst:torino}}, 
M.~Manteiga\orcit{0000-0002-7711-5581}\inst{\ref{inst:coruna}}, 
A.~Masip\inst{\ref{inst:ieec}}, 
M.~Mongui\'o\inst{\ref{inst:ieec}}, 
P.~Ramos\orcit{0000-0002-5080-7027}\inst{\ref{inst:ieec},\ref{inst:strasbourg}}, 
N.~Robichon\inst{\ref{inst:gepi}}, 
A.~C.~Robin\inst{\ref{inst:utinam}}, 
M.~Romero-G\'omez\inst{\ref{inst:ieec}}, 
A.~S\'aez\inst{\ref{inst:ieec}}, 
R.~Santove\~na\orcit{0000-0002-9257-2131}\inst{\ref{inst:citic}}, 
L.~Spina\inst{\ref{inst:padova}}, 
G.~Torralba Elipe\inst{\ref{inst:citic}}, 
M.~Weiler\inst{\ref{inst:ieec}} 
}

\institute{
Univ. Grenoble Alpes, CNRS, IPAG, 38000 Grenoble, France
\label{inst:IPAG}
\and
GEPI, Observatoire de Paris, Universit{\'e} PSL, CNRS, 5 Place Jules Janssen, 92190 Meudon, France
\label{inst:gepi}
\and
Dept. FQA, Institut de Ci\`encies del Cosmos, Universitat de Barcelona (IEEC-UB), Mart\'i i Franqu\`es 1, E-08028 Barcelona, Spain %\\
%\email{claus@fqa.ub.edu}
\label{inst:ieec}
\and
Kapteyn Astronomical Institute, University of Groningen, Landleven 12, 9747 AD Groningen, The Netherlands
\label{inst:kapteyn}
\and 
INAF - Osservatorio di Astrofisica e Scienza dello Spazio di Bologna, via Piero Gobetti 93/3, 40129 Bologna,  Italy                                        \label{inst:bologna}
\and
Institut UTINAM - UMR 6213 - CNRS - University of Bourgogne Franche Comt\'e, France, OSU THETA, 41 bis avenue de l’Observatoire, 25000, Besan\c{c}on, France 
\label{inst:utinam}
\and
M\'{e}socentre de Franche-Comt\'{e}, University of Franche-Comt\'{e}, 16 route de Gray, 25030 Besan\c{c}on Cedex, France
\label{inst:mesofc}
\and
%Universit\'e C\^ote d'Azur, Observatoire de la C\^ote d'Azur, CNRS,
%Laboratoire Lagrange, Bd de l'Observatoire, CS 34229, 06304 Nice cedex
%4, France\label{inst:nice}
%\and 
%IMCCE, Observatoire de Paris, PSL Research University, CNRS, Sorbonne
%Universit\'e, UPMC Univ. Paris 06, Univ. Lille, 77 av.
%Denfert-Rochereau, 75014 Paris, France\label{inst:imcce}
Harvard-Smithsonian Center for Astrophysics, 60 Garden St., MS 15, Cambridge, MA 02138, USA\label{inst:cfa}
\and
INAF, Osservatorio Astronomico di Padova, Vicolo Osservatorio, Padova, I-35131, Italy
\label{inst:padova}
%\and
%Observatoire de Gen\`eve, Universit\'e de Gen\`eve, CH-1290 Versoix, Switzerland
%\label{inst:Geneva}
%\and
%Institute of Astronomy, University of Cambridge, Madingley Road, Cambridge CB30HA, United Kingdom
%\label{inst:ioa}
\and 
CITIC, Department of Computer Science and Information Technologies, University of A Coru\~na, Campus de
Elvi\~na S/N, 15071 A Coru\~na, Spain
\label{inst:citic}
\and 
SYRTE, Observatoire de Paris, Universit\'e PSL, CNRS, Sorbonne Universit\'e, LNE, 61 avenue de l'Observatoire, 75014 Paris, France
\label{inst:SYRTE}
\and 
Instituto de Astrofísica e Ciências do Espaço, Universidade do Porto, CAUP, Rua das Estrelas, PT4150-762 Porto, Portugal
\label{inst:caup}
\and
INAF - Osservatorio Astrofisico di Torino, Via Osservatorio 20, 10025 Pino Torinese, Torino,  Italy                           \label{inst:torino}
\and
LESIA, Observatoire de Paris, Universit\'e PSL, CNRS, Sorbonne Universit\'e, Universit\'e Paris Cit\'e, 5 place Jules Janssen, 92195 Meudon, France
\label{inst:lesia}
\and
%CENTRA, Universidade de Lisboa, FCUL, Campo Grande, Edif. C8, 1749-016 Lisboa, Portugal
%\label{inst:FCUL}
%\and
Leiden Observatory, Leiden University, Niels Bohrweg 2, 2333 CA Leiden, The Netherlands
\label{inst:leiden}
\and
SRON Netherlands Institute for Space Research, Niels Bohrweg 4, 2333 CA Leiden, The Netherlands
\label{inst:leiden2}
\and
CITIC, Department of Nautical Sciences and Marine Engineering, University of A Coru\~na, Campus de
Elvi\~na S/N, 15071 A Coru\~na, Spain
\label{inst:coruna}
\and 
Observatoire astronomique de Strasbourg, Universit{\'e} de Strasbourg, CNRS, 11 rue de l’Universit{\'e}, 67000 Strasbourg, France
\label{inst:strasbourg}
\and
Max-Planck-Institut f{\"u}r Astronomie, K{\"o}nigstuhl 17, D-69117 Heidelberg, Germany
\label{inst:mpi}
%\and 
%INAF - Osservatorio Astronomico di Roma, Via di Frascati 33, 00078 Monte Porzio Catone (Roma), Italy
%\label{inst:roma}
%\and
%ASI Science Data Center, Via del Politecnico, Roma
%\label{inst:asdc}
%\and
%Laboratoire d'astrophysique de Bordeaux, Univ. de Bordeaux, CNRS, B18N, all{\'e}e Geoffroy Saint-Hilaire, 33615 Pessac, France
%\label{inst:Bordeaux}
%\and
%INAF - Osservatorio Astrofisico di Arcetri, Largo Enrico Fermi 5, I-50125 Firenze, Italy                                \label{inst:0024}
%\and Laboratoire Univers et Particules de Montpellier, Universit\'{e} Montpellier, CNRS, Place Eug\`{e}ne Bataillon, CC72, F-34095 Montpellier Cedex 05, France\relax \label{inst:montpellier}
}

   \date{ }

\abstract
  % context heading (optional)
{
The third \gaia\ data release (DR3) provides a wealth of new data products. The early part of the release, \gedrthree, already provided the astrometric and photometric data for nearly two billion sources. 
The full release now adds improved parameters compared to \gdrtwo for radial velocities, astrophysical parameters, variability information, light curves, and orbits for Solar System objects. The improvements are in terms of the number of sources, the variety of parameter information, precision, and accuracy. For the first time, \gdrthree also provides a sample of spectrophotometry and spectra obtained with the Radial Velocity Spectrometer, binary star solutions, and a characterisation of extragalactic object candidates.
}
  % aims heading (mandatory)
{
Before the publication of the catalogue, these data have undergone a dedicated transversal validation process. The aim of this paper is to highlight limitations of the data that were found during this process and to provide recommendations for the usage of the catalogue.
}
  % methods heading (mandatory)
{
The validation was obtained through a statistical analysis of the data, a confirmation of the internal consistency of different products, and a comparison of the values to external data or models. 
}
  % results heading (mandatory)
{
\gdrthree is a new major step forward in terms of the number, diversity, precision, and accuracy of the \gaia\ products. As always in such a large and complex catalogue, however, issues and limitations have also been found. Detailed examples of the scientific quality of the \gdrthree\ release can be found in the accompanying data-processing papers as well as in the performance verification papers. Here we focus only on the caveats that the user should be aware of to scientifically exploit the data. 
}
  % conclusions heading (optional), leave it empty if necessary 
{}

   
   \titlerunning{{\gdrthree} Catalogue Validation} 
   \authorrunning{C. Babusiaux et al.}
   
   \maketitle
%
%________________________________________________________________
%
%
%
%+++++++++++++++++++++++++++++++++++++++++++++++++++++++++++++++++++++++++++
\section{Introduction}
%+++++++++++++++++++++++++++++++++++++++++++++++++++++++++++++++++++++++++++

This paper describes the validation of the third \gaia\ data release, \gdrthree \citep{2016A&A...595A...1G,DR3-DPACP-185}. The validation of the astrometric and photometric content can be found in the {\gedrthree} validation paper \citep{EDR3-DPACP-126}. We focus here on the new products of \gdrthree, which are summarised in \cite{DR3-DPACP-185}. The main new products of \gdrthree are the radial velocities, as well as line broadening and  {\grvs} magnitude, astrophysical parameters, variable stars, Solar System objects, and for the first time, spectra (both from the spectrophotometric instrument and from the Radial Velocity Spectrometer (RVS)), non-single stars, and quasar (QSO) and galaxy candidates, and associated characterisation.
The processing papers\footnote{\label{dr3papers}\url{https://www.cosmos.esa.int/web/gaia/dr3-papers}} and the on-line documentation\footnote{\label{onlinedoc}\url{https://gea.esac.esa.int/archive/documentation/GDR3/index.html}} describe the data and their internal validation in detail. The performance verification papers\textsuperscript{\ref{dr3papers}} highlight the overall quality of the catalogue. In this paper, we focus on presenting the caveats that the \gdrthree users should be aware of. Although the scientific validation process has confirmed the high quality of the {\gedrthree} data, certain issues remain, and there are caveats. We focus this paper on highlighting them, and we provide advice to the users. 

The approach followed by the validation presented in this paper is a transverse analysis of the properties of the catalogue content. Tests are either internal (including overall statistics, correlations, and clustering analysis between catalogue entries) or use external data, a Galaxy model, or clusters.  
The comparison with a Galaxy model was made using the Gaia object generator (GOG20) as a reference model or the Gaia universe model snapshot (GUMS20, which contains the intrinsic properties of the objects generated by GOG). They are described in detail in the on-line documentation\footnote{\url{https://gea.esac.esa.int/archive/documentation/GEDR3/Data\_processing/chap\_simulated/}} and were released with the \egdr{3} set of 
catalogues\footnote{GOG is published in the \gaia\ archive in the table \tableName{gaiaedr3.gaia\_source\_simulation} and GUMS in the table \tableName{gaiaedr3.gaia\_universe\_model}}.

Although the validation tests were designed to be transverse, we organise the paper by product for convenience. We therefore discuss in turn the Radial Velocity Spectrometer products (Sect.~\ref{sec:rvs}: radial velocities, line broadening, {\grvs} magnitude, and RVS spectra),  the low-resolution (Blue and Red Photometers, BP and RP) spectra (Sect.~\ref{sec:xp}), the astrophysical parameters (Sect.~\ref{sec:AP}),  QSO and galaxy candidates (Sect.~\ref{sec:extragal}),  non-single stars (Sect.~\ref{sec:nss}), variables (Sect.~\ref{sec:vari}), and Solar System objects (Sect.~\ref{sec:sso}).

%+++++++++++++++++++++++++++++++++++++++++++++++++++++++++++++++++++++++++++
\section{Radial Velocity Spectrometer}\label{sec:rvs}
%+++++++++++++++++++++++++++++++++++++++++++++++++++++++++++++++++++++++++++

While only radial velocities were provided in \gdrtwo, more products from the Radial Velocity Spectrometer (RVS) are available in \gdrthree within the \tableName{gaia\_source} table: the radial velocities (\fieldName{radial\_velocity}) down to \grvs$<$14, the spectral line broadening (\fieldName{vbroad}), and the magnitude \grvs\ estimated using the RVS spectra flux (\fieldName{grvs\_mag}). Moreover, a subset of RVS spectra are available through the \tableName{rvs\_mean\_spectrum} datalink\footnote{\label{datalink}\url{https://geapre.esac.esa.int/archive/documentation/GDR3/Gaia\_archive/chap\_archive/}}.

\subsection{Radial velocity}

The radial velocity ($RV$) data are presented in detail in \cite{DR3-DPACP-159}. The radial velocities of hot stars are specifically targeted in \cite{DR3-DPACP-151}.

\begin{figure}
    \centering
    \includegraphics[width=0.45\textwidth]{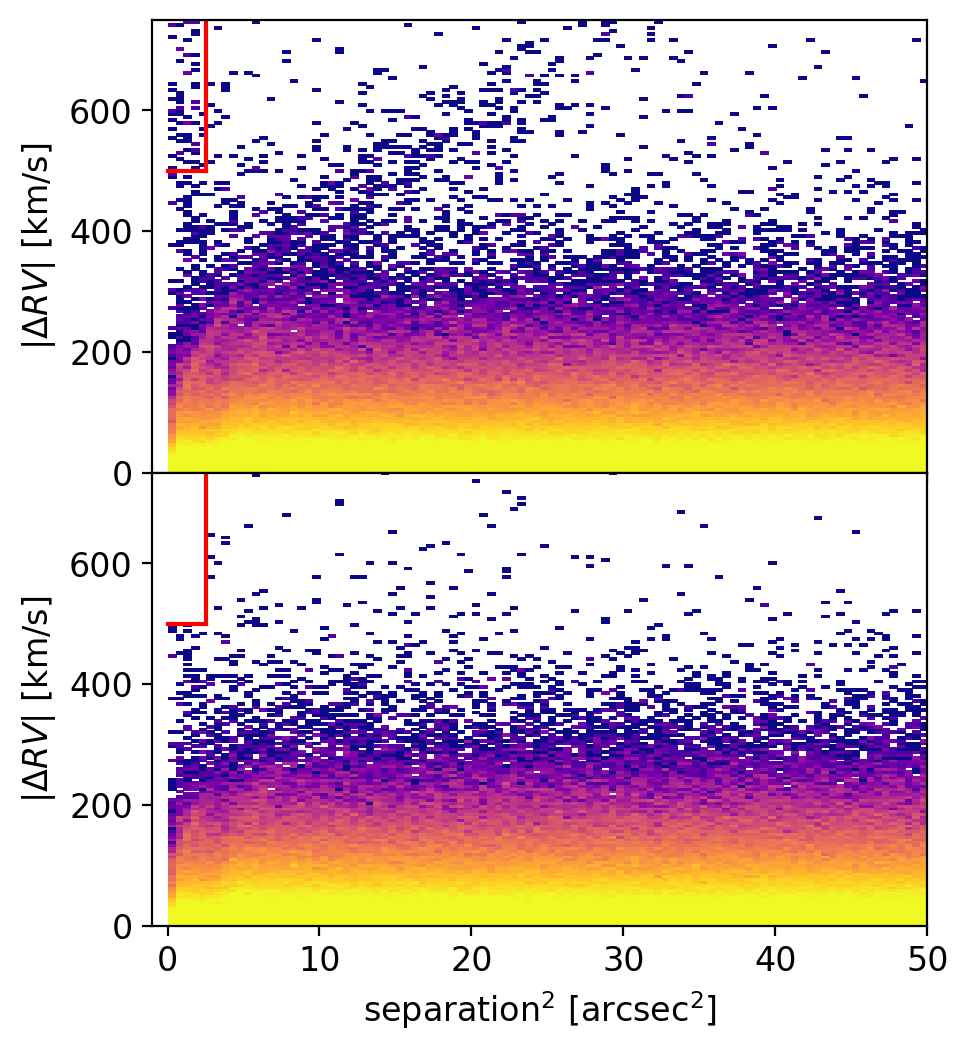}
     \caption{Differences in radial velocities of the members of close pairs of sources as a function of the square of the angular separation in arcsec$^2$ (top) and after filtering (bottom). The red lines enclose one of the criteria that were used to filter the problematic cases.}
    \label{contaminants}
\end{figure}

\subsubsection{Radial velocity contaminants}

During the process of the internal validation of a preliminary version of the
catalogue, we detected erroneous radial velocities due to nearby bright
contaminant sources, which was a well-known issue for \gdr{2} (\cite{2019MNRAS.486.2618B,2021A&A...653A.160S}). This is illustrated in Fig.~\ref{contaminants} (top), where
we took all pairs of sources whose components are closer than 10 arcsec and
plotted their difference in radial velocity (absolute value) versus their
angular separation (squared). 
In this plot, optical pairs should contribute a constant density of points at
a given ordinate, while physical binaries should contribute pairs with small
$RV$ differences. If in a given transit, the dispersion of the spectra is oriented close to
the line of separation of the two sources, the lines from both sources will be
present in both spectra. If in
addition, the lines from the neighbour source are confused with the lines of
the (fainter) target source, this will give the target an erroneous $RV$, which differs
from the $RV$ of its neighbour in proportion to the separation. This will
normally just result in an outlying observation, but if a particular scan
direction dominates, the final radial velocity difference will become 145\kms\
for each arcsecond of separation.  In the upper panel of \figref{contaminants}, we
see a rounded front, a parabola, closely matching the predicted
145\kms\,arcsec$^{-1}$\ dependence. This is a clear sign that source confusion
is occurring. We also note a second, weaker front below the first. It
corresponds to a similar effect, but the two sources are separated 1\farcs8 in
the direction perpendicular to the scan, which is the limit at which two
observations have independent data acquisition.

%We see a rounded front, indicating correlations between these two quantities, and even a second weaker front is observed. This is due to confusion between positions of the bright and faint sources of the pair in the process of deriving the radial velocity of the faint one ({\it rephrase?}). 
%Indeed, the shape of these fronts equals that of the expected error in radial velocity introduced at a certain separation given the radial velocity error per pixel. In general, this produces 
As a result, we have a population of 
false high radial velocities, but also sources with biased radial velocities that are not necessarily very high. Most of these problematic sources have been filtered out of the data released in {\gdrthree} based on the separation and magnitude difference of the pair for sources with $|RV|>200$\kms\ (see \cite{DR3-DPACP-159}). This filtering causes the fronts to almost disappear (Fig.~\ref{contaminants} bottom), although some hints of them still remain. 

In the top plot panel of Fig.~\ref{contaminants}, a vertical band of sources with very large radial velocity differences at very small separations is visible as well. From the pairs with separations smaller than 1.6 arcsec (corresponding to 2.56 arcsec$^2$ in these plots) with velocity differences above and below 500 and $-500$\kms\ (enclosed within the red lines), we filtered out the members with a higher $RV$, which in general correspond to the faint member of the pair. This is a total of 57 stars.

\begin{figure}
    \centering
    \includegraphics[width=0.5\textwidth]{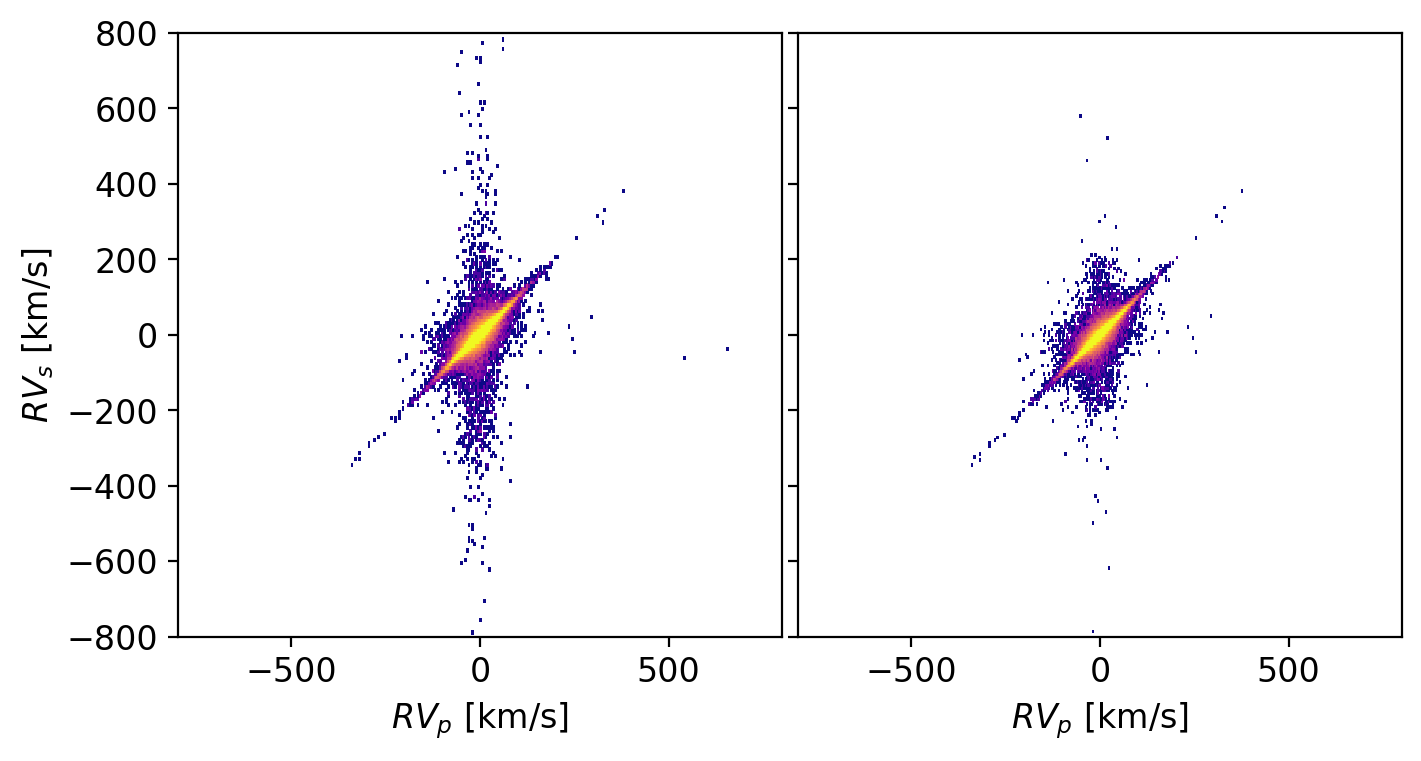}
     \caption{Radial velocity of the primary star against the secondary for stars of binary pairs from the \citet{2021MNRAS.506.2269E} catalogue before (left) and after (right) filtering. }
    \label{fighvb}
\end{figure}

We also used the binary catalogue by \citet{2021MNRAS.506.2269E}  to test the internal consistency of the radial velocities. More than 100\,000 pairs with a probability of 90\% of being bound according to this catalogue have a radial velocity for both members in \gdrthree. The comparison of their radial velocities before filtering (Fig.~\ref{fighvb} left) indicates that most of the pairs closely follow the one-to-one line (agreement of $RV$), but some sources have suspiciously high radial velocities, especially the secondary members. This sample also shows correlated velocity differences and separations in a plot similar to Fig.~\ref{contaminants}. Some problematic high radial velocities still remain in the catalogue after filtering (Fig.~\ref{fighvb} right panel). 

\begin{figure}\begin{center}
\includegraphics[width=0.95\columnwidth]{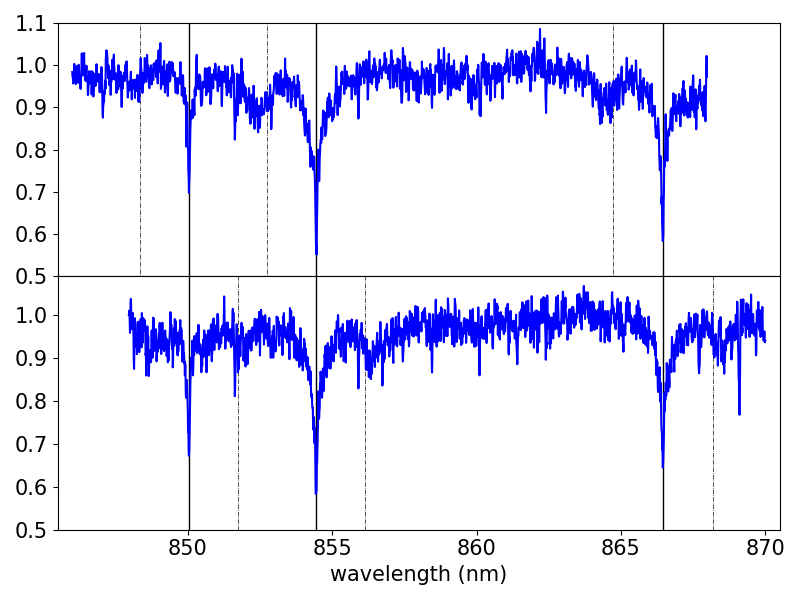}
\caption{Stacked spectra for all the sources with a radial velocity >600~\kms\ (421 sources, top) and <-600~\kms\ (349 sources, bottom). 
Solid vertical lines indicate the position of the calcium triplet, and dashed lines show the same lines shifted by 1.7\,nm, 
indicating where the spectral line would be if the radial velocity correction were incorrect by 600~\kms.}
\label{fig:hrv-stack}
\end{center}\end{figure}

%\textbf{Stacked spectra for HRV targets}\\
The |\fieldName{radial\_velocity}|>600~\kms\ for 770 sources. We expect most of them to be real high-velocity stars, 
but due to the low signal-to-noise ratio of most of the spectra, %(only 78 of them have rvExpectedSigToNoise>5)
it is difficult to know which fraction of the measurements is truly spurious. 
We can stack all of them, however, to improve the summed signal. % get an overall better signal. 
All the spectra were corrected for radial velocity. If the $RV$ value that was used was the correct value, the stacked spectra should have strong lines in the expected 
places such as the calcium triplet lines. This is shown in Fig.~\ref{fig:hrv-stack}. 
However, if the radial velocity that was used for the correction was incorrect, the triplet lines are shifted and appear in the wrong place: 
for $RV>600$~\kms\ , they are at least 1.7\,nm to the left of the expected position, while for $RV<-600$~\kms\ , they appear at least 1.7\,nm to the right (dashed lines). 
These secondary peaks are also seen in the figure. They are less sharp because the incorrect velocity corrections range from 600 to 900~\kms\ (i.e. between 1.7 and 2.6\,nm), 
so they do not all peak in the same place. 
In any case, the figure clearly shows that most measurements are good, that is, most of the sources are real high-velocity targets. 
%\sout{That is confirmed when we stacked the spectra after de-correcting from radial velocity: in that case we still see that the lines due to good RV values are sharper.}

\begin{figure}
    \centering
    \includegraphics[width=0.4\textwidth]{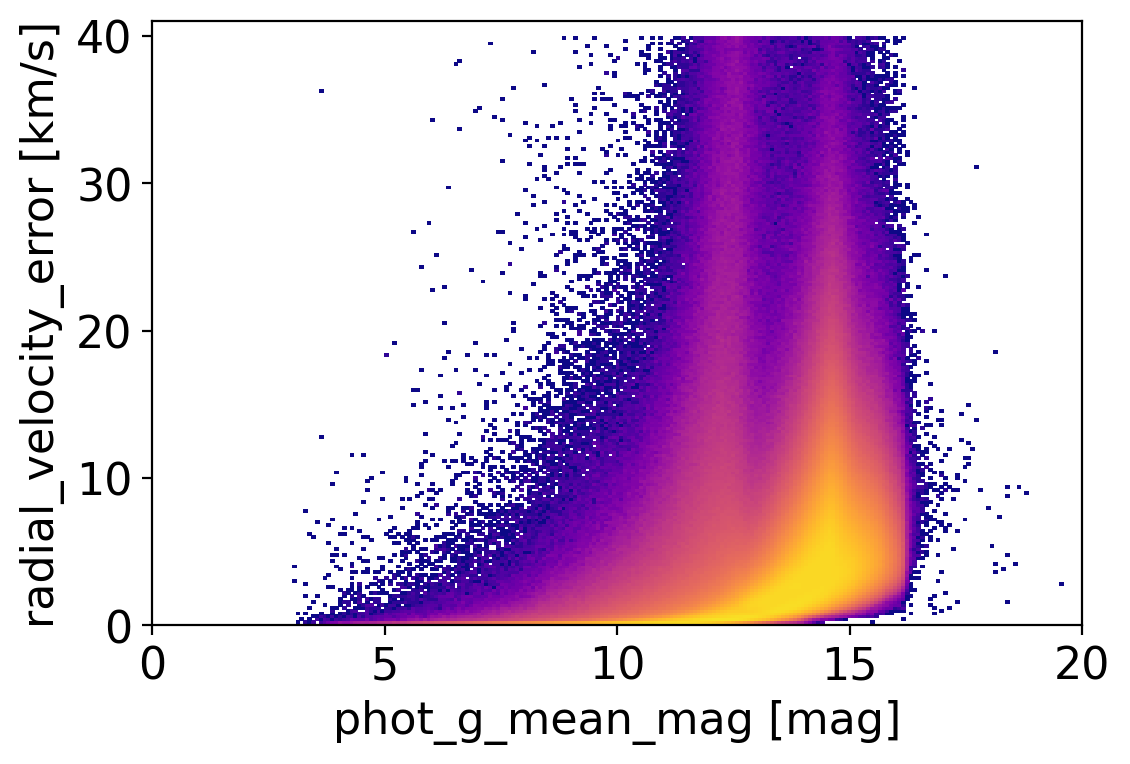}
     \caption{Uncertainty in radial velocity as a function of $G$.}
    \label{fige}
\end{figure}

%\textbf{Faint $G$ sources with Radial velocity}\\
% \sout{The global distribution of radial velocity against $G$ band magnitudes is shown in Fig.~\ref{fig:faint_rvs_dist}. While the distribution is complete down to $G_{\rm RVS}\sim14$, there are a few sources fainter than $G=16$, that are also assigned a radial velocity. For these faint sources ($16<G<19$), the discrepancy between the two $G$ bands is quite large, and in Fig.~\ref{fig:faint_rvs}, we show the 874 sources where $(G_{\rm RVS} - G) < -3$. For a few sources, there is as much as a 6 magnitude discrepancy. The radial velocity distribution for these sources is fairly $Gaussian$-like, and thus not unusual. Additionaly, on the sky the sources are mostly confined to the midplane and towards the Galactic Center, but otherwise uniformly distributed above and below the Galactic plane. However, Fig.~\ref{fig:faint_rvs}(e) shows that, compared to $DR2$, many of these sources were assigned a pseudocolor (astrometric\_params\_solved=95) in $EDR3$. A large $(G_{\rm RVS} - G)$ could indicate contamination from nearby bright stars (affecting the RV estimation) but also (for the faintest $G_{\rm RVS}$) an undersubtraction of the background. While these 874 sources have not been filtered from the $RVS$ sample, we recommend users urge caution while using radial velocity where $(G_{\rm RVS} - G) < -3$.
% }
Radial velocities are provided down to $G_{\rm RVS}=14$. A few sources are still fainter than $G>16$ (see Fig.~\ref{fige}), however.  A high $(G_{\rm RVS} - G)$ could indicate contamination from nearby bright stars (affecting the $RV$ estimation), but also (for the faintest $G_{\rm RVS}$) an under-subtraction of the background.
We recommend caution for radial velocities with $(G_{\rm RVS} - G) < -3$. 
% \begin{figure}
% \includegraphics[width=1.\columnwidth]{figures/sk_rvs_mags_dist.png}
% \caption{\sout{Global distribution on radial velocity against $G_{\rm RVS}$ (panel a), and $G$ (panel b). The RVS is complete down to $G_{\rm RVS}\sim14$, but extends to fainter magnitudes in $G$ for a few potentially spurious sources. \textbf{Extending Fig.\ref{fige} to fainter magnitude could replace this fig.}}}
% \label{fig:faint_rvs_dist}
% \end{figure}

% \begin{figure*}
% \includegraphics[width=2\columnwidth]{figures/sk_rvs_rogue_stars.png}
% \caption{\sout{Faint stars ($16<G<19$), with $(G_{\rm RVS} - G) < -3$ difference, but with a finite radial velocity assigned.}}
% \label{fig:faint_rvs}
% \end{figure*}

\subsubsection{Radial velocity systematics}

\begin{figure}\begin{center}
\includegraphics[width=0.6\columnwidth]{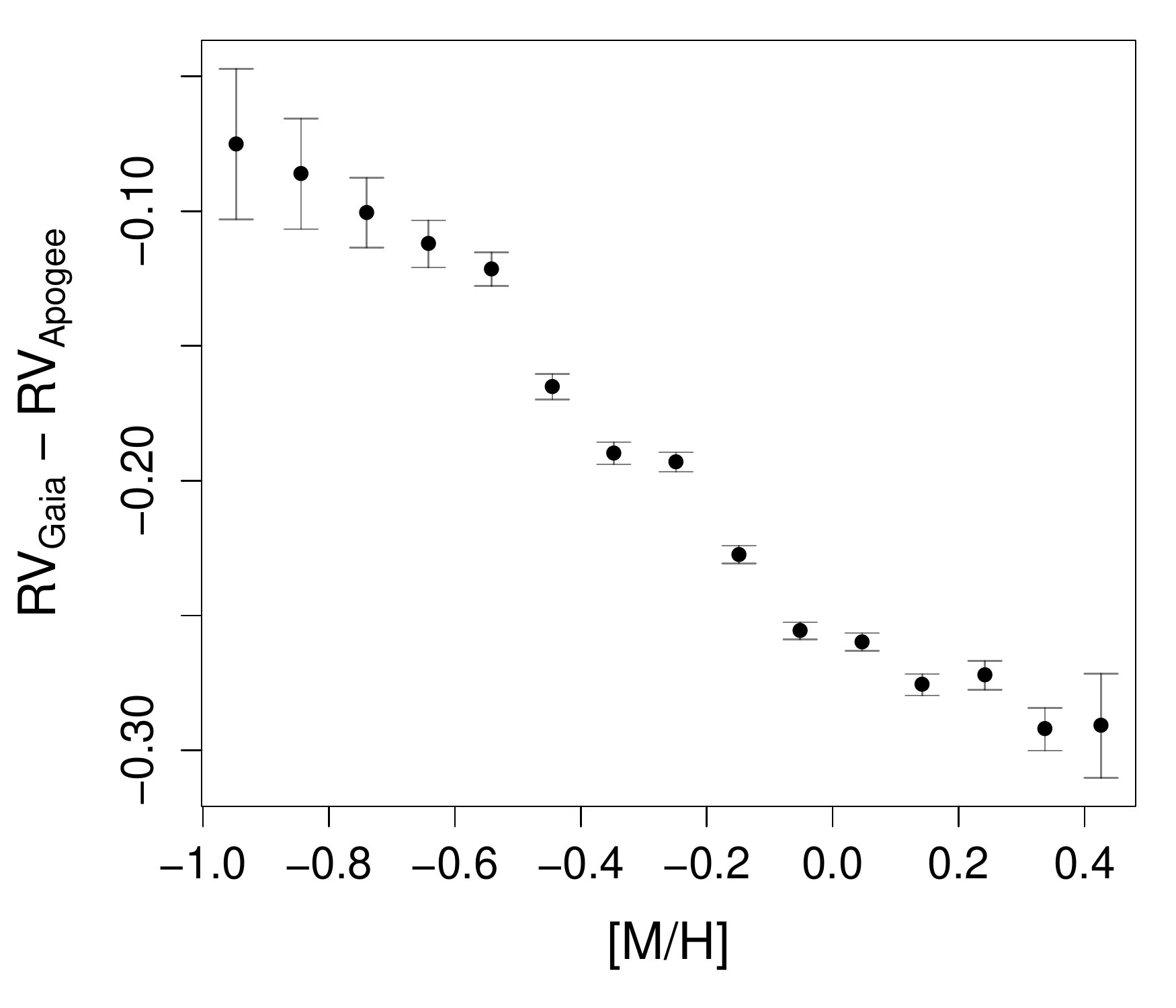}
\caption{Variation in radial velocity difference with APOGEE DR16 as a function of the APOGEE metallicity.}
\label{fig:rv_mh}
\end{center}\end{figure}

%\textbf{Comparison with external catalogues}
Comparison with external catalogues shows that the zero-point of the radial velocities is lower than 0.1~\kms\ than in the radial velocity standard catalogue of \cite{2018A&A...616A...7S}, Carmenes \citep{2020A&A...636A..36L}, and SIM \citep{2015MNRAS.446.2055M}, but it is about -0.2~\kms\ for GALAH DR3 \citep{2021MNRAS.508.4202Z}, APOGEE DR16 \citep{2020ApJS..249....3A}, and GES DR3 \citep{2012Msngr.147...25G}. The number of 5$\sigma$ outliers is smaller than 3\%. The radial velocity zero-point shows a decrease with metallicity in all surveys illustrated in Fig.~\ref{fig:rv_mh}. The global change in radial velocity with magnitude is not consistent across the surveys. However, \cite{DR3-DPACP-159} used subsamples on which they found a consistent trend between APOGEE and GALAH and propose a magnitude-term correction. For stars with \fieldName{rv_template_teff}$>$8500~K, the correction derived in \cite{DR3-DPACP-151} is to be used instead. 
 
\begin{figure}
\begin{center}
\includegraphics[width=8cm]{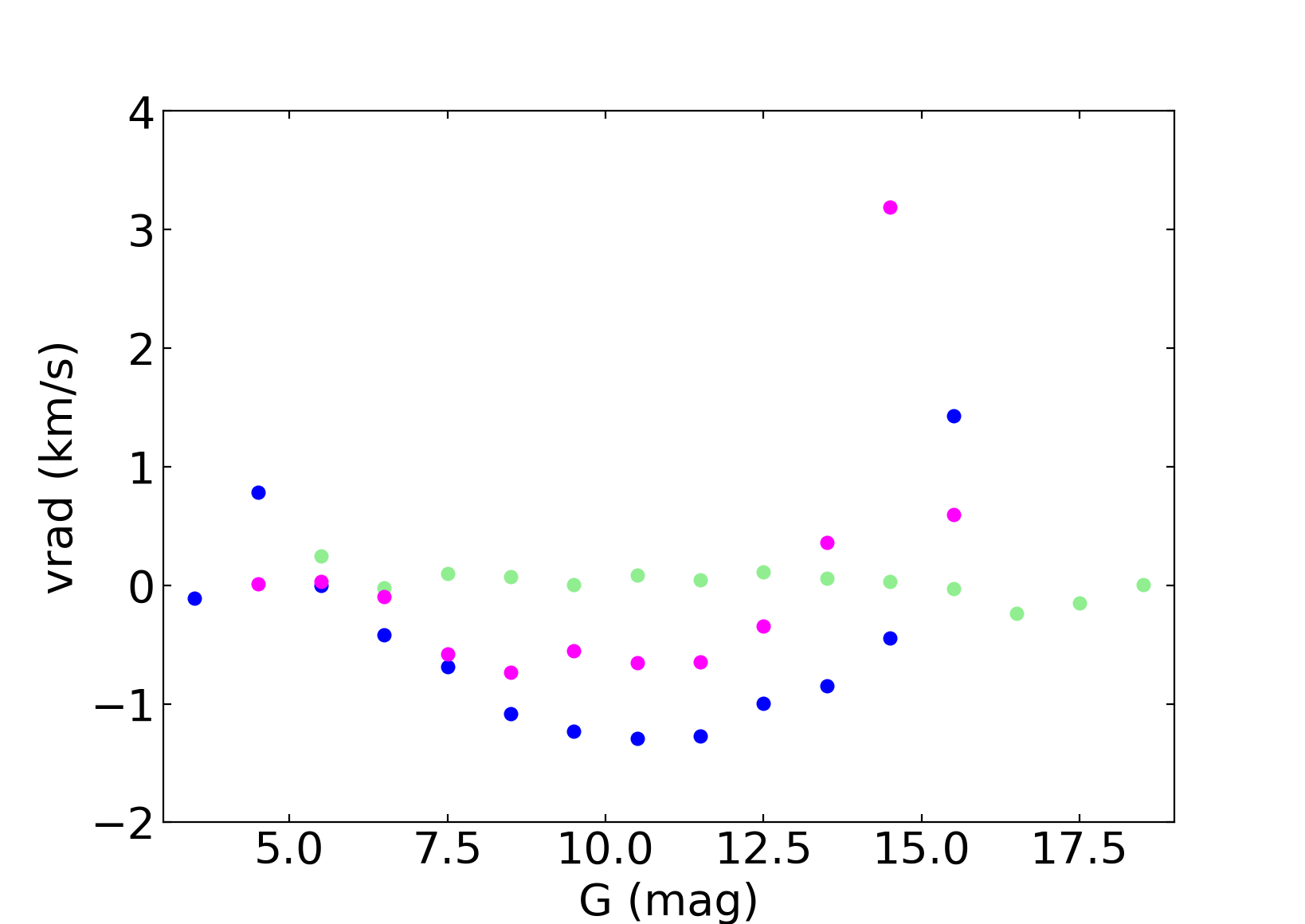}
\caption{Radial velocities averaged over the whole sky as a function of $G$ magnitude for DR3 (blue), GOG20 (green), and EDR3 (pink).}
\label{fig:vrad-mag}
\end{center}
\end{figure}

%\textbf{Comparison with model}
The comparison of the median radial velocity with the GOG model does not show any systematic significant difference throughout the sky, at least none that can be attributed to the data themselves. 
Figure~\ref{fig:vrad-mag} shows the median value of the radial velocity throughout the whole sky and per magnitude bin for DR3, EDR3, and GOG20. 
From $G=4$ to $G=15$, the DR3 values agree with GOG20 at the level of 1.5\kms\ or lower. We note that it was at the level of 1\kms\ or lower with EDR3.
The EDR3 values are not reliable at $G>13$ because there are too few stars. This limit is pushed to $G>15$ for DR3.
The dependence on the $G$ magnitude seen in the data is not predicted by the model and might indicate some systematics in the data or in the model that are not yet understood.

\subsubsection{Radial velocity uncertainties\label{sec:rverrors}}

The different methods that were used to compute the radial velocity (indicated by \fieldName{rv\_method\_used}, see \citealt{DR3-DPACP-159}) lead to different error distributions as a function of magnitude (Fig.~\ref{fige}). 
In particular, the limit of using one method or the other is \grvs = 12, which produces the plume of large errors at $G\sim12$. 

\begin{figure}
 \begin{center}
\includegraphics[width=0.95\columnwidth]{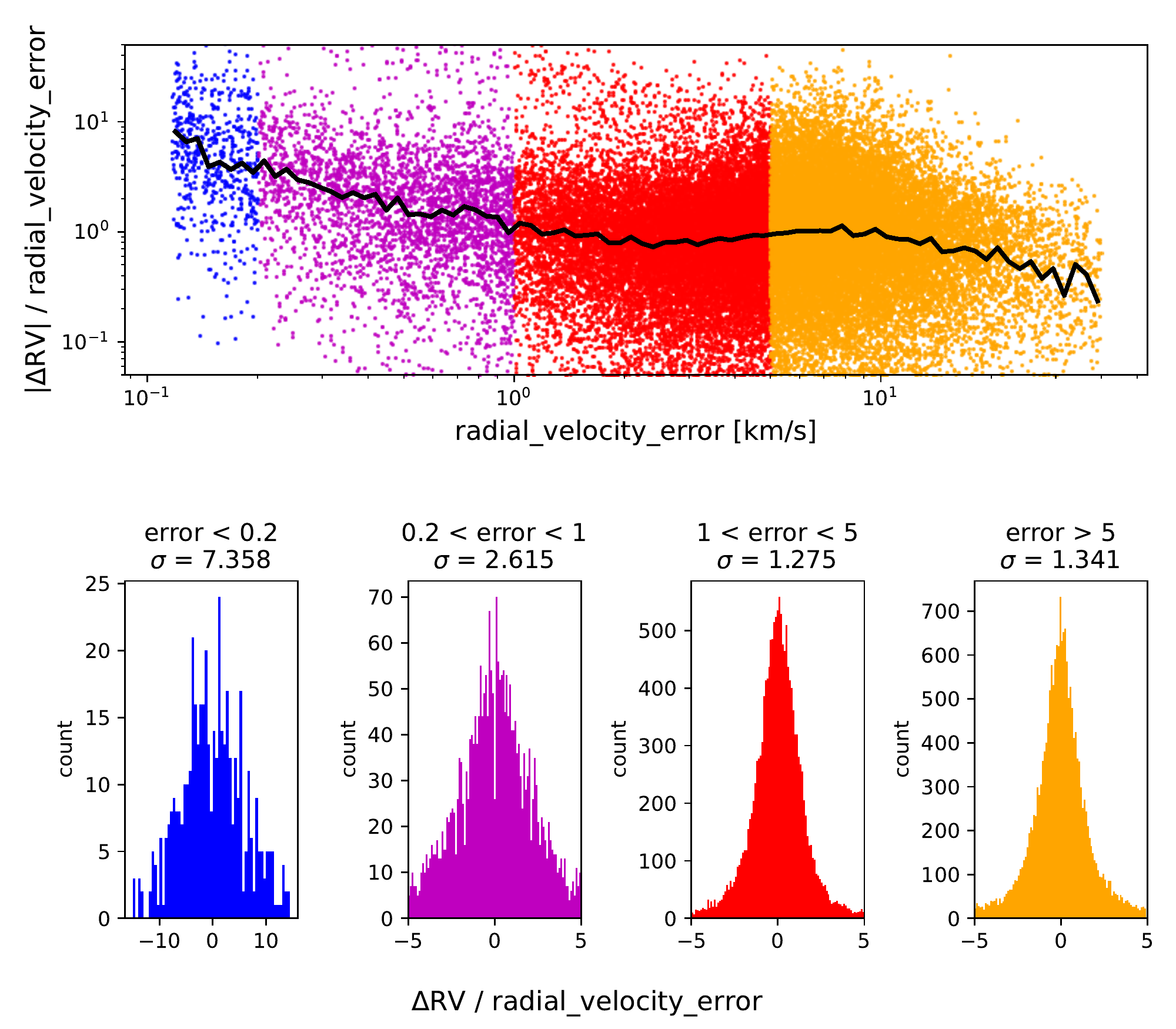}\\
\end{center}
\caption{Radial velocity uncertainties tested with open clusters. Top panel: Absolute value of the  difference between the radial velocity of a star and its cluster median $|\Delta RV|$ normalised by the \fieldName{radial\_velocity\_ error}. The black line is the lowess (locally weighted scatterplot smoothing). 
%The lowess  $\sim 1$ for errors $>1-2 $ km/s indicated that the errors are correctly estimated. 
The slope of the lowess for lower values of radial\_velocity\_error indicates that the errors can be underestimated at the bright end (but see the text for a discussion). Bottom panel: Difference between the radial velocity of a star and its cluster median $\Delta RV$ normalised by the radial velocity error in different radial velocity error bins. 
%Difference between the radial velocity of a star and it's cluster median $\Delta RV$ normalised by the radial velocity error in %different radial velocity error bins. 
%\textbf{Difference between the radial velocity of a star and it's cluster median $\Delta RV$}. top panel: \textbf{radial velocity difference normalised by the error vs radial velocity error}. The black line is the lowess. Bottom: distribution of the \textbf{normalised radial velocity difference} in different error bins indicated by the different colors. 
%\comment{using |DeltaRV| is confusing as it makes a quick look see an offset with magnitude. Top plot removed.}
% Color code is the same of the top panel.
%\comment{labels hard to read}
}\label{fig:rvuncertainties3}
\end{figure}

%clusters
%\comment{Referee: The section I had most difficulty with is 2.1.3, where cluster stars are used to assess errors on radial velocities. I find the number of 804 clusters used in this context unexpectedly large. It is not clear if these are only open clusters or also globular clusters. In my experience, the distinction of cluster members based on parallax and proper motion degrades rapidly beyond about 2 kpc, in particular for open clusters. When membership is based on a minimum of 10 stars, are these ten stars all brighter than Magn.14, as used in RVS? If not, what is the mean radial velocity based on for these clusters? For globular clusters the practical limit is more like 10 to 15 kpc, when crowding as well as the limiting magnitude cause restrict studies. Even though internal velocity dispersion is mentioned in the text, it is not recognized as a contributor to the data as displayed in fig.7. In open clusters this amounts to 0.5 to 1.0 km/s typically, in a globular cluster! it can easily get as high as 20 to 30 km/s.}
We tested the uncertainties on radial velocity given by \fieldName{radial\_velocity\_error} using 48 944 stars in 804 open clusters in which  at least ten members were brighter than \grvs$<$14 with radial velocities. Membership was derived using parallaxes and proper motions. The sample used for this test includes the open clusters catalogued by \citet{2020A&A...640A...1C} and new clusters in \citet{2021arXiv211101819C}.
These clusters  are typically closer and more populated than average: 70\%
are located within 2~kpc of the Sun, and 50\% of them have more than 140 identified members.
We computed the difference $\Delta RV$  between the radial velocity  of each star and the bulk cluster radial velocity (defined as the median of RVS radial velocities), and we compared this value with the nominal uncertainty of each star. The results are shown in  Fig.\ref{fig:rvuncertainties3}.
In the ideal case,  $\Delta RV$/\fieldName{radial\_velocity\_error} should follow a normal distribution (centred on zero and with a dispersion of 1),  but Fig.~\ref{fig:rvuncertainties3} shows that bright stars (with high \fieldName{rv\_expected\_sig\_to\_noise} and  small \fieldName{radial\_velocity\_error}) tend to have a much broader dispersion than faint stars. 
It should be noted that several effects can broaden the distribution, such as the gravitational redshift and convective blueshift which affect stars of  different spectral types differently, \sout{and also affect} unrecognised binaries, non-members that are still present in the distribution, and the intrinsic internal dispersion of the clusters \sout{in different ways}.  While some of the effects are difficult to quantify,  the internal velocity dispersion is about 0.5-1 \kms \citep[see e.g.][]{2021ApJ...921..117T}.  It is difficult to produce diagnostics on a per-cluster
basis because the effect is revealed in a statistical way. However, the
pattern seems to be identical for all clusters, and this is in favour of an instrumental effect.

%This is not due to a resolved velocity dispersion, since only distant clusters are selected. 
%In practice, the stars with very small nominal uncertainties are 5 to 10 error bars away from the cluster velocity.
%The nominal uncertainties for stars with errors from 2 km/s to 10 km/s seem realistic as it is clear from Fig.\ref{fig:rvuncertainties3} (top panel), where the lowess (Locally Weighted Scatterplot Smoothing) is close to 1.
%A similar analysis using rv\_mean\_of\_error \comment{???} shows that uncertainties are undestimated for G$< 11.5$ and overestimated for G$>12.8$.

\begin{figure}
    \centering
    \includegraphics[width=0.49\columnwidth]{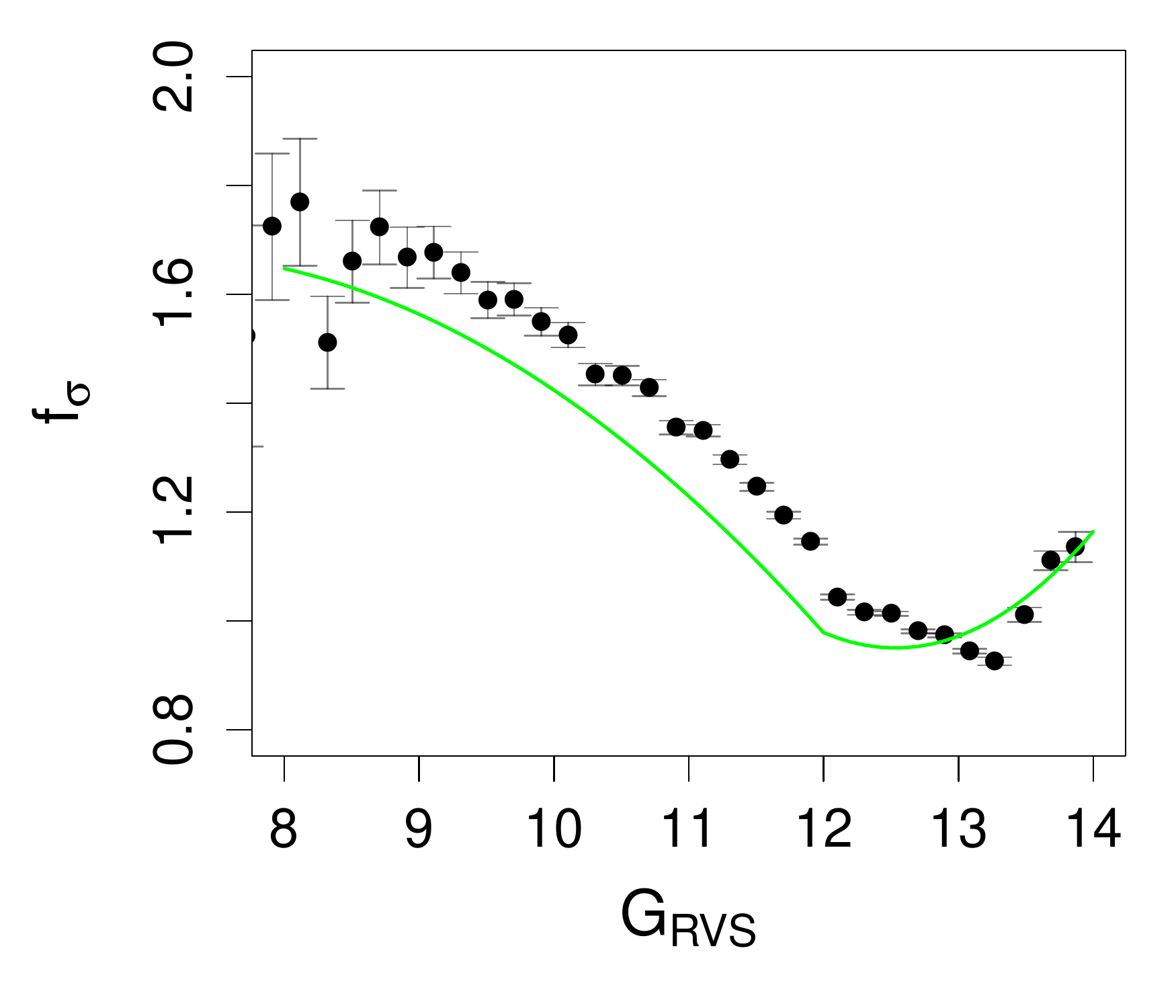}
    \includegraphics[width=0.49\columnwidth]{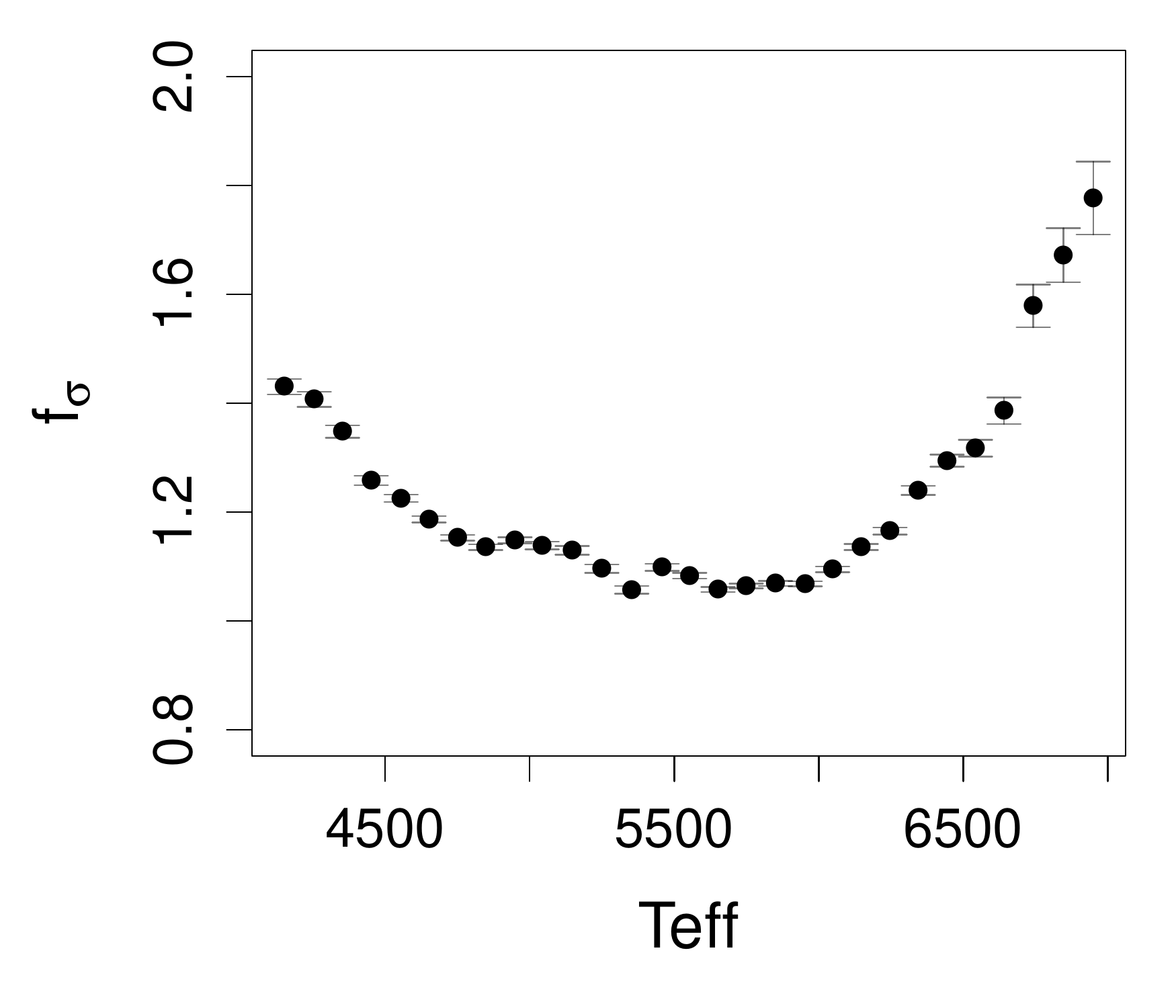}
     \caption{Standard error factor $f_\sigma$ that should be applied to \fieldName{radial_velocity_error} as a function of magnitude (left) and temperature (right), estimated from the comparison with GALAH. In green we over-plot $f_\sigma$ estimated from the wide binaries (Eq.~\ref{eq:rv_err} and Table~\ref{tab:rvserrorfactor}).}
    \label{fig:rvserrorfactor}
\end{figure}

Comparison with external catalogues and the wide binary catalogue of \cite{2021MNRAS.506.2269E} confirms that the errors are underestimated for \grvs$<12,$ but also for \teff$<4500$ and \teff$>6000$~K (Fig.~\ref{fig:rvserrorfactor}). 
As the external catalogues provide different error underestimation estimates due to their own error estimation uncertainties, we used the wide binaries to estimate a correction. To limit the impact from the gravitational redshift, we selected stars with similar colours and magnitudes, using a difference of 0.1~mag in \bprp\ and $G$. To avoid the additional dependence on the temperature, we selected only systems in which both components lay within 4500$<$\fieldName{rv\_template\_teff}$<$6000~K. We further removed 5$\sigma$ radial velocity outliers. This led to a total of 2452 systems that could be used. According to the APOGEE and GALAH comparison,  we split the fit into bright and faint regimes at \grvs=12~mag (which corresponds to the magnitude separation between the different methods that were used to derive the radial velocity) and fitted a second-order polynomial to the factor $f_\sigma$ to apply to the standard deviation, 
 \begin{equation}
  f_\sigma(G_\mathrm{RVS}) = a + b~G_\mathrm{RVS} + c~G_\mathrm{RVS}^2
  \label{eq:rv_err}
 ,\end{equation}
 by maximising the product of the likelihoods of 
 \begin{equation}
 \frac{RV_1-RV_2}{\sqrt{(f_\sigma(G_\mathrm{RVS1}) \sigma_\mathrm{RVS1})^2+(f_\sigma(G_\mathrm{RVS2}) \sigma_\mathrm{RVS2})^2}}
 \end{equation} 
 to be normally distributed. The coefficients we obtained are illustrated in Fig.~\ref{fig:rvserrorfactor} and are provided in Table~\ref{tab:rvserrorfactor}. The bright side is not constrained, so that it should not be extrapolated beyond \grvs$<8$~mag. Based on the comparison with \cite{2018A&A...616A...7S}, the value at \grvs=8~mag seems to be a good estimate for \grvs$<8$~mag. The \teff ranges showing a strong departure in Fig.~\ref{fig:rvserrorfactor} correspond to systematic offsets between \fieldName{rv\_template\_teff} and  the GALAH temperature. For cool stars, the deviation with APOGEE is found only for $\teff<4000$~K. The effect of the random temperature template mismatch is included in the correction provided in Table~\ref{tab:rvserrorfactor}, but not the systematics as we used an internal comparison. On the range 4500$<$\fieldName{rv\_template\_teff}$<$6000~K, the median absolute deviation between \fieldName{rv\_template\_teff} and the GALAH \teff is 250~K. 
 When the correction of Table~\ref{tab:rvserrorfactor} was applied to wide binaries with a \teff template that was hotter and cooler, no correlation of an additional factor with \fieldName{rv\_template\_teff} was detected.
 
 \begin{table}
 \footnotesize
 \caption{Coefficients to derive the standard error factor $f_\sigma$ that should be applied to \fieldName{radial_velocity_error} according to Eq.~\ref{eq:rv_err} for $G_\mathrm{RVS}>8$~mag.}
  \begin{tabular}{lll|lll}
  \multicolumn{3}{c}{$G_\mathrm{RVS}<12$} & \multicolumn{3}{c}{$G_\mathrm{RVS}>12$} \\
  \hline
  a & b & c & a & b & c \\
  0.318 & 0.3884 & -0.02778 & 16.554 & -2.4899 & 0.09933 \\
  \end{tabular}
  \label{tab:rvserrorfactor}
 \end{table}

%%%%%%%%%%%%%%%%%%%%%%%%%%%%%%%%%%%%%%%%%%

\subsection{Vbroad}

\begin{figure}\begin{center}
\includegraphics[width=0.8\columnwidth]{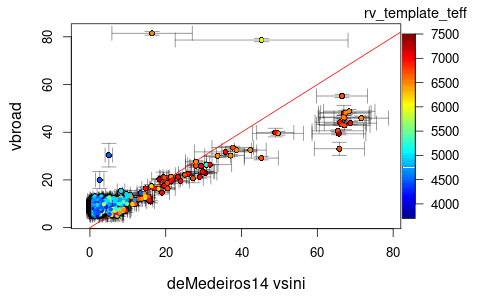}
\caption{Comparison of the spectral line broadening parameter with the \cite{2014A&A...561A.126D} catalogue of FGK stars, colour-coded by the template temperature.}
\label{fig:vbroad_ext}
\end{center}\end{figure}

The estimation of the line-broadening parameter, \fieldName{vbroad}, is detailed in \cite{DR3-DPACP-149}.
The comparison of the spectral line-broadening parameter \fieldName{vbroad} with external catalogues shows that values lower than $\sim10$\kms\ are systematically overestimated,
while higher values tend to be underestimated for FGK stars, as illustrated in Fig.~\ref{fig:vbroad_ext}, which shows a similar behaviour as the comparison with GALAH \citep{2021MNRAS.508.4202Z}. More details about the validation of the spectral line-broadening parameter can be found in \cite{DR3-DPACP-149}.

\subsection{Grvs magnitude}

\begin{figure}\begin{center}
\includegraphics[width=0.8\columnwidth]{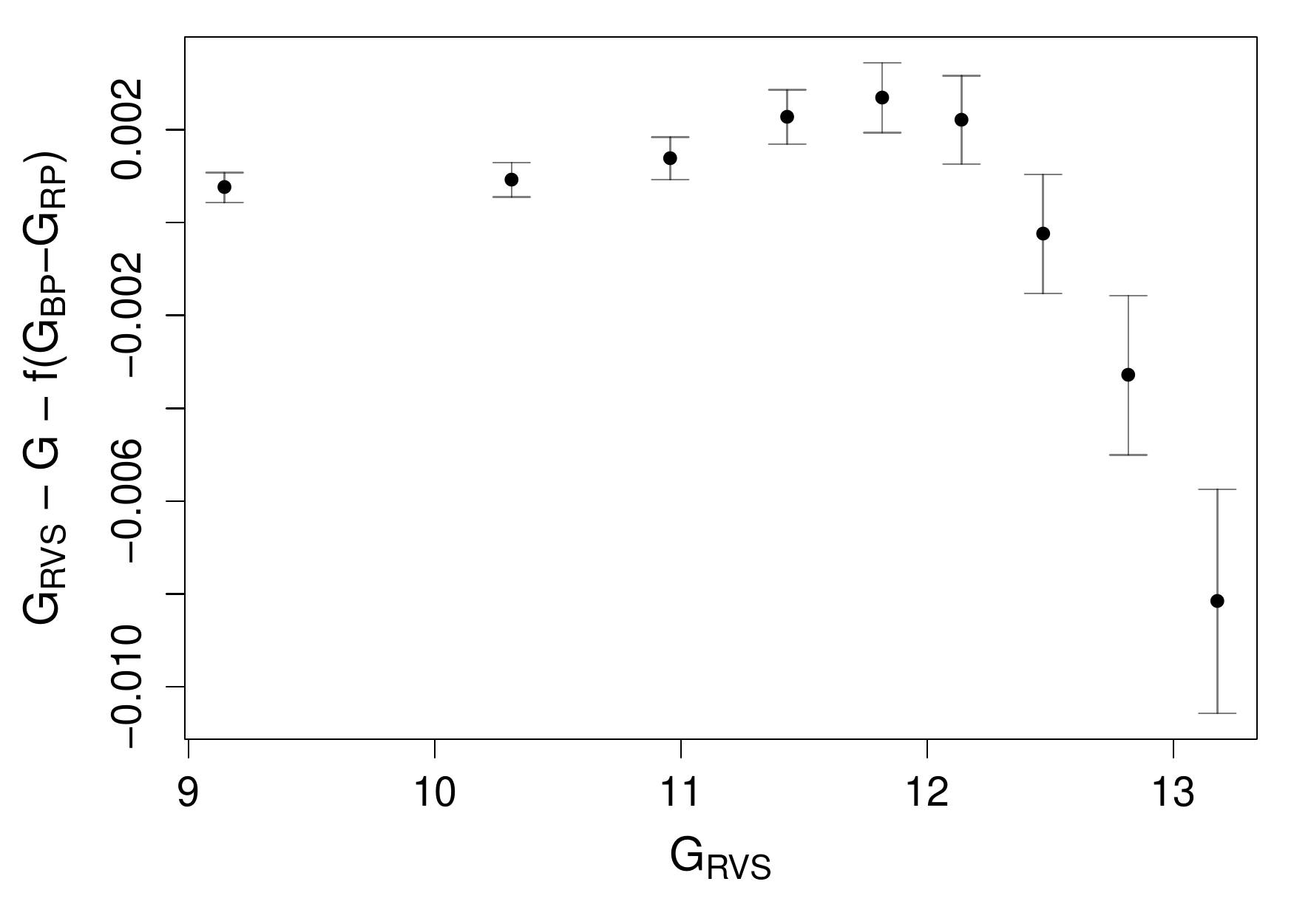}
\caption{Residuals from a global relation $G_{\rm RVS}-G=f(G_{\rm BP}-G_{\rm RP})$  for a sample of APOGEE low-extinction solar metallicity dwarfs.}
\label{fig:grvs_systematics}
\end{center}\end{figure}

The estimation of the {\grvs} magnitude, \fieldName{grvs\_mag}, is detailed in \cite{DR3-DPACP-155}.
The comparisons of {\grvs} with the Hipparcos magnitude and Tycho2 colours indicate no saturation issues. 
The comparison of {\grvs} with {\gaia} $G$ magnitude and \gbp-\grp\ colour shows a change in behaviour at \grvs$>12$. To illustrate this, we used here a sample of solar metallicity dwarfs selected from APOGEE DR16 \citep{2020ApJS..249....3A} that had low extinction ($A_0<0.05$~mag according to \cite{2019A&A...625A.135L}). An empirical robust spline regression was derived to model the global relation of $G$-\grvs\ versus \gbp-\grp\ . The residuals from this spline are plotted as a function of magnitude in Fig.~\ref{fig:grvs_systematics}. The effect appears to be much larger than the internal variations observed with the $G$, \gbp\ , and \grp\ magnitudes \citep{EDR3-DPACP-126}, but are still only at the 10~mmag level. 
%\comment{Referee: I think that part of the problem here is that G(RVS) and (BP-RP) are not uncorrelated, and a spline function fit of G(RVS)-G as a function of BP-RP will show in the residuals the positioning of the knots when plotted as a function of G(RVS), as it would show when plotted as a function of BP-RP.}

\begin{figure}
\begin{center}
\includegraphics[width=0.8\columnwidth]{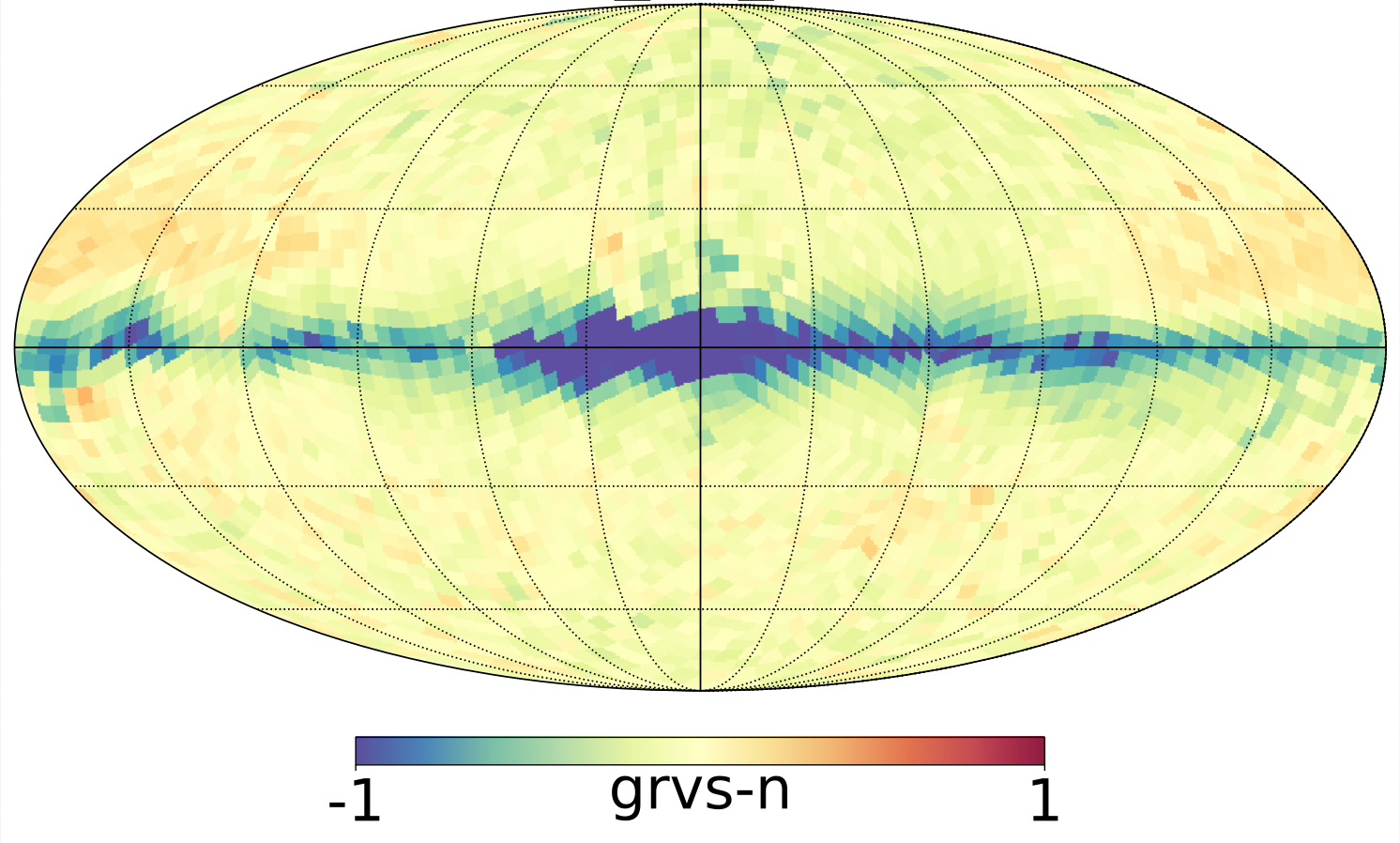} 
\caption{Relative difference of the number of stars with a {\grvs} value between DR3 and GOG20 (DR3-GOG20)/DR3 in the magnitude range $12<G<13$ in Galactic coordinates. -1 (+1) corresponds to a deficit (an excess) of 100\% in DR3 data with regard to the GOG20 model.}
\label{fig:grvs-sky}
\end{center}
\end{figure}

Figure~\ref{fig:grvs-sky} shows the relative difference between DR3 and GOG20 in the $G=12$ to 13 magnitude range. The agreement is very good, except in the bulge and Galactic plane, where the excess of simulated stars is large. This is expected as only sources with unblended spectra were used to estimate {\grvs} \citep{DR3-DPACP-155}. Exploring these maps at other magnitude bins shows that the completeness for {\grvs} measurements is still high at $G$ = 14 outside of the Galactic plane, and at fainter magnitudes, the star counts start to drop. In the Galactic plane, the data already start to be incomplete at $G$ = 11. 

\subsection{RVS spectra}

\begin{figure}\begin{center}
\includegraphics[width=0.8\columnwidth]{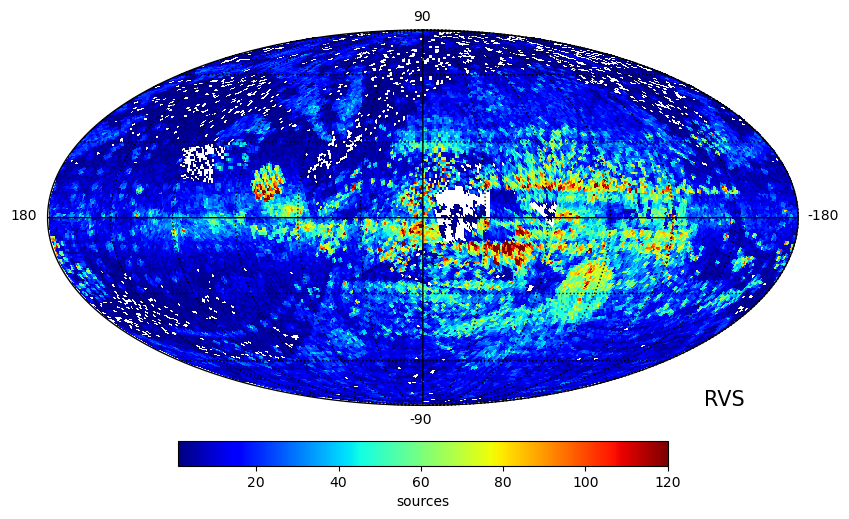}
\caption{Galactic distribution of the sources for which RVS spectra are available in the HEALPix map of order 6. White patches are regions without sources.}
\label{fig:spec_lb}
\end{center}\end{figure}

The main properties of the RVS spectra, available through the \tableName{rvs\_mean\_spectrum} datalink table, are described in \cite{DR3-DPACP-154}.
The sky distribution of the sources with spectra, presented in \figref{fig:spec_lb}, is non-uniform. There are patches with a higher density of sources, and some regions are basically empty. 
More details about how this sample was selected are given in \cite{DR3-DPACP-154}.

\begin{figure}\begin{center}
\includegraphics[width=0.98\columnwidth]{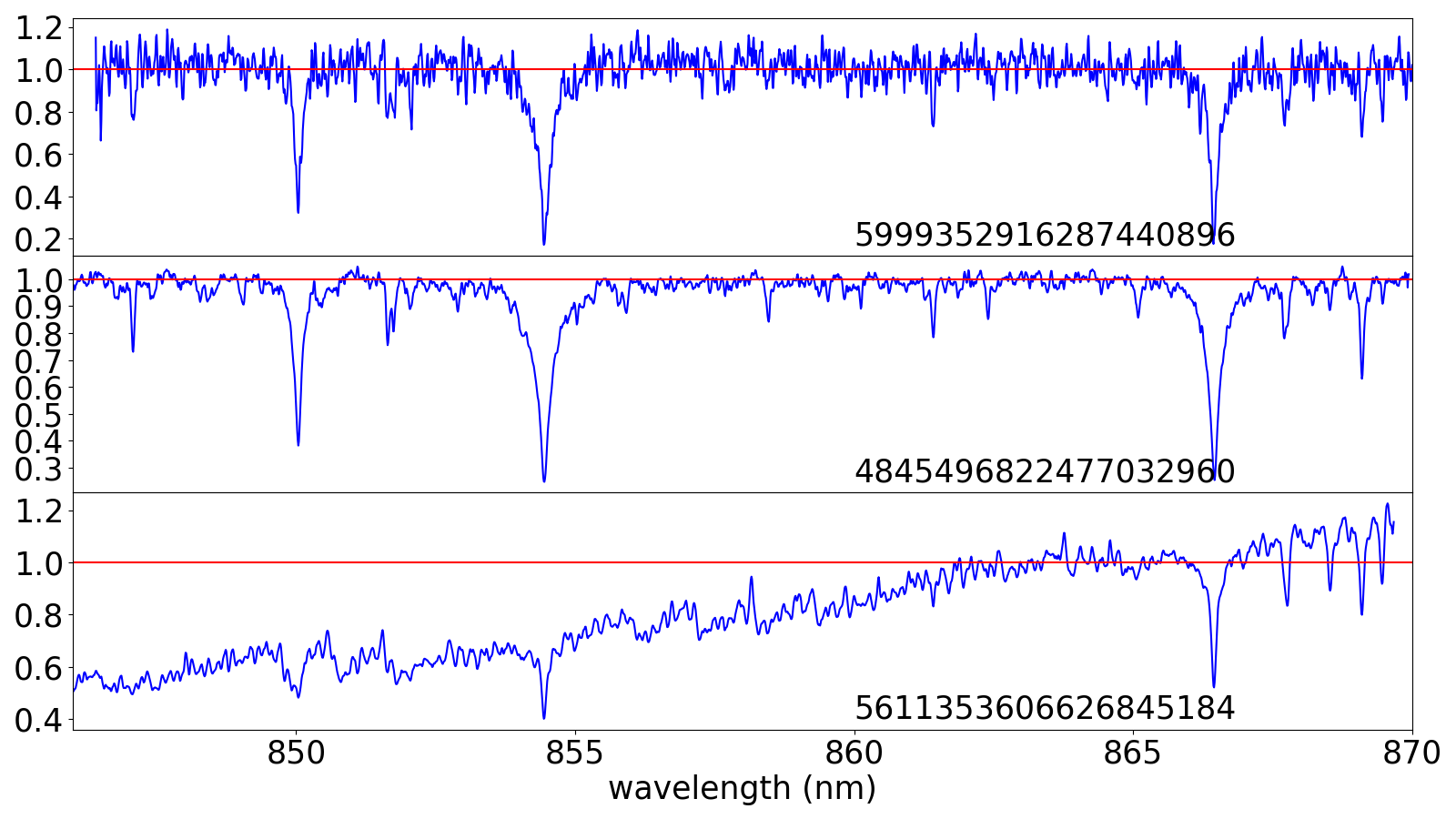}
\caption{Three example spectra with three different continuum levels.}
\label{fig:spec_cont}
\end{center}\end{figure}

The continuum calibration of the spectra was performed using different methods, which resulted in different continuum levels. 
For faint targets, \grvs\ $>12$ (or {\tt rv\_method\_used}=2), the method set the median value of the flux at 1, which forces the continuum to be slightly above this value 
(see \figref{fig:spec_cont} top). For brighter targets (or {\tt rv\_method\_used}=1), the continuum is slightly below 1 (see \figref{fig:spec_cont} middle). When the target is red, the continuum is not
flattened, and a positive slope is visible (see \figref{fig:spec_cont} bottom).

%+++++++++++++++++++++++++++++++++++++++++++++++++++++++++++++++++++++++++++
\section{Spectrophotometry}\label{sec:xp}
%+++++++++++++++++++++++++++++++++++++++++++++++++++++++++++++++++++++++++++

% DESCRIPTION OF THE PUBLISHED XP SPECTRA DATASET
% ===============================================

{\gdrthree} provides low-resolution spectral data for about 220 million sources for the first time. These data consist of two sets of coefficients with the corresponding uncertainties and correlation matrices, available in the \tableName{xp_continuous_mean_spectrum} table through the datalink interface\textsuperscript{\ref{datalink}}. One set of coefficients is for the BP instrument, and the other set is for the RP instrument. The only exception is DR3 \fieldName{source\_id}$=5405570973190252288$. This very red and faint source only has an RP spectrum. The coefficients are the development of a spectrum in basis functions for the internal spectrum in units of electrons per second per pseudo-wavelength within the {\gaia} aperture as a function of pseudo-wavelength  \citep{DR3-DPACP-118}. Externally calibrated spectra can be obtained through the GaiaXPy tool\footnote{https://gaia-dpci.github.io/GaiaXPy-website/} (see also the cosmos pages\footnote{https://www.cosmos.esa.int/web/gaia/auxiliary-data} for the configuration files that allow producing these externally calibrated spectra).
For a subset of the spectra with $G<15$, these externally calibrated sampled spectra are available directly in the \tableName{xp\_sampled\_mean\_spectrum} table through the datalink interface. However, the direct usage of the coefficients is strongly recommended \citep{DR3-DPACP-118}. GaiaXPy can also be used to transform an external spectrum into the Gaia XP (shortcut for BP and/or RP) continuous representation.
A detailed description of the data is provided by \cite{DR3-DPACP-118} for the internal spectra, and by \cite{DR3-DPACP-120} for the external spectra. Tests were performed to ensure the validity of the spectrophotometric data, both for internal and external spectra. The tests are described below.

\subsection{Ensemble properties of the spectrophotometric data}

\begin{figure}[h]
        \centering
        \includegraphics[width=0.9\columnwidth]{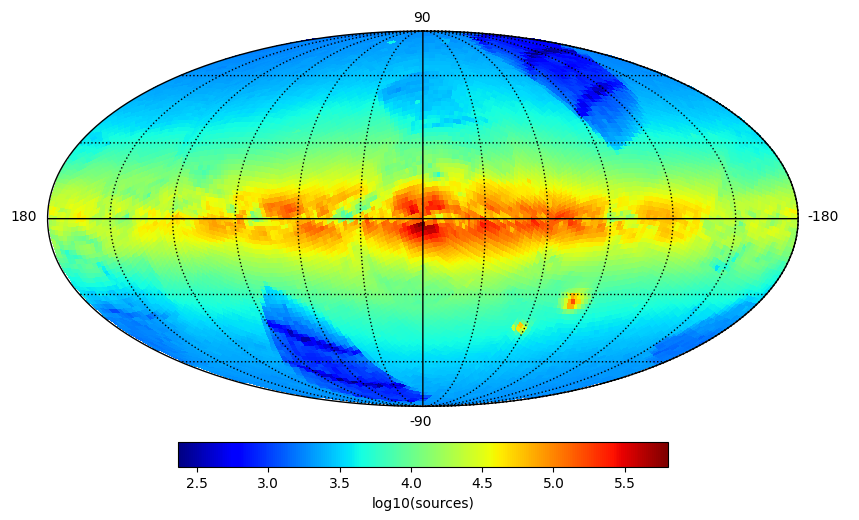}
        \caption{Density of {\gdrthree} sources in the sky with available \tableName{xp_continuous_mean_spectrum} in Galactic coordinates.
        \label{fig:wp942_650_000_skyplot}
        }
\end{figure}

Figure~\ref{fig:wp942_650_000_skyplot} shows the sky density distribution for all sources with \tableName{xp_continuous_mean_spectrum} in {\gdrthree}.
In addition to the natural variation in source density, several distinct regions with a lower source density are seen. The natural variation includes high densities along the Galactic plane and in particular towards the Galactic centre, and decreasing densities towards the Galactic poles.
These artificial patterns in the sky distribution of the sources result from the selection process of spectra to be published in {\gdrthree}, in particular, the requirement of at least 15 observations.
\par
\begin{figure}[h]
        \centering
        \includegraphics[width=0.9\columnwidth]{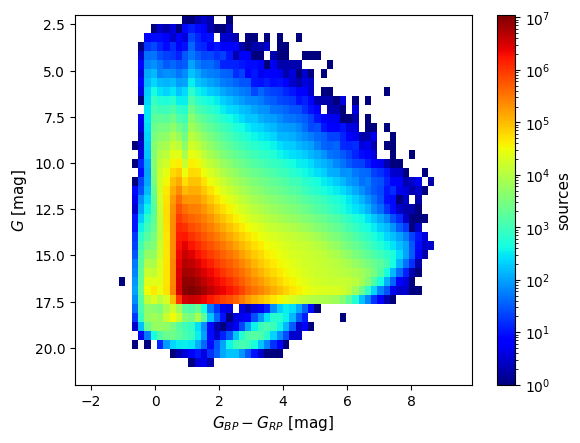}
        \caption{Magnitude-colour diagram for sources for which \tableName{xp_continuous_mean_spectrum} is available in {\gdrthree}.
        \label{fig:wp942_650_000_GvsBPRP}
        }
\end{figure}

Figure \ref{fig:wp942_650_000_GvsBPRP}  shows the distribution of the sources with XP spectra in the colour -- apparent magnitude diagram. The natural increase in the number of sources at fainter apparent magnitudes is clearly visible. In addition to this, artificial structures are superimposed. For a G magnitude brighter than 17.65, all available XP spectra are included in \gdrthree, while for fainter sources, only a subset is included, with a focus on red sources. This results in the break in the distribution at G = 17.65, and in the generally smaller number of sources at larger magnitudes, with  a larger proportion of red sources. A detailed description of the selection process is provided by \cite{DR3-DPACP-118}.

\subsection{Tests of source coefficients}

% COEFFICIENTS ABSOLUTE VALUE
% ===========================
The first test we performed on the coefficients of the \xp spectra determined the stability of the representation of internal spectra. The internal BP and RP spectra are represented by a linear combination of basis functions, and the integrated flux of a source is thus a linear combination of the integrals of the basis functions over the entire real axis. The absolute values of these integrals are about one, therefore we might expect the absolute values of the coefficients of the spectrum of any particular source to not be significantly higher than the integrated flux of the source.
Coefficients that are very large compared to the integrated flux would indicate that the source spectrum is a linear combination of basis functions that mostly cancel each other out, and would thus be an indicator of an unstable representation of the internal spectrum. We compared the absolute values of all coefficients of all sources with the integrated flux, and they are all lower than 3.8 times the integrated flux. In most cases, the values are significantly lower. We therefore see no indications for excessively large contributions from different basis functions to internal spectra that are cancelling each other out.\par

\begin{figure}
        \centering
        \includegraphics[width=0.9\columnwidth]{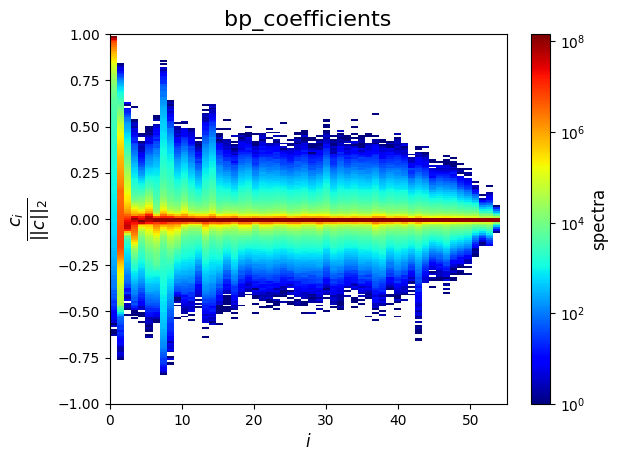}
        \includegraphics[width=0.9\columnwidth]{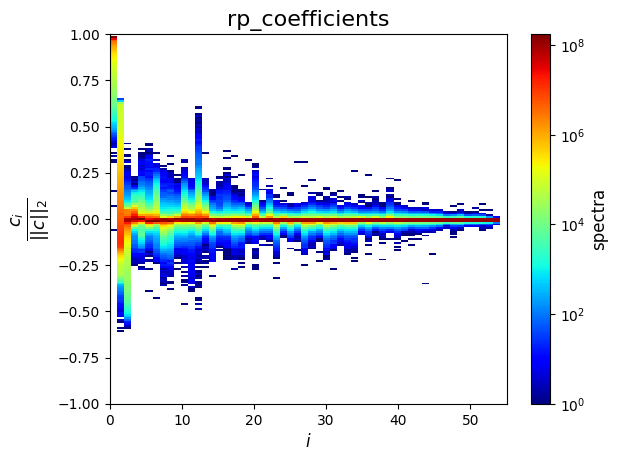}
        \caption{Source mean spectrum coefficients for all sources in the \tableName{xp_continuous_mean_spectrum} table for BP (top) and RP (bottom). The colour index indicates the source density.}
        \label{fig:wp942_650_510_decreasing_coefsXP}
\end{figure}

% DECREASING COEFFICIENTS
% =======================
The basis functions used to represent BP and RP spectra are constructed such that they are efficient in representing typical stellar spectra \citep{Carrasco2021,DR3-DPACP-118}. As a consequence, the broad structure of a spectrum is represented by the low-order basis functions, and detailed spectral patterns are represented by higher-order basis functions. The absolute values of the {\xp} source coefficients should therefore decrease in general with the order of the coefficient. \figureref{fig:wp942_650_510_decreasing_coefsXP} shows the distributions of the BP and RP coefficients, normalised with respect to the $L_2$-norm, for all sources in \gdrthree. In both instruments, most coefficients of the majority of \xp spectra are close to zero. To study this further, we compared the sum of absolute values for the first five coefficients with the sum of the remaining higher-order coefficients. We computed the difference between the first and the second sum, with the uncertainty on the difference, and considered sources for which the difference was smaller than five times the error. Of all the sources with \xp spectra in \gdrthree, 26037 sources have BP and 5470 sources have RP spectra that fulfil this criterion. 
The majority of these sources are concentrated in the Galactic plane and towards the direction of the Galactic centre. These sources may therefore be affected by crowding, resulting in a contamination of the spectra by flux from nearby sources and thus unexpected spectral shapes that require an unusual combination of basis functions. The larger number of sources as compared to RP may result from the larger number of faint sources in BP.

\begin{figure}
        \centering
                \includegraphics[width=0.9\columnwidth]{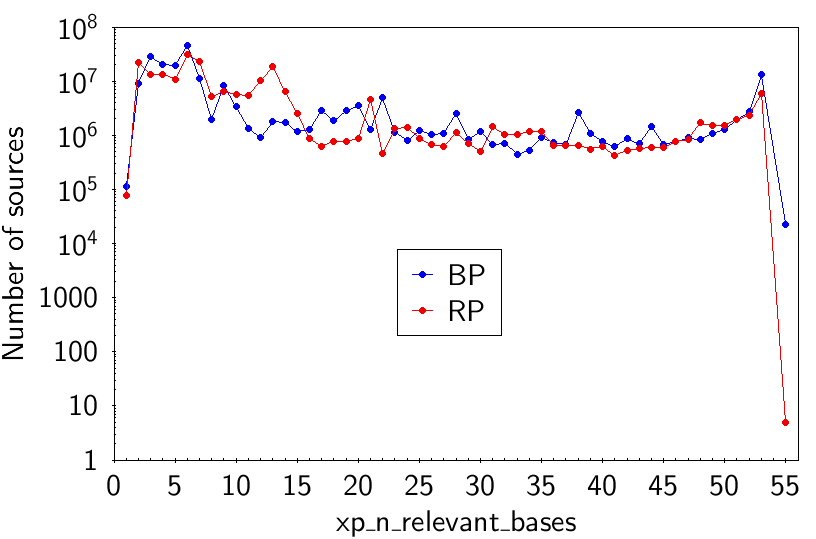}

        \caption{Number of relevant bases in the \tableName{xp_summary} table for BP (blue) and RP (red). 
        \label{fig:wp942_651_010_Nrelevant}
        }
\end{figure}

% Relevant basis
% ==============
The \tableName{xp_summary} table contains two parameters specifying the number of relevant coefficients for each source in BP and RP, respectively (\fieldName{bp\_n\_relevant\_bases} and \fieldName{rp\_n\_relevant\_bases}). All coefficients with indices larger than the specified number are considered to be consistent with being zero. \figureref{fig:wp942_651_010_Nrelevant} shows the histogram of the number of relevant bases for BP and RP.

No source has a relevant number of coefficients zero and 54. In the first case, this is caused by assigning 55 coefficients as relevant if no coefficients are found to be relevant, and the spectrum therefore agrees with consisting of random noise alone. The lack of 54 relevant coefficients arises because only one last coefficient prevents the computation of the standard deviation \citep{DR3-DPACP-118}.

\subsection{Tests of the spectral shape}

\begin{figure}
        \centering
        \includegraphics[width=1\columnwidth]{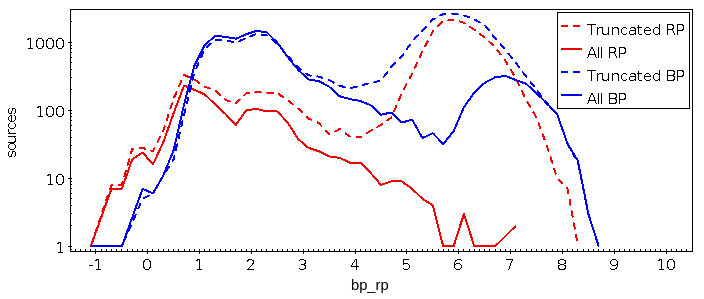}
        \caption{Comparison of the number of sources failing the small wings test in BP (blue/cyan) and RP (red/magenta) when all source coefficients are considered (solid lines) or only truncated coefficients are taken into account (dashed lines) as a function of colour.
        \label{fig:wp942_650_510_smallwings_truncated}
        }
\end{figure}

% SMALL WINGS
% ===========
Due to the lower instrumental response at its edges, the flux values of the sampled internal spectra should be lower in the outer samples than in the central samples for sources with significant flux. These regions with low fluxes at either side are referred to as the wings of the spectra. To evaluate the behaviour of the spectra at the wings at either side, we integrated the fluxes over the pseudo-wavelength ranges $[-\infty,0]$ and $[0,5]$ and compared them to the integrated flux over the interval $[5,10]$. Analogously, on the other side of the spectra, the integrals over the pseudo-wavelength intervals $[60,\infty]$ and $[55,60]$ were compared with the integral over the interval $[50,55]$. For the comparison, the difference between the integrals was computed and normalised with respect to its uncertainty.  
The four normalised differences are smaller than five for 204 to 796 RP sources and for 88 to 7411 BP sources. As was already the case for the test of the decreasing coefficients, a small part of these sources is homogeneously distributed in the sky, while the majority is concentrated in the Galactic plane and in the direction of the Galactic centre. This indicates crowding and the resulting contamination of the \xp spectra with flux from nearby stars as a reason for the non-decreasing spectral wings. %, while in particular the homogeneously distributed sources in this test might result from random noise. 
The larger number of BP spectra in this test that do not meet the threshold as compared to RP spectra may be a result of the larger number of faint spectra in BP. Figure \ref{fig:wp942_650_510_smallwings_truncated} shows the distribution of sources with normalised differences larger than five when all \xp coefficients are used and when the representation is truncated at \fieldName{xp\_n\_relevant\_bases}. The truncation results in an increase in the number of sources above the chosen threshold, in particular, for very red sources. A possible reason might be that the truncation results in an underestimated error.

\begin{figure}[t]
        \centering
        \includegraphics[width=0.45\textwidth]{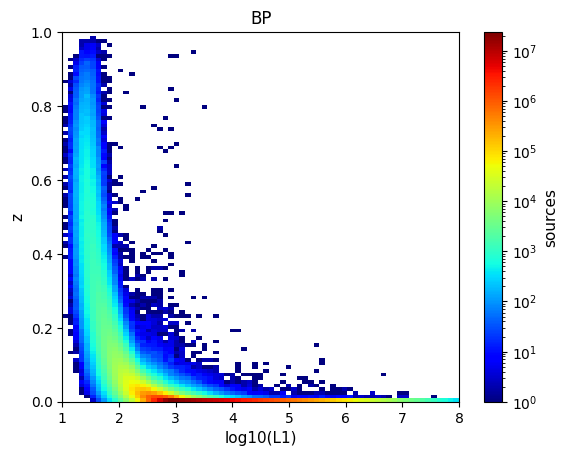}
        \includegraphics[width=0.45\textwidth]{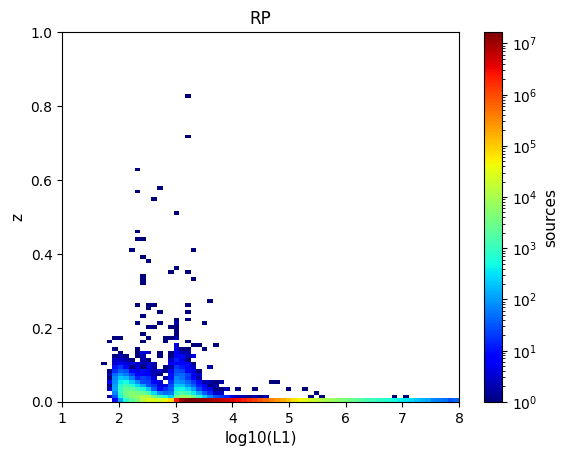}   
        \caption{Distribution of sources in the $z$--$L^1$ norm plane for BP (top panel) and RP (bottom panel). 
}
        \label{fig:wp942_650_510_negativity}
\end{figure}

% NEGATIVITY
% ==========
Noise may cause parts of the spectrum to be negative, in particular, for faint sources. In order to determine the number of negative values in the sampled \xp spectra, we defined the negativity of a spectrum as
\begin{equation*}
z = \frac{\int\limits_{-\infty}^{\infty} \left|f(u)\right|\, {\rm d}u - \int\limits_{-\infty}^{\infty} f(u)\, {\rm d}u}{2\, \int\limits_{-\infty}^{\infty} \left|f(u)\right|\, {\rm d}u.}
\end{equation*}
This measure for negativity $z$ is zero if the sampled \xp spectrum is positive at all values of the pseudo-wavelength $u$, and one if it is negative at all values of $u$. 
Figure \ref{fig:wp942_650_510_negativity} shows the distribution of sources as a function of the $L^1$ norm of the spectrum and the value of $z$ for BP and RP. The majority of sources follows a general trend of low negativity for large $L^1$ norms and increasing negativity and a wider spread in the distribution as the norm decreases. The latter case corresponds to faint sources with increasing negativity due to noise. The more pronounced tail of sources with small $L^1$ norms in BP results from the larger number of faint BP spectra as compared to RP. Only a small fraction of sources clearly lies beyond the general relation between the $L^1$ norm and $z$. These outliers in general result from an over-subtraction of the background in the spectra, shifting its overall flux level towards negative values.

\subsection{Wiggling patterns}

We tested whether truncation efficiently removes unnecessary wiggling patterns in the XP spectra. Here we considered 6377 main-sequence star members \citep{2020A&A...640A...1C} of 17 open clusters \footnote{IC 4651, Melotte 20, Melotte 22, NGC 2099, NGC 2168, NGC 2287, NGC 2506, NGC 2516, NGC 2632, NGC 3114, NGC 3532, NGC 3766, NGC 457, NGC 6405, NGC 6475, Stock 2, and Trumpler 19.}. By considering only the members of these open clusters, we ensured that our testing sample is composed of stars with metallicities similar to the solar value. We employed XP spectra that were externally calibrated by GaiaXPy with the default constant wavelength step used for the \tableName{xp_sampled_mean_spectrum} table.

First, we defined a coefficient that measures the wiggling level in XP spectra. For each ith$^{\rm }$ wavelength sampled portion of the spectrum, we calculated

\begin{equation}
        \delta^{n}_{i} = \frac{|f_i - \mathrm{mean}(f_{i-n},f_{i+n})|}{\fieldName{phot\_g\_mean\_flux}},
\end{equation}

where f$_{i}$ is the flux associated with the ith$^{\rm }$ wavelength sample. The wiggling coefficient is thus defined as the average of the $\delta^{n}_{i}$ across the entire spectrum or a portion of it,

\begin{equation}
\label{wigg_coeff}
        w_{n} = \sum_{i} \frac{ \delta^{n}_{i}}{N},
\end{equation}
where N is the number of wavelengths over which $\delta^{n}_{i}$ is determined.

The wiggling coefficient w$_{3}$ was calculated for each star within the spectral range [450,900]~nm. This coefficient is higher when the spectra contain more undesired wiggles, but also for later spectral types, whose spectra typically contain more molecular bands. In order to ensure that we probed the wiggling and not real spectral features, we therefore defined a differential coefficient 
\begin{equation}
\label{diff_wigg_coeff}
        \Delta w_{3}=\log_{10} (w_{3}) - \log_{10}(\overline{w_{3}})
,\end{equation}

where w$_{3}$ is the coefficient defined in Eq.~\ref{wigg_coeff} and measured for the jth$^{\rm }$ star, while $\overline{w_{3}}$ is the coefficient measured on the average spectrum calculated over a sample of 100 dwarf stars with 
%\fieldName{phot\_g\_mean\_mag} = \fieldName{phot\_g\_mean\_mag}$_{j}$$\pm$0.01 mag
$\vert$\fieldName{phot\_g\_mean\_mag} - \fieldName{phot\_g\_mean\_mag}$_{j}\vert<0.01$~mag
%and \fieldName{bp\_rp}=\fieldName{bp\_rp}$_{j}$$\pm$0.005. 
and $\vert$\fieldName{bp\_rp}-\fieldName{bp\_rp}$_{j}\vert<0.005$.
By averaging spectrum stars with very similar \fieldName{phot\_g\_mean\_mag} and \fieldName{bp\_rp} , we obtained a single spectrum that contained the typical absorption features that can be found in spectra of stars similar to the jth$^{\rm }$ star and cleaned from wiggles. Therefore, $\Delta$w$_{3}$ is truly representative of the actual wiggling shown by the jth$^{\rm }$ spectrum, without the contribution from molecular bands or any other spectral feature.

\begin{figure}\begin{center}
\includegraphics[width=1.\columnwidth]{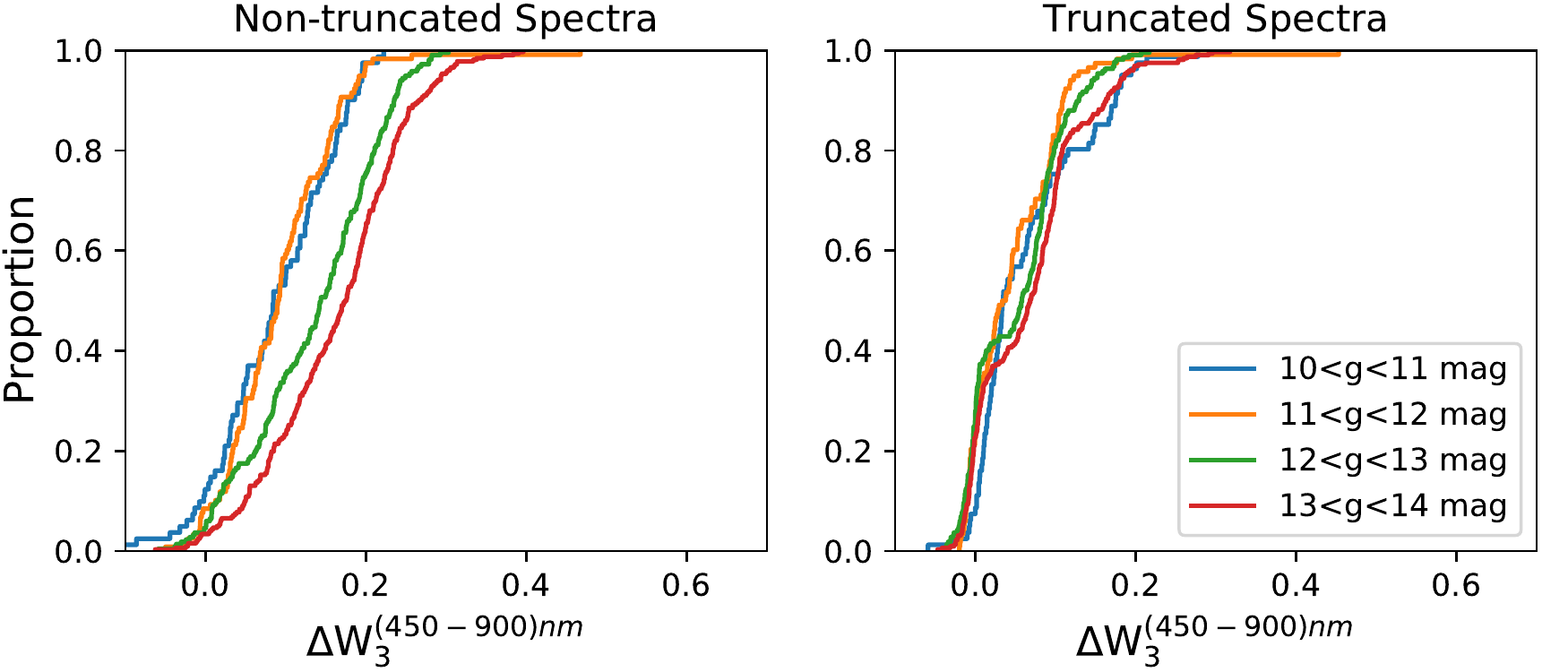}
\caption{Cumulative histogram of the differential wiggling coefficient $\Delta$w$_{3}$ measured for stars with 0.5$<$\fieldName{bp\_rp}$<$0.7 and within different bins of \fieldName{phot\_g\_mean\_mag}. The left panel shows $\Delta$w$_{3}$ values measured in the non-truncated spectra, and the right panel shows these values for truncated spectra.}
\label{fig:diff_wiggl_cumulative_g_bins}
\end{center}\end{figure}

Figure~\ref{fig:diff_wiggl_cumulative_g_bins} shows the cumulative histograms of the differential wiggling coefficients $\Delta$w$_{3}$ derived for stars in different bins of \fieldName{phot\_g\_mean\_mag}. The coefficients derived from non-truncated spectra are plotted in the left panel, and those from truncated spectra are shown in the right panel. While fainter stars tend to have larger $\Delta$w$_{3}$ in their non-truncated spectra due to the lower signal-to-noise ratio, this dependence is significantly reduced by truncation.

%\begin{figure}\begin{center}
%\includegraphics[width=1.\columnwidth]{figures/diff_wiggl_cumulative_br_bins.pdf}
%\caption{\sout{Cumulative histogram of the wiggling factor $\Delta$w$_{3}$ measured for stars with 13$<$\texttt{phot\_g\_mean\_mag}$<$15 mag and within different bins of \texttt{bp\_rp}. The left panel shows $\Delta$w$_{3}$ values measured in the non-truncated spectra, while the right panel is relative to truncated spectra.}}
%\label{fig:diff_wiggl_cumulative_br_bins}
%\end{center}\end{figure}

%\subsection{Wiggling and strong spectral features}
%\label{Sec:wigg_features}

Wiggling in XP spectra might be enhanced by strong spectral features. It is especially important to test this possibility in spectra of young accretors, which are typically characterised by strong H$\alpha$ emission lines. Therefore, we used XP spectra from 197 members of the star-forming regions Chamaeleon~I, IC~348, Lupus, NGC~2024, NGC~2068, ONC, Ophiucus, and R~Coronae~Australis. 
%The spectra were externally calibrated by GaiaXP with a standard sampling. (already said)

We defined a coefficient that measures the height of the H$\alpha$ line,

\begin{equation}
        H_{H\alpha} = \frac{f_{H\alpha}}{\mathrm{mean}(f_{(626-636)\cup(676-686)})},
\end{equation}
where f$_{H\alpha}$ is the flux measured at 656nm, corresponding to the centre of the H$\alpha$ line, while the mean flux at the denominator is measured at the base of the H$\alpha$ line, that is, at all wavelengths within 626-636nm and within 676-686nm.

\begin{figure}\begin{center}
\includegraphics[width=1.\columnwidth]{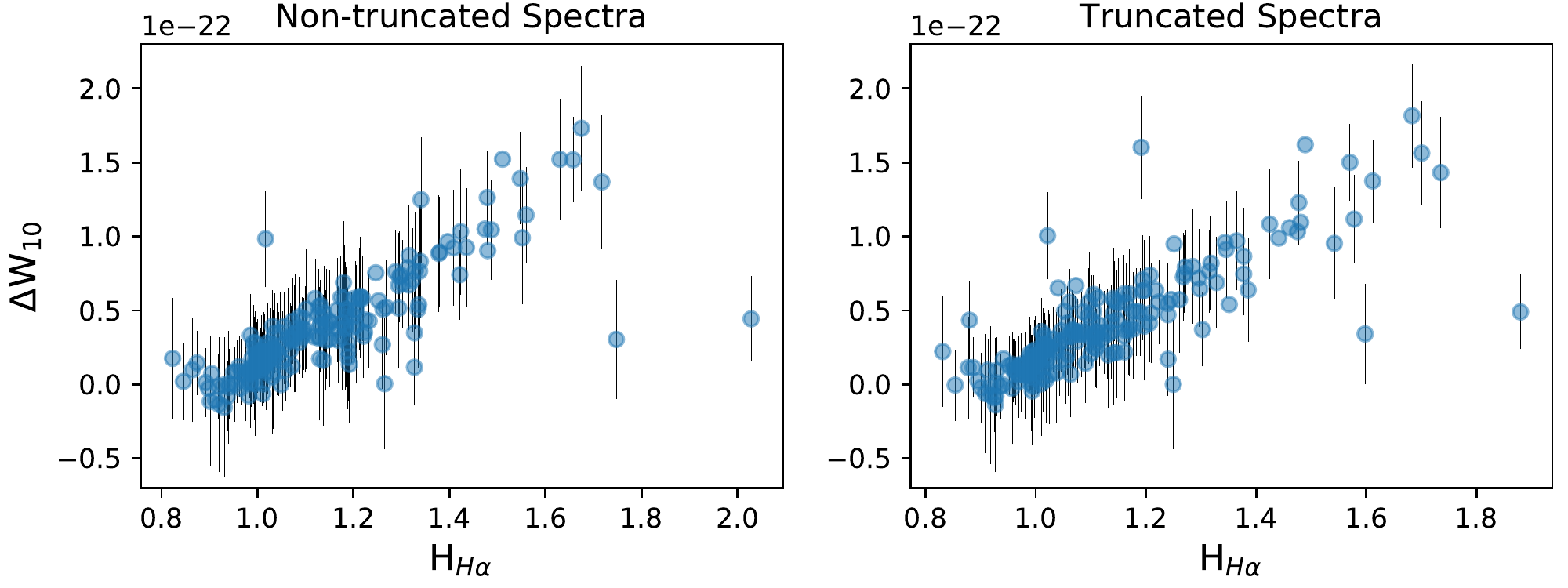}
\caption{Differential wiggling $\Delta$w$_{10}$ as a function of H$_{H\alpha}$. Non-truncated spectra are shown on the left, and truncated spectra are shown on the right.}
\label{fig:wiggling_Ha_one}
\end{center}\end{figure}

The differential wiggling coefficient $\Delta$w$_{10}$ (see Eq. \ref{wigg_coeff} and \ref{diff_wigg_coeff}) is shown in Fig.~\ref{fig:wiggling_Ha_one} (left panel) as a function of H$_{H\alpha}$ for non-truncated spectra. Each error bar represents the standard deviation of the coefficient $\overline{w_{10}}$ measured on the comparison sample of 100 dwarfs. The plot shows that wiggling increases with the height of the H$\alpha$ line. Therefore, we conclude that the presence of strong spectral features enhance wiggling in XP spectra.

In order to test whether truncation is able to fix or alleviate the problem, we repeated the experiment on truncated XP spectra. The results are shown in the right panel of Fig.~\ref{fig:wiggling_Ha_one}, which indicates that truncation does not significantly remove the additional wigging produced by strong H$\alpha$ lines.
Instead, we found that the height of the H$\alpha$ line is affected by truncation. %\sout{In Fig.~\ref{fig:reduced_Ha} we show the relative variation in H$_{H\alpha}$ height measured in the truncated and non-truncated spectra as a function of the H$_{H\alpha}$ from non-truncated spectra:} 
We observe that for 23$\%$ of stars with H$_{H\alpha}^{non-trunc}$$>$1.1,  H$_{H\alpha}$ is reduced by more than 5$\%$ by truncation.

%\begin{figure}\begin{center}
%\includegraphics[width=1.\columnwidth]{figures/reduced_Ha.pdf}
%\caption{\sout{Relative variation of the H$_{H\alpha}$ between truncated and non-truncated spectra as a function of the H$_{H\alpha}$ measured on non-truncated spectra.}}
%\label{fig:reduced_Ha}
%\end{center}\end{figure}

\subsection{Tests of the integrated fluxes}

% INTEGRATED FLUXES
% =================
The calibration of {\xp} integrated photometry and the spectra follow different calibration procedures that only have low-level processing steps in common. Although some differences might occur between the integrated fluxes from the XP spectra and the integrated photometry, mainly because of potential differences in passband calibration and noise, we expect to have comparable results among these two different processes in principle. In order to test this, we computed the ratio of the photometric and spectrum flux.

\begin{figure*}
        \centering
        \includegraphics[width=0.4\textwidth]{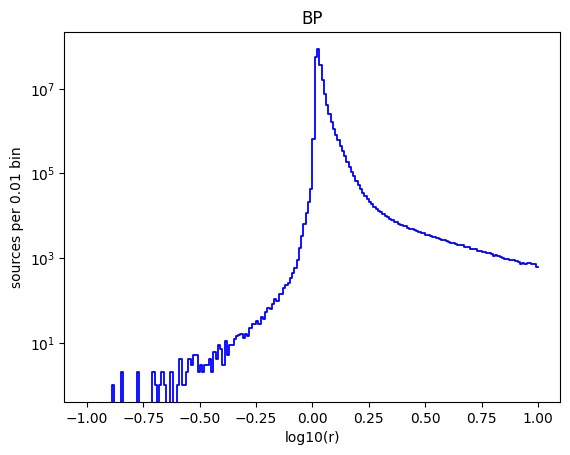}
        \includegraphics[width=0.4\textwidth]{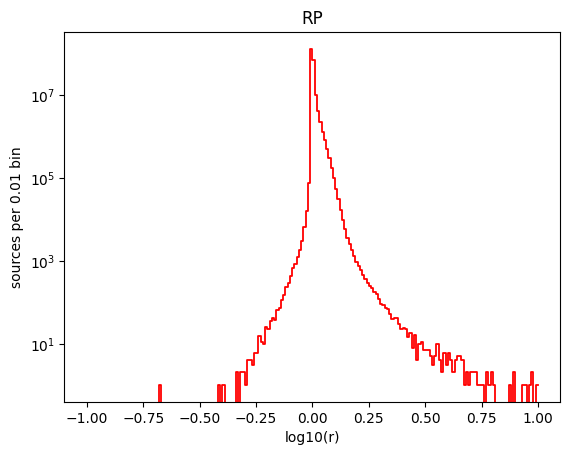}        
        \caption{Histogram of the ratio $r$ of the photometric and spectrum flux in BP (left) and RP (right).}
        \label{fig:wp942_650_510_intflux}
\end{figure*}

The distribution of this ratio shows that most sources have a value close to one (\figref{fig:wp942_650_510_intflux}). For BP, however, there is a significant population of sources with values higher than one. This might be a result of a threshold of one electron/s that was applied in the selection of transits in the integrated photometry. Transits with a flux below this threshold were excluded from the computation of the mean flux, resulting in a biased mean flux for faint sources \citep{riello2021}. This threshold was not applied in the computation of the mean spectra, thus avoiding the bias towards too high fluxes and leading to a better behaviour at low BP fluxes for the integrated flux from the spectra.

\begin{figure}[t]
        \centering
        \includegraphics[width=0.45\textwidth]{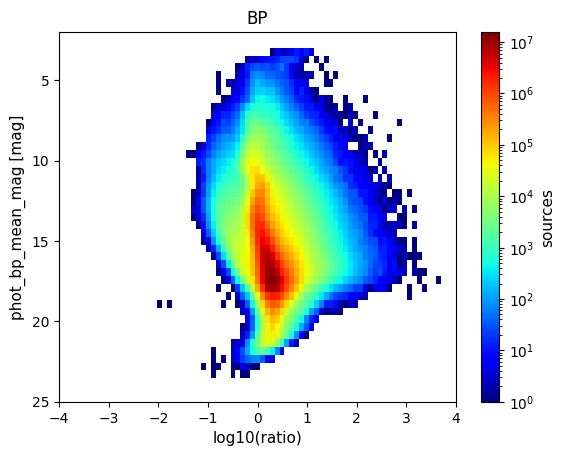}
        \includegraphics[width=0.45\textwidth]{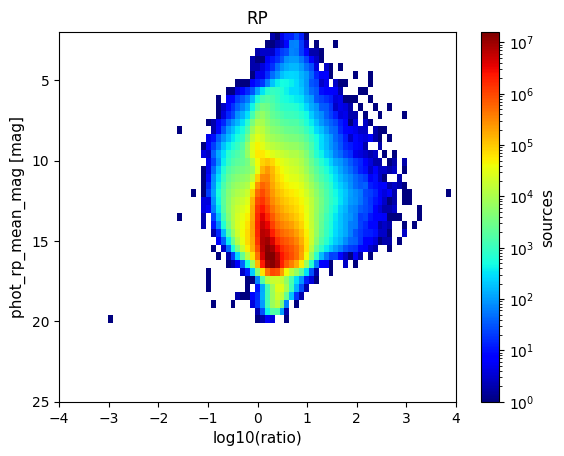} 
        \caption{Distribution of sources in decadic logarithm of the ratio of photometric and spectroscopic uncertainty and \xp magnitude for BP (top panel) and RP (bottom panel). 
}
        \label{fig:wp942_650_510_error_phot_spec}
\end{figure}

% INTEGRATED FLUX UNCERTAINTY
% ===========================
We compared the flux error uncertainties derived from photometric fluxes and those derived from the spectra. Although these flux uncertainties are similar, those derived from the photometric calibration tend to be slightly larger than those derived from the spectra. The ratio of the uncertainty in \xp fluxes and on the flux resulting from integrating the spectra is shown in Fig.~\ref{fig:wp942_650_510_error_phot_spec} for BP and RP as a function of BP and RP photometric magnitude. The shift towards larger photometric errors is clear, together with a dependence on the source magnitude. This behaviour might result from underestimated uncertainties, in particular for low-order coefficients in the source representation \citep{DR3-DPACP-118} to which the integration of the spectra is particularly sensitive. The distribution of the uncertainty ratio has strong tails towards extreme values. The photometric uncertainties of hundreds of sources are 100--1000 times larger than those derived from the spectra. 

\subsection{Uncertainties of the XP coefficients}
The analysis that we performed on the coefficients of the XP spectra tested whether their uncertainties were evaluated correctly. To do this, we compared pairs of stars using a chi-square,
 
\begin{equation}
\label{chi2}
        \chi^2 = (X_1-X_2)^{T}(C_1+C_2)^{-1}(X_1-X_2)
,\end{equation}

where $X_1$ and $X_2$ are the coefficients of the two stars in either the BP or RP channel, while $C_1$ and $C_2$ are the associated covariance matrices. In order to ensure that we compared stars with the same metallicity and reddening, we applied Eq.~\ref{chi2} only to pairs of stars belonging to the same open cluster (i.e. membership probability$\geq$0.7 from \citealt{2020A&A...640A...1C}). We also excluded all stars with \fieldName{ruwe}$>$1.4 from the comparison (to remove binaries) and stars belonging to open clusters younger than 100~Myr. This latter selection was necessary to avoid contamination due to differential extinction. Furthermore, the two stars must have {\gmag} magnitudes and {\bprp} colours that were consistent within their uncertainties. By applying all these selection criteria, we obtained a controlled sample of 1560 stellar pairs whose $\chi^2$ values we were able to derive. 

\begin{figure} \begin{center}
\includegraphics[width=1.\columnwidth]{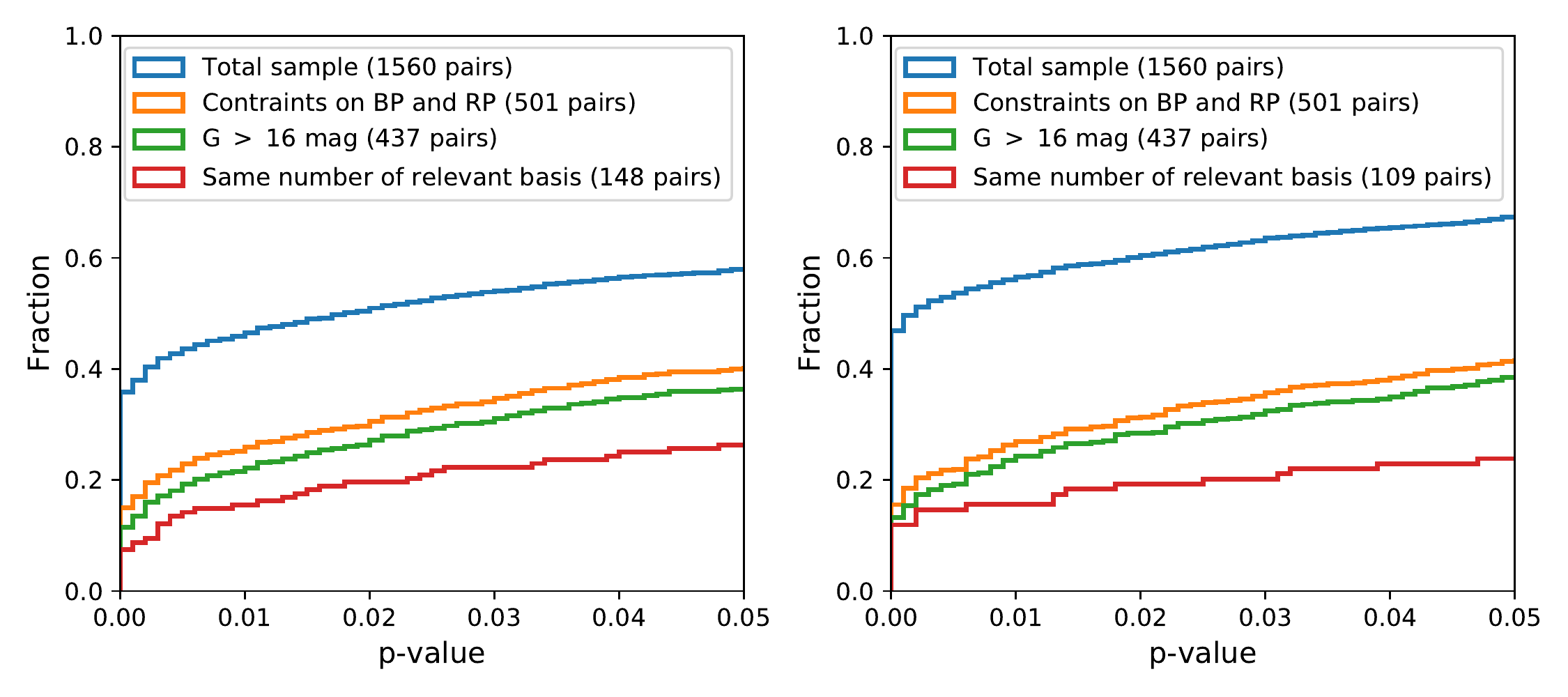}
\caption{Zoom on the p-value distributions obtained for the two bands BP (left) and RP (right). The pairs with a p-value below 0.01 failed the test. 
}
\label{fig:val_dist}
\end{center} \end{figure}

The $\chi^2$ values were then used to calculate the associated p-values (the null hypothesis being that $\chi^2$ follows the expected chi-square distribution, the degree of freedom being the number of coefficients) separately for the BP and RP channels. If the $\chi^2$ indeed followed a chi-square distribution, the p-values should be distributed uniformly. The cumulative histograms of the p-value distributions of the 1560 stellar pairs are shown in Fig.~\ref{fig:val_dist} (blue lines). The p-values of 49$\%$ and 56$\%$ of the pairs are below 0.01 in BP and RP, respectively. This indicates that the uncertainties of the XP coefficients are underestimated.

As a second test, we applied a more stringent criterion in the selection of the pairs that were to be tested. Specifically, we imposed that the {\gmag}, {\gbp} and {\grp} magnitudes of the two stars must be consistent within their uncertainties. This reduced our sample to 501 pairs. The relative p-value distributions are plotted as orange lines in Fig.~\ref{fig:val_dist}: 25$\%$ and 26$\%$ of these pairs fail our test in BP and RP, respectively.

 When we applied the further condition that the magnitudes $G_1$ and $G_2$ must be fainter than 16 mag, we further reduced the sample to 437 pairs. The difference between this new sample, which is plotted as green lines in Fig.~\ref{fig:val_dist}, and the previous sample is too small to observe significant effects in the p-value distribution. The fraction of pairs that fails our test now decreased to 22$\%$ and 24$\%$ in BP and RP, respectively.

Finally, we studied the p-value distributions obtained from the pairs composed of stars with the same number of \fieldName{bp\_n\_relevant\_bases} and \fieldName{rp\_n\_relevant\_bases}. In this way, we compared spectra with similar wiggling levels. Applying these criteria for the two bands separately, we obtained a sample of 148 and 109 pairs for BP and RP, respectively. The cumulative distributions are plotted in Fig.~\ref{fig:val_dist} as red lines. The fraction of stars that do not pass the test decreases to 21$\%$ for BP and 15$\%$ for RP, but it is still significant.

\begin{figure} \begin{center}
\includegraphics[width=1.\columnwidth]{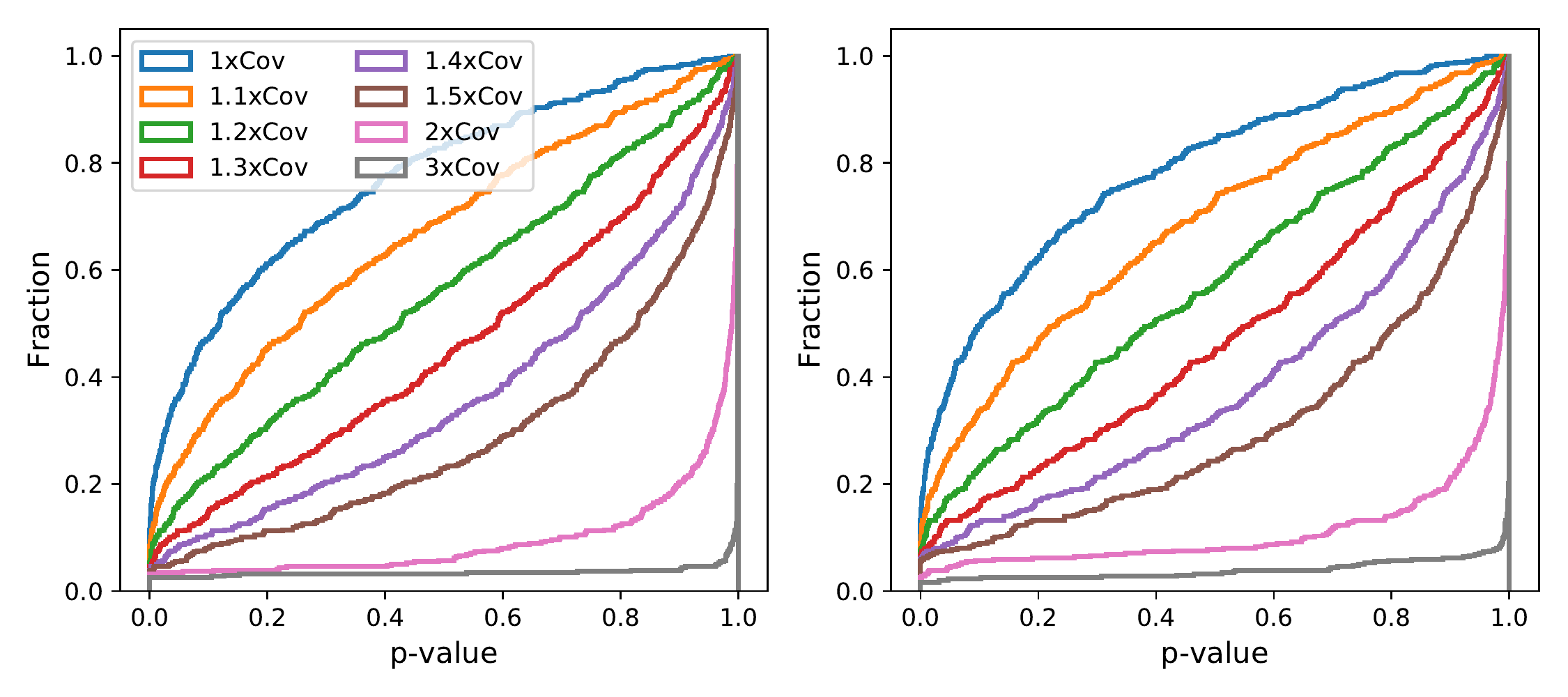}
\caption{Same as Fig.~\ref{fig:val_dist} for the 437 pairs that are fainter than G=16 mag, but have a covariance matrix (Cov) of one to three times its original value for the two bands BP (left) and RP (right). 
}
\label{fig:p-val_dist}
\end{center} \end{figure}

In order to estimate how much the errors are underestimated, we multiplied the covariance matrix by various factors and then repeated the experiment on the 437 pairs that are fainter than G=16 mag. The resulting p-value distributions are shown in  Fig.~\ref{fig:p-val_dist}. The figure shows that the variances are underestimated by a factor that is between 1.2 and 1.5. 

However, a detailed study of the coefficient error underestimation is presented in \cite{DR3-DPACP-118} by dividing the data into two groups of transits for the same source and comparing the obtained values. They show that the error underestimation depends on the coefficients. The lower-order coefficients lead to the highest underestimation.

\subsection{Comparison with external spectra}

\begin{figure}\begin{center}
\includegraphics[width=0.8\columnwidth]{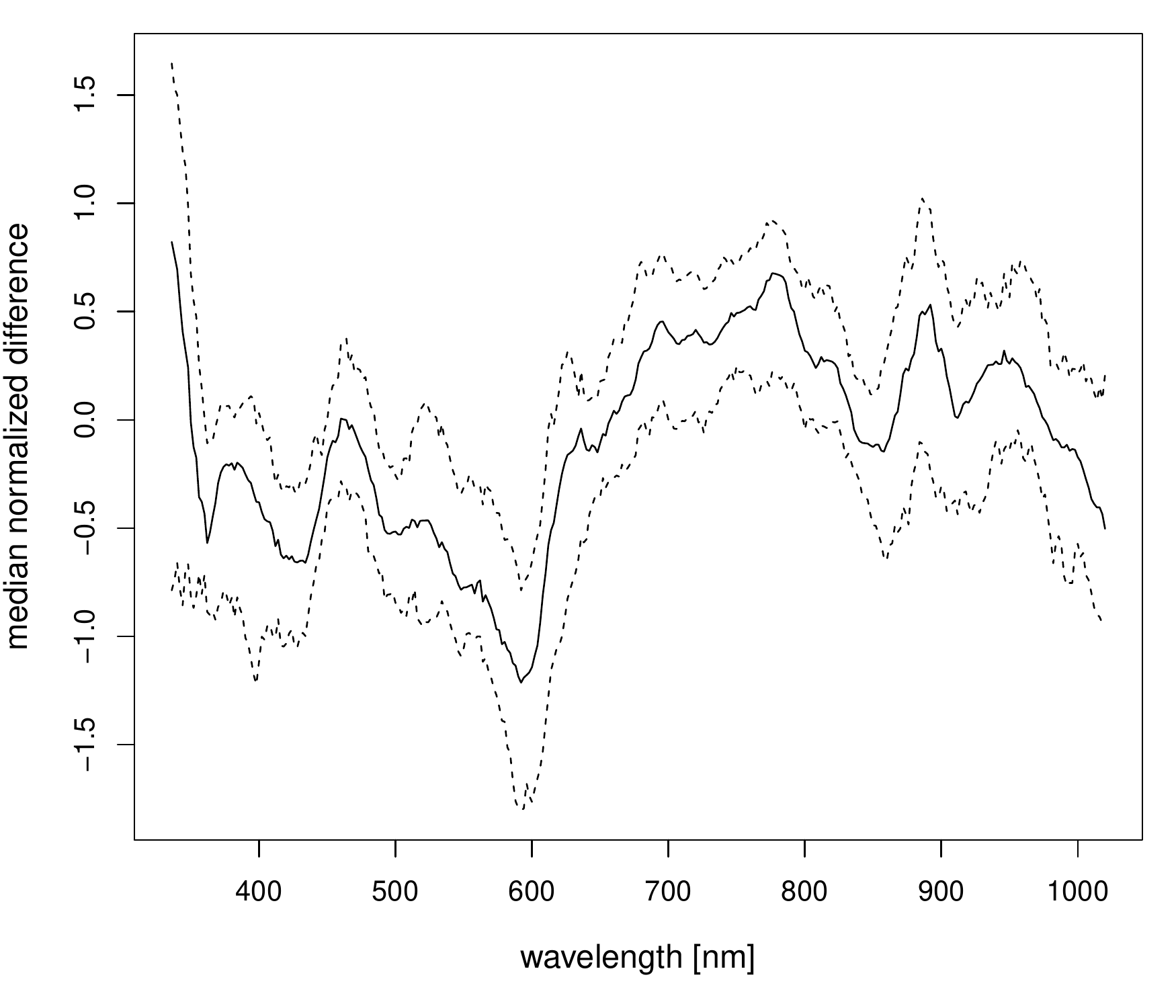}
\caption{Median flux difference normalised by the errors between the XP sampled spectra and the CALSPEC spectra normalised by the errors as a function of wavelength. Dotted lines correspond to the 1$\sigma$ confidence interval.}
\label{fig:xpspec_calspec}
\end{center}\end{figure}

\begin{figure}\begin{center}
\includegraphics[width=0.8\columnwidth]{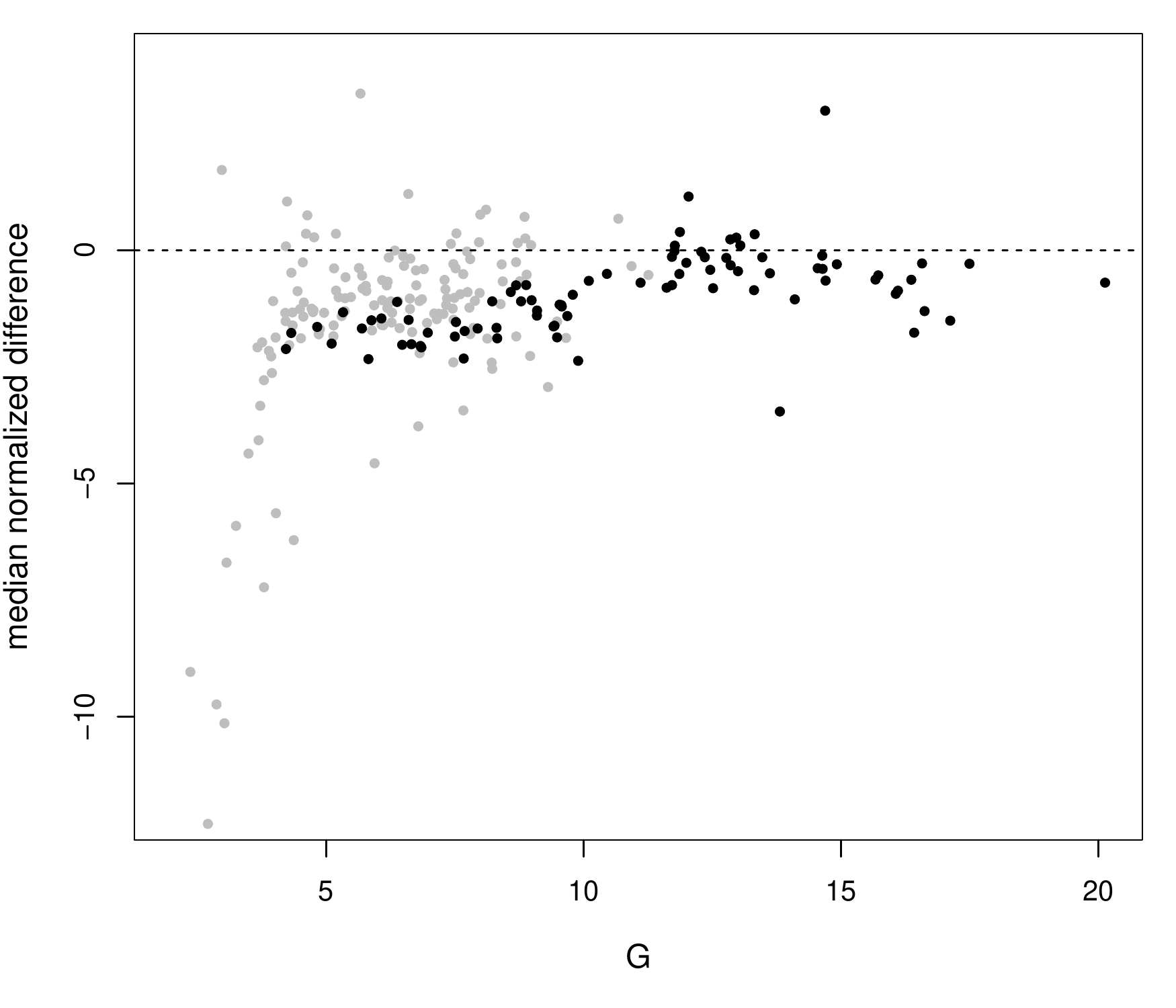}
\caption{Median flux difference normalised by the errors within $560<\lambda<620$~nm between the XP sampled spectra and the CALSPEC (black dots) or NGSL (grey dots) spectra normalised by the errors as a function of magnitude.}
\label{fig:xpspec_dip}
\end{center}\end{figure}

\begin{figure}\begin{center}
\includegraphics[width=0.8\columnwidth]{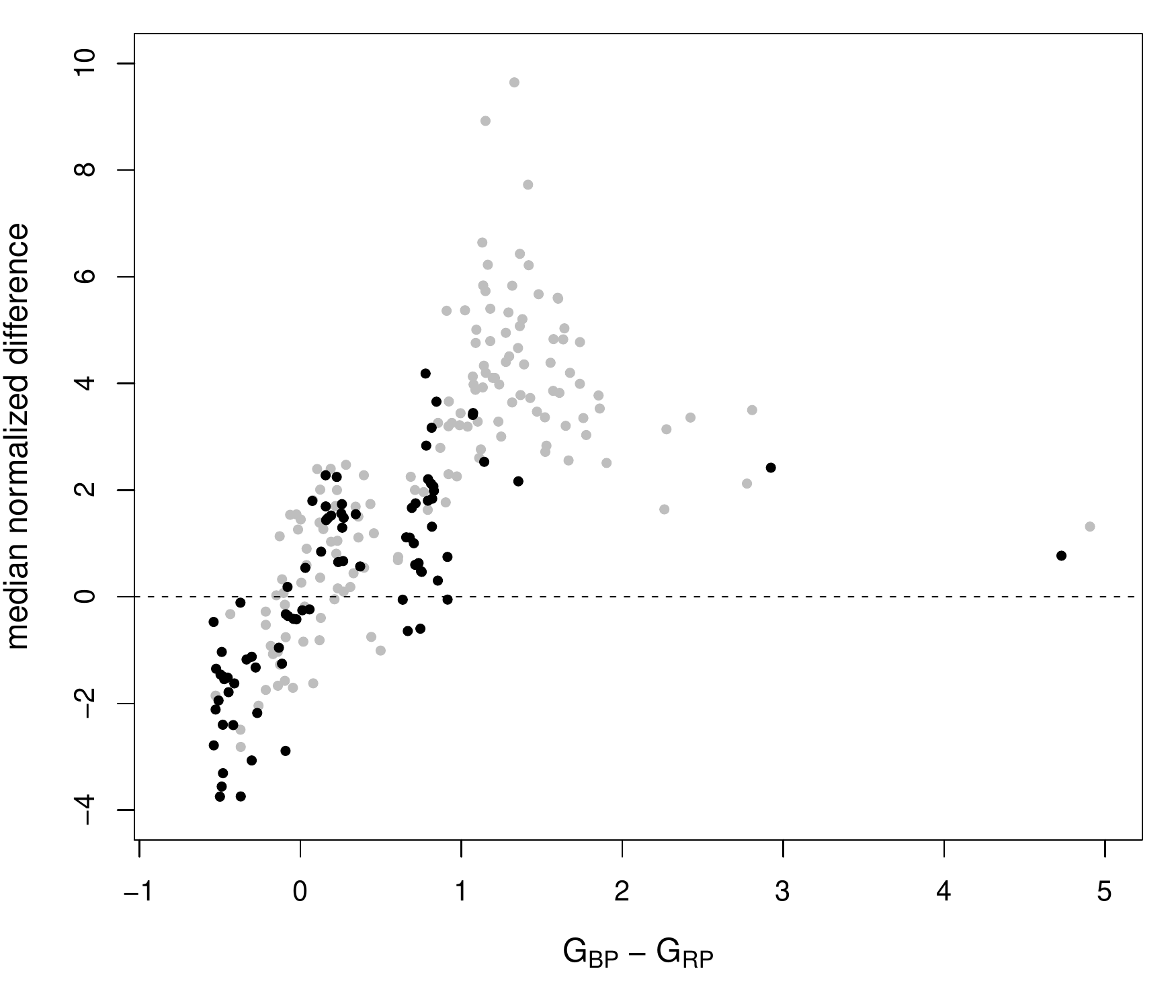}
\caption{Median flux difference within $\lambda<350$~nm between the XP sampled spectra and the CALSPEC (black dots) or NGSL (grey dots) spectra normalised by the errors as a function of magnitude.}
\label{fig:xpspec_bluecol}
\end{center}\end{figure}

Figure~\ref{fig:xpspec_calspec} shows the median flux difference, normalised by the errors, between the XP sampled and the CALSPEC\footnote{2021 March update} spectra \citep{2014PASP..126..711B} for the sources in common. A dip at $\sim$600~nm is visible. Figure~\ref{fig:xpspec_dip} presents the median normalised flux difference within $560<\lambda<620$~nm with both CALSPEC and NGSL \citep{2016ASPC..503..211H} as a function of magnitude. It shows that the strength of this dip is magnitude dependent and has a saturation effect. Figure~\ref{fig:xpspec_calspec} seems to suggest a difference in flux level between BP and RP, but it is not statistically significant in the CALSPEC or the NGSL sample. However, when the MILES library is used \citep{2011A&A...532A..95F} and the MILES spectra are normalised to the absolute flux of the XP spectra in the common wavelength range, this difference in flux level becomes significant. The bluest wavelengths show a colour-dependent trend that is illustrated in Fig.~\ref{fig:xpspec_bluecol}. See also \cite{DR3-DPACP-120} for a discussion of these features.

%Tests transforming the CALSPEC spectra into XP continuous coefficients through GaiaXPy indicates that those are not consistent with the published XP coefficients. This is either an issue with GaiaXPy simulation mode \footnote{tested GaiaXPy version is 0.0.6} or due to the coefficient uncertainties under-estimation (see previous section). 

%+++++++++++++++++++++++++++++++++++++++++++++++++++++++++++++++++++++++++
\section{Stellar astrophysical parameters}\label{sec:AP}
%+++++++++++++++++++++++++++++++++++++++++++++++++++++++++++++++++++++++++

An overview of the \gdrthree astrophysical products produced by 13 different modules\footnote{DSC: Discrete source classifier\\GSP-Phot: General stellar parametriser from photometry\\GSP-Spec: General stellar parametriser from spectroscopy\\FLAME: Final luminosity age mass estimator\\ESP-CS: Extended stellar parametriser for cool stars\\ESP-UCD: Extended stellar parametriser for ultra-cool dwarfs\\ESP-HS: Extended stellar parametriser for hot stars\\ESP-ELS: Extended stellar parametriser for emission line stars\\MSC: Multiple star classifier\\QSOC: QSO classifier\\UGC: Unresolved galaxy classifier\\OA: Outlier analysis\\TGE: Total Galactic extinction} is presented in \cite{DR3-DPACP-157}. The non-stellar content part of the astrophysical parameters is discussed in Sect.~\ref{sec:extragal}. This section focuses on the stellar content, which is presented in detail in \cite{DR3-DPACP-160}. 

The astrophysical parameters are available in two tables: \tableName{astrophysical\_parameters} , and \tableName{astrophysical\_parameters\_supp}. We present here only the validation results of the main parameters. In particular, the specialised modules \citep[ESP][]{DR3-DPACP-157} are almost not discussed here.  The Outlier Analysis tables ({\tt oa\_neuron\_information, oa\_neuron\_xp\_spectra}) are not discussed here either. They were successfully checked for internal but not external consistency. Tests on the ESP and OA modules can be found in \cite{DR3-DPACP-160}. GSP-Phot parameters \citep{DR3-DPACP-156} were derived using several spectral libraries (MARCS, PHOENIX, A, and OB). The values obtained with these different libraries are presented in \tableName{astrophysical\_parameters\_supp} , while \tableName{astrophysical\_parameters} contains the values that were obtained with what was selected as the best library indicated in the field \fieldName{libname_gspphot}. We mainly discuss the best library results of GSP-Phot here. 

In this section, we compare the Galaxy model using GUMS as a reference. In contrast to GOG, GUMS contains most of the astrophysical parameters, but they are error free.

\subsection{DSC}

The development of the discrete source classifier (DSC) was mainly driven by extragalactic source completeness (see the \href{https://gea.esac.esa.int/archive/documentation/GDR3/Data\_analysis/chap\_cu8par/}{on-line documentation} and \cite{DR3-DPACP-158}). The purity for QSOs and galaxies is discussed in Sec.~\ref{sec:extragal}.
We did not find correlations between the \fieldName{classprob\_dsc\_binarystar} probabilities with Multiple Source Classifier (MSC) results or with known binaries. White dwarfs are also often confused with hot main-sequence stars. %  (see Fig.\ref{fig:947-DSC1})
We therefore advise against using the physical binary and white dwarf class probabilities\footnote{they are needed to be able to adapt the DSC probabilities to a new prior, however; see Sect.~11.3.2.7 of the on-line documentation}.

% \begin{figure}[h!]
%  \begin{center}
%  \includegraphics[width=0.85\columnwidth]{figures/WDProbComb.png}
%  \end{center}
%  \caption{\sout{Color-magnitude diagram of all the stars in clusters in our sample. The color indicates  classprob\_dsc\_combmod\_wd.  In green the stars classified as WDs with probability=1}}
%  \label{fig:947-DSC1}
%  \end{figure}

\subsection{Extinction}

The extinction is provided as the monochromatic extinction $A_0$ at 541.4~nm by GSP-Phot ({\fieldName{azero_gspphot}), the hot star module ESP-HS ({\fieldName{azero_esphs}), and the multiple source module MSC ({\fieldName{azero_msc}). GSP-Phot and ESP-HS also provide $A_G$ and \ebpminrp. 

\begin{figure}\begin{center}
\includegraphics[width=0.8\columnwidth]{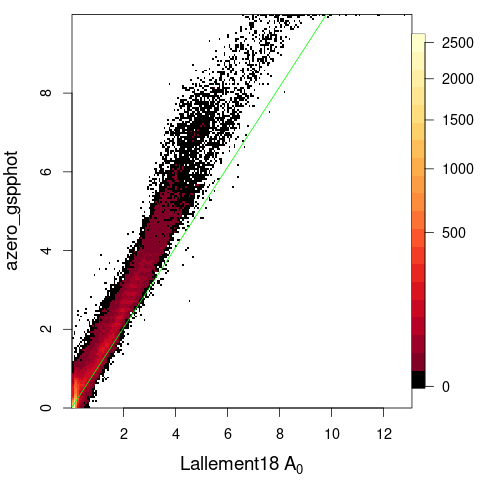}
\caption{Density plot of the comparison of the monochromatic extinctions of GSP-Phot with those derived by \cite{2018A&A...616A.132L}. The green line corresponds to the 1.02 relation that is expected given the slight wavelength difference between the two $A_0$.}
\label{fig:azeroVlallement18}
\end{center}\end{figure}

The well-known and expected temperature - extinction degeneracy is discussed in \cite{DR3-DPACP-156}, as is the effect of imposing the extinction to be positive on the mean of low-extinction regions. 
\cite{DR3-DPACP-156} also showed that the GSP-Phot extinction values \fieldName{azero\_gspphot} are globally overestimated with a saturation at 10~mag (by construction) in their comparison with Bayestar19 \citep{2019ApJ...887...93G}. We find the same trend in our comparison with the monochromatic extinction $A_0$ at 550~nm derived from APOGEE, {\gaia,} and 2MASS by \cite{2018A&A...616A.132L} shown in Fig.~\ref{fig:azeroVlallement18}. The \cite{2018A&A...616A.132L} $A_0$ are consistent with the $A_V$ values from StarHorse provided with APOGEE DR16 \citep{2020A&A...638A..76Q} with only the expected deviation for large extinctions between $A_V$ and $A_0$ (See Sect.~11.2.3.1.4 of the on-line documentation). We further confirmed this overestimation of the GSP-Phot extinction values with the bstep extinctions provided with GALAH DR3 \citep{2021MNRAS.506..150B} and the \cite{2019A&A...625A.135L} 3D extinction map. It is also confirmed with clusters in \cite{DR3-DPACP-160}.
For nearby stars, the extinction stays low because of the ad hoc extinction prior \citep{DR3-DPACP-156}. However, at high Galactic latitudes, 22\% of the stars have \fieldName{azero\_gspphot\_lower}$>$0.16, the highest extinction value expected according to the map of \cite{1998ApJ...500..525S}. As illustrated in Fig.~\ref{fig:gsphota0_hrd}, high extinction values occur at the bottom of the main sequence 
% (especially for low-mass YSOs and binaries) CB: no this is for all main sequence stars
for red giants (which can be confused with extincted hot stars), but also for some stars near \bpminrp$\sim$0.2 and $\sim$2, which also have an impact on the temperature that is estimated for these stars. 

\begin{figure}\begin{center}
\includegraphics[width=0.8\columnwidth]{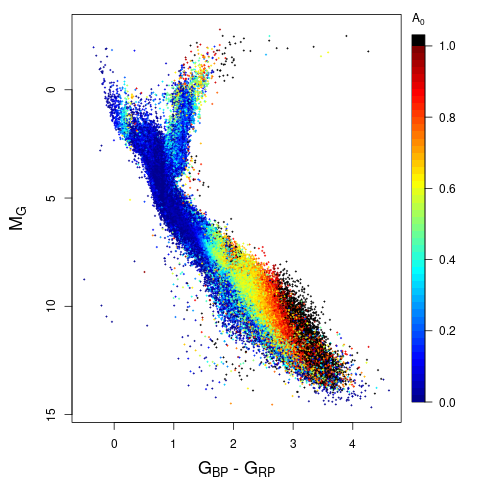}
\caption{Hertzsprung-Russel diagram of low-extinction stars ($A_0<0.05$~mag according to \cite{2019A&A...625A.135L}) with a parallax relative precision lower than 10\%, colour-coded with the mean extinction \fieldName{azero_gspphot}. The colour is saturated as black for values higher than 1~mag. In this low extinction sample, $M_G$ is simply $G+5+5\log(\varpi/1000)$.}
\label{fig:gsphota0_hrd}
\end{center}\end{figure}

The global overestimation of GSP-Phot extinction values naturally leads to an overestimation of the total Galactic extinctions provided in the table \tableName{total\_galactic\_extinction\_map\_opt}. This is shown in the comparison with Planck \citep{2016A&A...596A.109P} in Fig.~\ref{fig:tge_planck}. 
As \fieldName{a0\_uncertainty} provides the error on the mean and can become very small when the number of stars becomes high, we recommend using \fieldName{a0\_uncertainty}$\times \sqrt{\fieldName{num_tracers_used}}$ instead.
While the overall appearance of the Galactic extinction map is as expected (large-scale dust filaments are clearly visible in the approximate expected relative intensity; see \citealt{DR3-DPACP-158}), the $A_0$ estimates are systematically overestimated in a large region of around 20 deg around the Galactic centre. At high Galactic latitudes, the uncertainties are large enough for the $A_0$ overestimation to be not visible in Fig.~\ref{fig:tge_planck}. 
%\sout{We also note that the resolution of the map is quite low in many sky areas compared to other 2D Galactic extinction maps} (e.g. \citealt{1998ApJ...500..525S, 2019A&A...625A.135L, 2019ApJ...887...93G}).

\begin{figure}\begin{center}
\includegraphics[width=0.95\columnwidth]{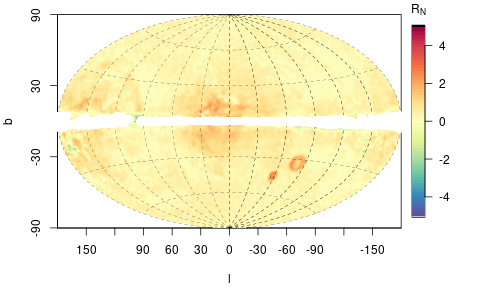}
\caption{Comparison of the total galactic extinction map (\fieldName{a0}) with Planck $E(B-V)$ normalised by the error $R_N=(\fieldName{a0}-E(B-V)\times3.1)/(\fieldName{a0_uncertainty}\times \sqrt{\fieldName{num_tracers_used}})$. The white area corresponds to locations with \fieldName{status}$>$0.} % Values outside 5$\sigma$ are saturated as black or grey (15\% of the map)
\label{fig:tge_planck}
\end{center}\end{figure}

The multiple source classifier (MSC) finds 37\% of its lower extinction \fieldName{azero\_msc\_lower}$>0.05$ for the nearby star sample ($\varpi>20$~mas) for which no significant extinction is expected. When compared to \cite{2020A&A...638A.145T}, the MSC extinction is also globally overestimated. We advise that the GSP-Phot extinctions be preferred even for binary stars. 

For the hot-star module, extinction \fieldName{azero\_esphs} can reach very high values for white dwarfs that are treated as hot main-sequence stars. We recommend to filter them out using a colour-absolute magnitude diagram. 

Another estimate of the extinction is provided through the diffuse interstellar band (DIB) at 862~nm that is present in the RVS spectra: \fieldName{dibew\_gspspec}. The details of the measurement are presented in \cite{DR3-DPACP-186}, and details of its performance are reported in \cite{DR3-DPACP-144}.
The DIB equivalent width correlates well with the extinction \citep{DR3-DPACP-144}. The large number of outliers in wavelength seen with wide binaries (Appendix~\ref{sec:widebinaries}) is due to wavelength clusters around 862.5 and 861.8~nm, most of which are removed when only sources are kept that have \fieldName{dibqf\_gspspec}$<2$. This selection criterion is globally recommended for the DIB parameters  \citep{DR3-DPACP-144}.

\subsection{Teff, logg, metallicity, and abundances\label{sec:APatmo}}

The comparison of the astrophysical parameters with the GUMS model is satisfactory within the model uncertainties outside the Galactic plane. The exception is the GSP-Phot metallicity, which follows extinction patterns, as expected from the extinction-temperature degeneracy.

% \begin{itemize}
%  \item GSP-Phot MCMC does not cover the upper/lower bounds (C9VALSCI-179), MSC MCMC versus inflated errors (C9VALSCI-207), edge effects (C9VALSCI-224), gspspec abundances discretization (C9VALSCI-201)
%  \item \textcolor{blue}{\bf MRG: PLOT DONE} GSP-Phot/GSP-Spec comparison
%  \item external catalogues
%  \item \textcolor{blue}{\bf MRG: DONE} cluster comparison (C9VALSCI-226 , C9VALSCI-229) (not clear if it sholud go in the next section: Mg: C9VALSCI-230)
% \end{itemize}

%% TEFF 
Stars with $3000 <\teff<8000 K$ are analysed by GSP-Phot using the MARCS and PHOENIX spectral libraries. Comparing the values of \teff, we find 
a median difference \teff(MARCS-PHOENIX)=-63\,K (median absolute deviation, MAD=145\,K) due to the different temperature scale.
\teff estimated using the best library compilation presents  nonphysical clustering of points on the Hertzsprung-Russel (HR) diagram 
%(see \figref{fig:cu8par_qa_nss_AllCluHRDlib}) 
due to edge effects at the library borders. This is most evident for the OB library border at 15000~K.

% \begin{figure}
%     \centerline{
%         \includegraphics[width=0.7\textwidth]{figures/cu8par_qa_nss_AllCluHRDLib.png}
%     }
%     \caption{\teff vs. \logg diagram (aka Kiel diagram) 
%     from \teff\_gspphot. We color coded the stars by their ``best'' library.}
%     \label{fig:cu8par_qa_nss_AllCluHRDlib}
% \end{figure}

%\sout{Fig.~\ref{fig:Kielgspphot} shows the Kiel diagram for a random sample of one million sources. The colour code shows the different gspphot libraries, which are responsible for the edge effects in the diagram. The recomendation is to use the best library specified in \fieldName{libname\_gspphot}.}
% \begin{figure}\begin{center}
% \includegraphics[width=0.9\columnwidth]{figures/Kiel_gspphot.png}
% \caption{Kiel diagram for a random subset of sources with gspphot. The different colours show the results for the different libraries: Marcs (Grey), Phoenix (Red), A (Yellow) and OB (Blue).}
% \label{fig:Kielgspphot}
% \end{center}\end{figure}

Figure~\ref{fig:gspphotgspspec} shows the comparison of the atmospheric parameters for the sources derived by both GSP-Phot \citep{DR3-DPACP-156} and GSP-Spec \citep{DR3-DPACP-186}. The agreement between the two methods for the temperature is reasonable, has an offset on the surface gravity for small \logg, and a very large dispersion for the global metallicity.
\begin{figure*}\begin{center}
\includegraphics[width=0.52\columnwidth]{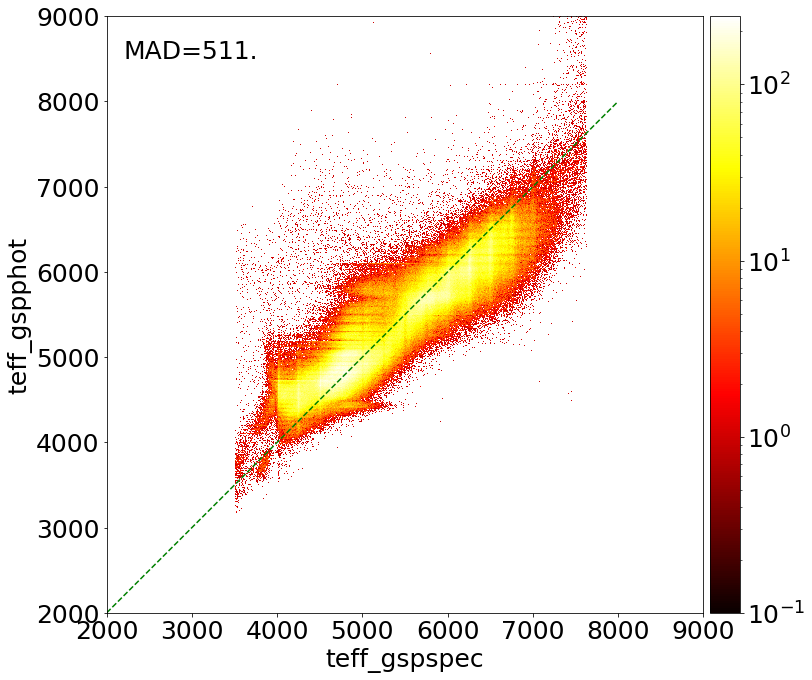}
\includegraphics[width=0.49\columnwidth]{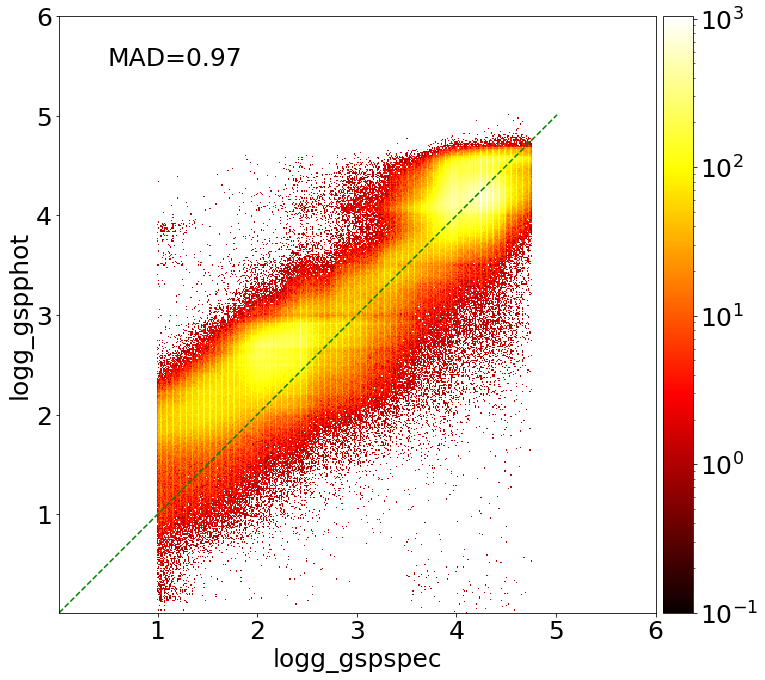}
\includegraphics[width=0.51\columnwidth]{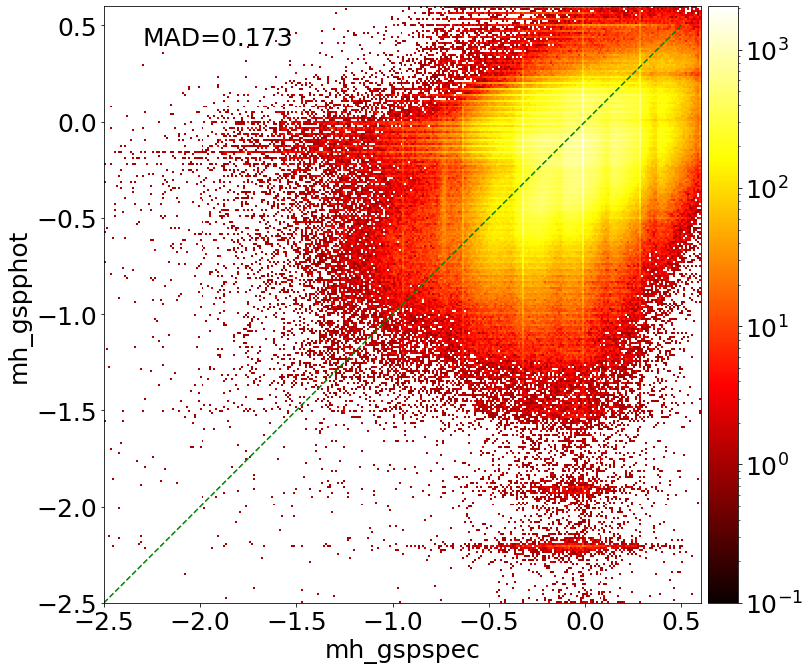}
\caption{Density plot of the comparison of the temperature (left), surface gravity (middle), and global metallicity (right) provided by GSP-Phot (y-axis) and GSP-Spec (x-axis). GSP-Phot has been filtered with \fieldName{parallax\_over\_error}$>$5 and \fieldName{teff\_gspphot}$<$10000. GSP-Spec parameters have been filtered with \fieldName{flags\_gspspec}[1,4,8,13]=0 for \teff, \fieldName{flags\_gspspec}[2,5,8,13]=0 for \logg,\ and \fieldName{flags\_gspspec}[3,6,8]=0 for [M/H]. The dashed green line shows the one-to-one correspondence. The median absolute deviation is indicated in each panel.}
\label{fig:gspphotgspspec}
\end{center}\end{figure*}

Figure~\ref{fig:mh_apogee} shows the comparison of the atmospheric parameters derived by GSP-Phot and GSP-Spec with APOGEE DR16 \citep{2020ApJS..249....3A}. The plots look similar to GALAH DR3. The large dispersion of the GSP-Phot metallicity explains Fig.~\ref{fig:gspphotgspspec}. The GSP-Spec \teff is slightly overestimated for Teff$>5500$ versus APOGEE and GALAH, while GSP-Phot is not. 
%GSP-Spec \teff is under-estimated for hot stars which have less spectral lines, which is confirmed by clusters \citep{DR3-DPACP-160}.
GSP-Phot \logg\ for small \logg\ is overestimated. GSP-Spec \logg\ is globally underestimated, and a correction is proposed in \cite{DR3-DPACP-186}. GSP-Spec median offsets and dispersion versus external catalogues after proposed corrections are quite good and are provided in \cite{DR3-DPACP-186}.
We also caution the user against using GSP-Spec \logg\ values for AGB stars.
\begin{figure}\begin{center}
\includegraphics[width=0.49\columnwidth]{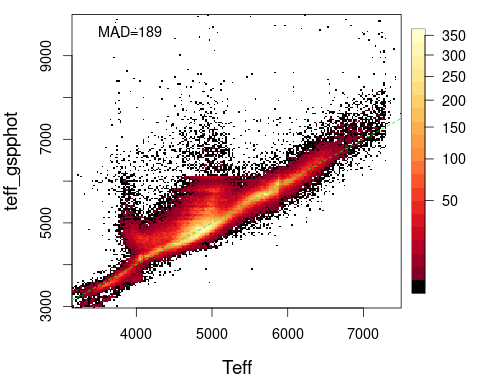}
\includegraphics[width=0.49\columnwidth]{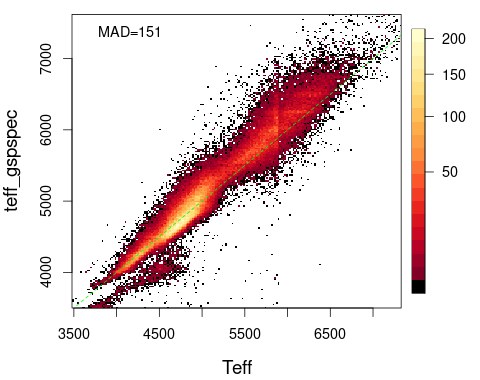}
\includegraphics[width=0.49\columnwidth]{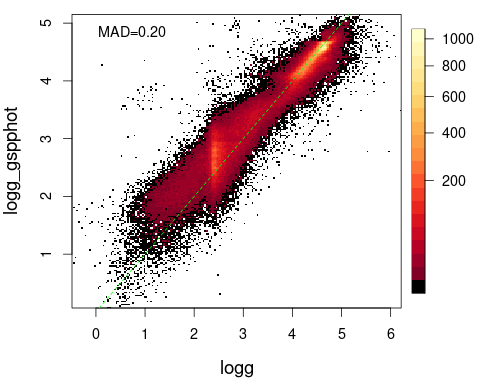}
\includegraphics[width=0.49\columnwidth]{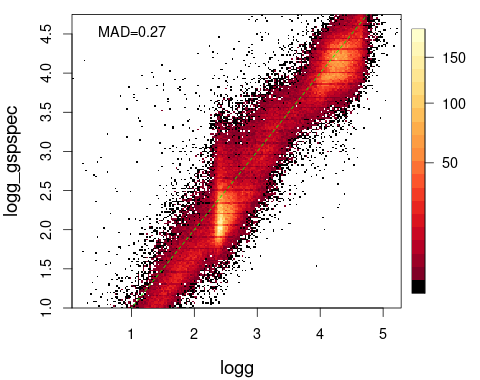}
\includegraphics[width=0.49\columnwidth]{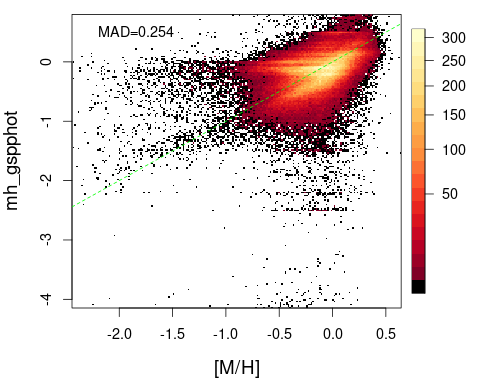}
\includegraphics[width=0.49\columnwidth]{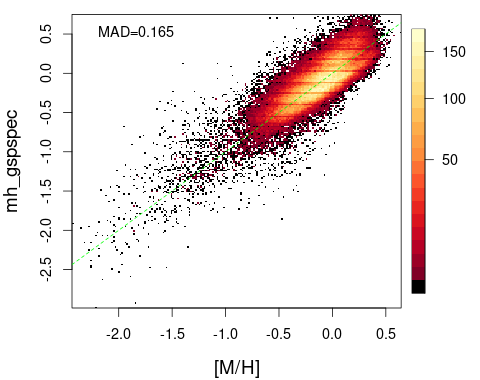}
\caption{Density plot of the comparison of the temperature (top), surface gravity (middle), and global metallicity (bottom) provided by GSP-Phot (left) and GSP-Spec (right) with APOGEE DR16. GSP-Phot has been filtered with \fieldName{parallax\_over\_error}$>$10, and \fieldName{teff\_gspphot}$<$10000. GSP-Spec parameters have been filtered with \fieldName{flags\_gspspec}[1,4,8,13]=0 for \teff, \fieldName{flags\_gspspec}[2,5,8,13]=0 for \logg,\ and \fieldName{flags\_gspspec}[3,6,8]=0 for [M/H]. The RVS spectrum signal-to-noise ratio was not filtered.}
\label{fig:mh_apogee}
\end{center}\end{figure}

The shift in metallicity for GSP-Phot shown in Fig.~\ref{fig:gspphotgspspec} and Fig.~\ref{fig:mh_apogee} is larger than the literature values for open clusters. In Fig.~\ref{fig:clusters} we averaged (median)  the \fieldName{mh\_gspphot} inside each cluster. We used only open clusters (i.e of about solar metallicity).  We find a trend of the difference with literature values (Gaia-literature) versus  [M/H] literature. A zero-point difference of $-0.55$  is found. GSP-Phot metallicities should be used with caution, and ideally, with a calibration (see \citealt{DR3-DPACP-156}).
\begin{figure}\begin{center}
\includegraphics[width=0.9\columnwidth]{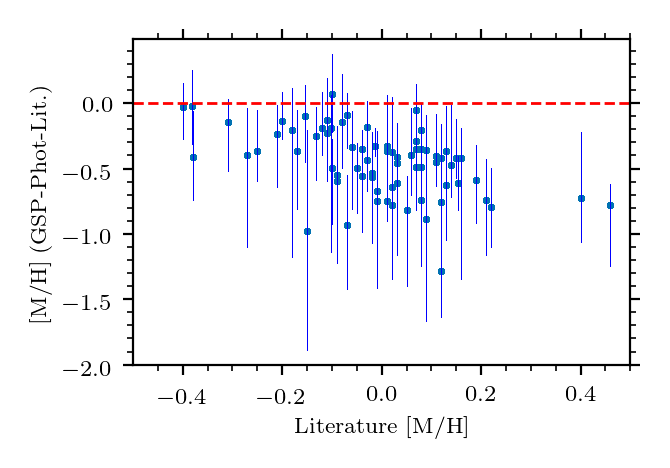}
\caption{Comparison of \fieldName{mh_gspphot} with literature values from open clusters.  We plot the median value for each cluster.  The error bars show the dispersion around the median. The red line indicates the zero value.}
\label{fig:clusters}
\end{center}\end{figure}

All the GSP-Spec parameters are found to be correlated with magnitude and metallicity. Figure~\ref{fig:gspspec_correlations} illustrates this correlation for \fieldName{teff\_gspspec} using APOGEE DR16. The plot is similar to that for GALAH DR3. This correlation with magnitude and metallicity leads to other unexpected correlations with extinction or sky position that are shown in the comparison with the GUMS model.
\begin{figure}\begin{center}
\includegraphics[width=0.49\columnwidth]{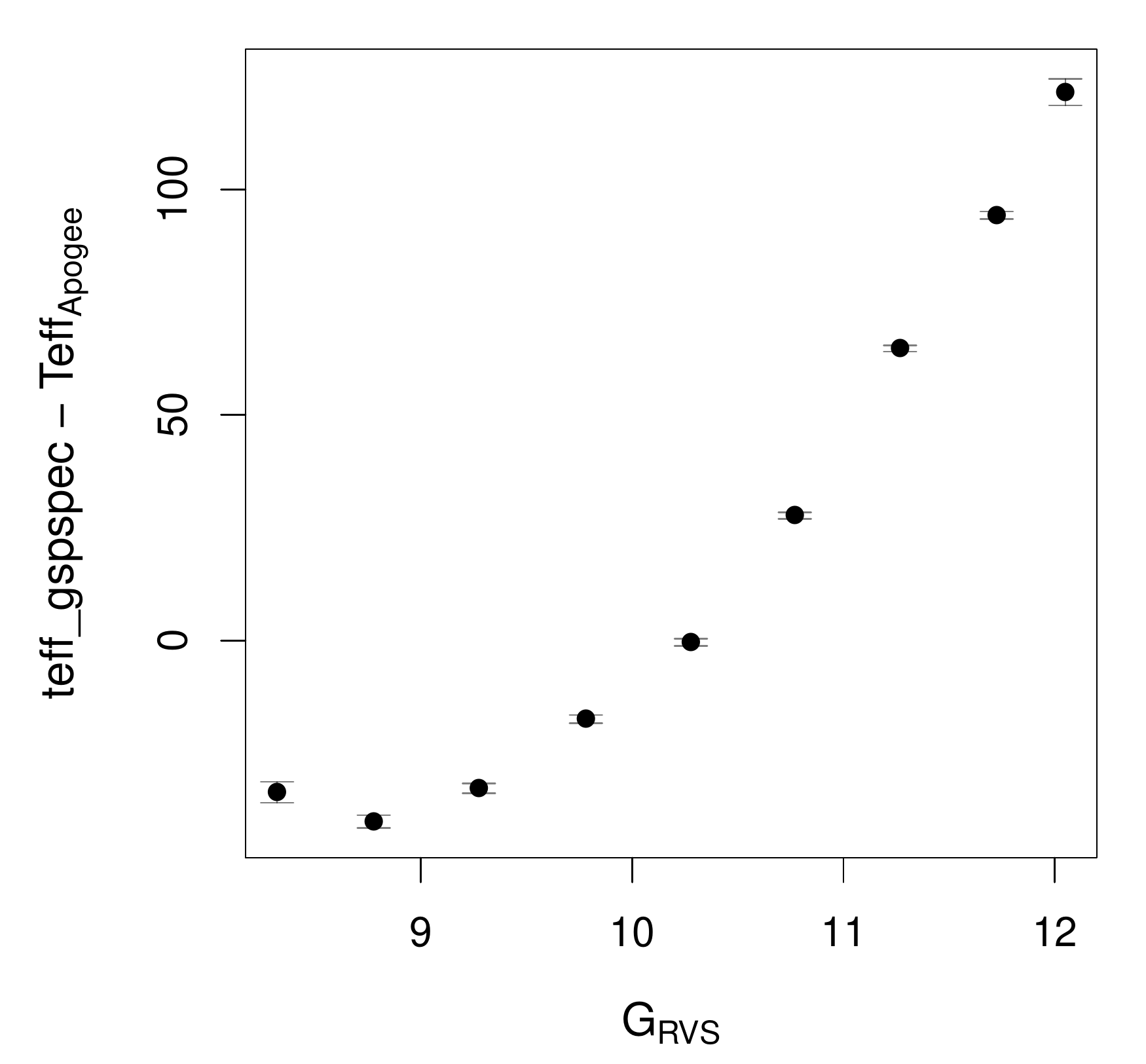}
\includegraphics[width=0.49\columnwidth]{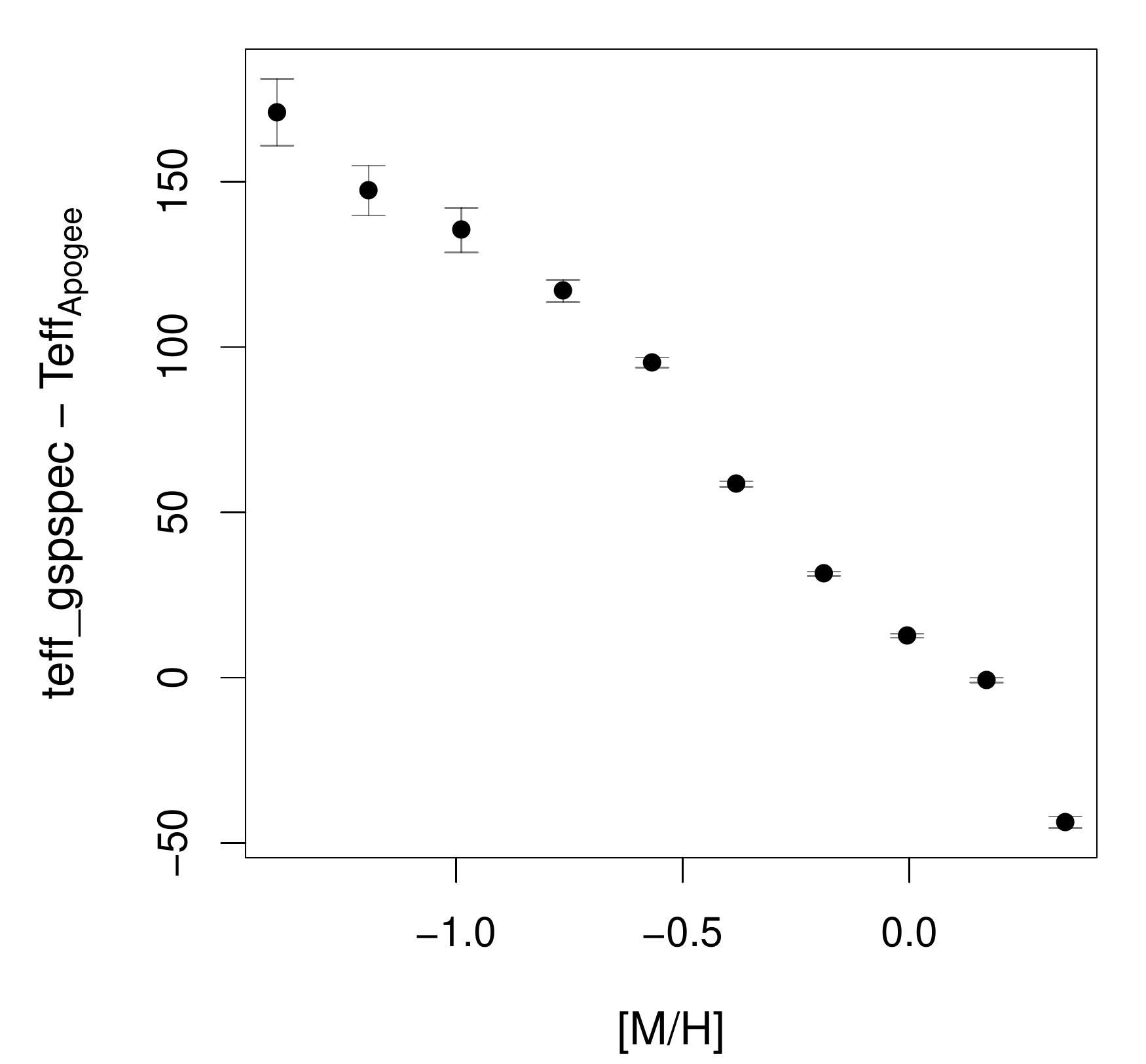}
\caption{Correlation between the GSP-Spec parameters and magnitude (left) and metallicity (right) illustrated here with the temperature residuals compared to APOGEE DR16.}
\label{fig:gspspec_correlations}
\end{center}\end{figure}

We also tested the correlation of the abundances with magnitude, temperature, \logg,\ and \fieldName{logchisq_gspspec}  with open clusters and present it for NGC~7789 in Fig.~\ref{fig:trends_mh}.
%FlagVsiniT==0 $\&$ FlagVsiniG==0 $\&$ FlagVsiniM==0 $\&$ FlagVradT==0 $\&$ FlagVradG==0 $\&$ FlagVradM==0 $\&$ FlagNoise==0 $\&$ FlagExtraP$<$=2 $\&$ FlagNegFlux==0 $\&$ FlagNaNFlux==0 $\&$ FlagEmiss==0 $\&$ FlagNullUnc==0 $\&$ FlagKM$<$=0. 
The plots show clear positive trends between \fieldName{mh_gspspec} and the two stellar parameters \fieldName{teff\_gspspec} and \fieldName{logg\_gspspec}. This correlation is similar to that obtained with magnitude, as expected for a cluster in which temperature and gravity are correlated with magnitude. It is also consistent with what is seen in the comparison with external catalogues (Fig.~\ref{fig:gspspec_correlations}). For \fieldName{alphafe_gspspec}, we observe correlations of the opposite sign.  The [M/H] and [$\alpha$/Fe] calibrations proposed in \cite{DR3-DPACP-186} alleviate these trends, but they do not remove them completely.
\begin{figure}\begin{center}
\includegraphics[width=0.95\columnwidth]{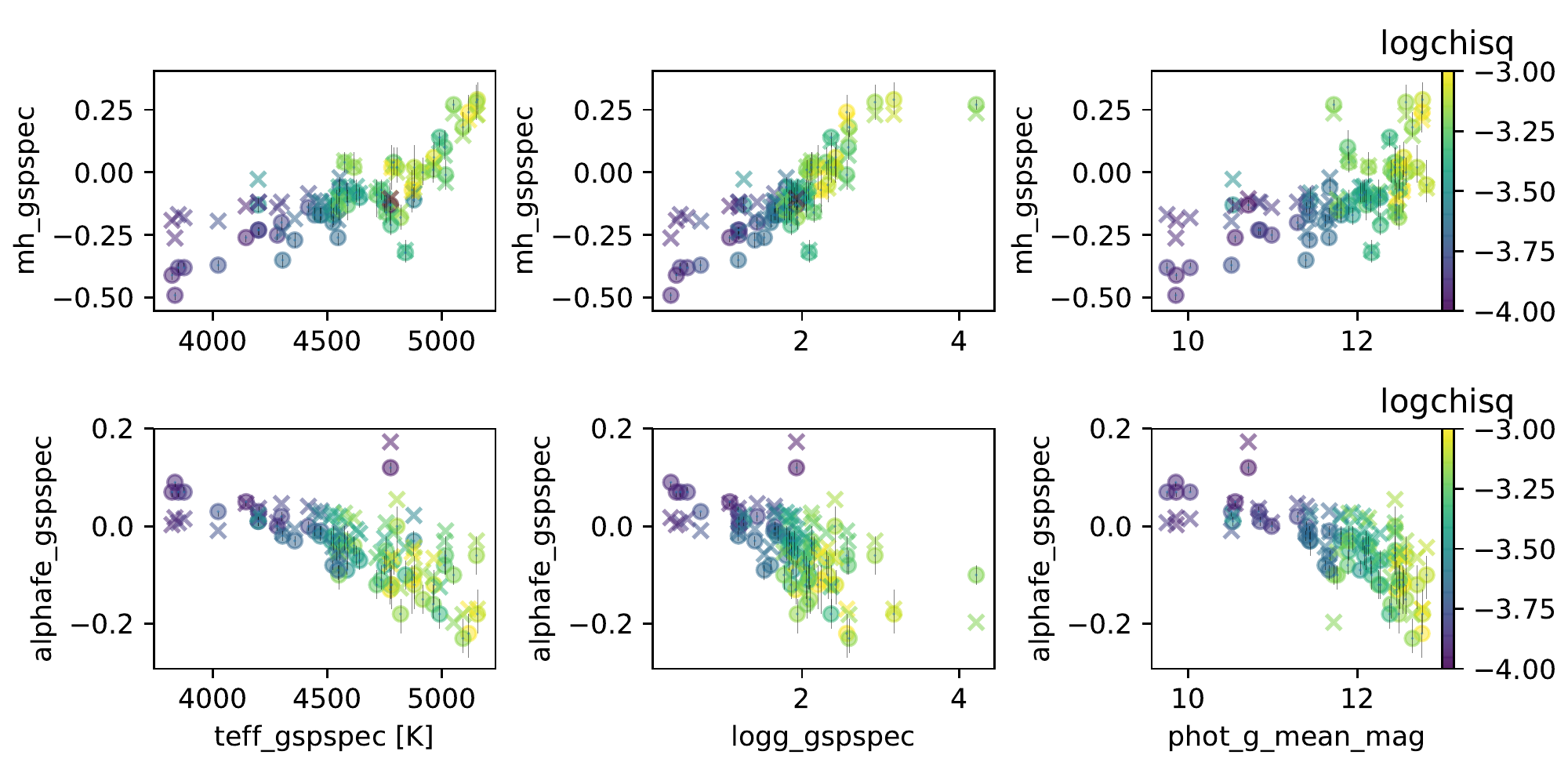}\\
\caption{Abundance trends for \fieldName{mh_gspspec} (top) and \fieldName{alphafe_gspspec} (bottom) as a function of \fieldName{teff\_gspspec} (left column), \fieldName{logg\_gspspec} (middle column), and  \fieldName{phot\_g\_mean\_mag} (right column) for the stellar members of NGC~7789. The symbols are colour-coded as a function of \fieldName{logchisq\_gspspec}. Circles indicate the uncorrected \fieldName{alphafe_gspspec} values, and crosses represent the calibrated \fieldName{mh_gspspec} and \fieldName{alphafe\_gspspec} values. The parameters have been filtered with \fieldName{flags_gspspec}[1:7,9:13]=0 and \fieldName{flags_gspspec}[8]$<=$2.}
\label{fig:trends_mh}
\end{center}\end{figure}

The GSP-Spec abundances do not correlate very well with external catalogue values in general (\cite{DR3-DPACP-186}). However, calibration formulas are proposed in \cite{DR3-DPACP-186}, and \cite{DR3-DPACP-104} showed that they allow retrieving the expected chemo-kinematical correlations in the disc.

GSP-Spec ANN metallicities (available in \tableName{astrophysical\_parameters\_supp}) have underestimated uncertainties (Table~\ref{tab:widebinaries}) and are offset by $\sim$-0.2 with respect to APOGEE DR16. See \cite{DR3-DPACP-186} for a proposed calibration of the GSP-Spec ANN parameters as well.

The MSC metallicity and gravity are overestimated compared to those obtained with GALAH \citep{2020A&A...638A.145T}. Hot stars are found to be assigned a temperature of about 7500~K due to the empirical calibration based on APOGEE. The poor convergence of the MSC values can be flagged as low \fieldName{logposterior_msc} values. The test using wide binaries (Table~\ref{tab:widebinaries}) indicates a low number of outliers for the metallicity, but the errors are so large that most of the possible values are covered. We advise using the MSC parameters with caution in general (see also \cite{DR3-DPACP-160} and the on-line documentation).  

% uncertainties
Tests using wide binaries (Table~\ref{tab:widebinaries}) and open clusters shows a strong underestimation of the errors for all GSP-Phot parameters with an associated large number of outliers. For GSP-Spec, they show an underestimation of the errors for \fieldName{mh\_gspspec}, \fieldName{alphafe\_gspspec}, \fieldName{cafe\_gspspec}, and \fieldName{crfe_gspspec}, while for some other elements, the uncertainties are slightly overestimated.
GSP-Spec values were discretised at two decimals, except for \fieldName{dibew_gspspec} , which was discretised at three decimals, and Teff, which was  stored as an integer. This might cause some parameters to have similar upper and lower values, in which case the discretisation step should be used as an uncertainty estimate. 

% MCMC
The published GSP-Phot MCMC  samples contain 2000 points for  $G< 12$ , but only the last 100 points are made available  for $G>12$, except for a random 1\% subset that was given the full 2000 points. When only 100 points are available, the upper or lower values of the GSP-Phot parameters, which were determined on the full 2000 steps, may not be fully consistent with the MCMC. This inconsistency is an indication of convergence issues. However, failed convergence usually does not relate to strong outliers, which are cases when the MCMC has converged to a very different solution.
We find that 18\% of the 2000-point chains present some problems such as multiple solutions, local maxima of the posterior probability, or edge effects.
The MSC inflated their errors in post-processing, therefore the MSC MCMCs are not consistent with the provided upper or lower values.

\subsection{Distance and absolute magnitude} 

\begin{figure}\begin{center}
\includegraphics[width=0.9\columnwidth]{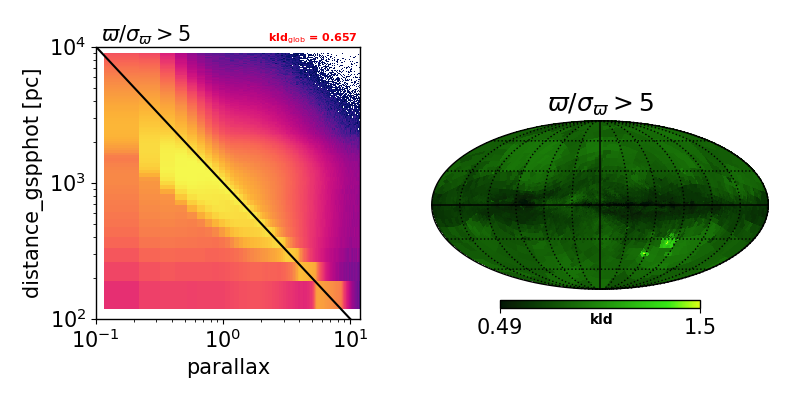}
\caption{Global distribution of \fieldName{distance_gspphot} against \fieldName{parallax} for sources with $\varpi/\sigma_{\varpi} >5$. The solid black line represents the 1/parallax relation. The map ($l,b$) on the right shows the sky distribution of the clustering between the two parameters.}
\label{fig:inverseplx}
\end{center}\end{figure}

The global distribution of GSP-Phot distances against parallax is shown in Fig.~\ref{fig:inverseplx}, where we consider the sources with $\varpi/\sigma_{\varpi} >5$. While a large fraction of sources follows the inverse parallax curve, 37\% are 5$\sigma$ outliers (considering only the parallax error). We measured the clustering in this space using the Kullback-Leibler divergence (KLD; \citealt{KLD}, \citealt{EDR3-DPACP-126}), which is higher away from the plane and in particular around the large and small Magellanic clouds (LMC and SMC).

Strong outliers in the GSP-Phot distances are seen in the comparison of the estimates of the two wide binary components (Appendix~\ref{sec:widebinaries}), while the relative precision of the parallax is better than 20\% in this sample. The distances are shown to be systematically underestimated at large distances, and the relative parallax precision is poor when the known cluster distances of \cite{DR3-DPACP-160} are used. This is also confirmed with the APOGEE DR16 red clump sample. This seems to be due to a too strong prior \citep{DR3-DPACP-156}. The MSC distances \fieldName{distance\_msc} are shown to present a higher dispersion than the GSP-Phot distances even for known binaries by \cite{DR3-DPACP-160}.

The GSP-Phot absolute magnitude estimate \fieldName{mg\_gspphot} is compared to the absolute magnitude computed directly from the parallax for a sample of stars with negligible extinction in Fig.~\ref{fig:mgbias}. It shows the combination of distance outliers (leading the strong outliers) and extinction overestimation for stars with $M_G\gtrsim7$ (see Fig.~\ref{fig:gsphota0_hrd}, leading to a bias). Moreover, \fieldName{mg\_gspphot} is not correctly estimated in  stars farther away than 1-2 kpc as an effect of the underestimated distance. This is clearly illustrated in  Fig.\ref{fig:cu8par_qa_nss_MG1039}, where the distance modulus (m-M) is derived from \fieldName{distance\_gspphot}.

\begin{figure}\begin{center}
\includegraphics[width=0.9\columnwidth]{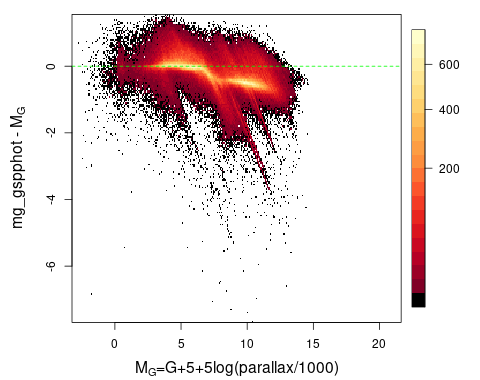}
\caption{Density plot of the difference between \fieldName{mg\_gspphot} and the absolute magnitude computed directly with the parallax for a sample of stars with negligible extinction (\a0$<0.05$ according to \cite{2019A&A...625A.135L}) and a $\fieldName{parallax_over_error}>10$.}
\label{fig:mgbias}
\end{center}\end{figure}

 \begin{figure}[h!]
\begin{center}
\includegraphics[width=0.9\columnwidth]{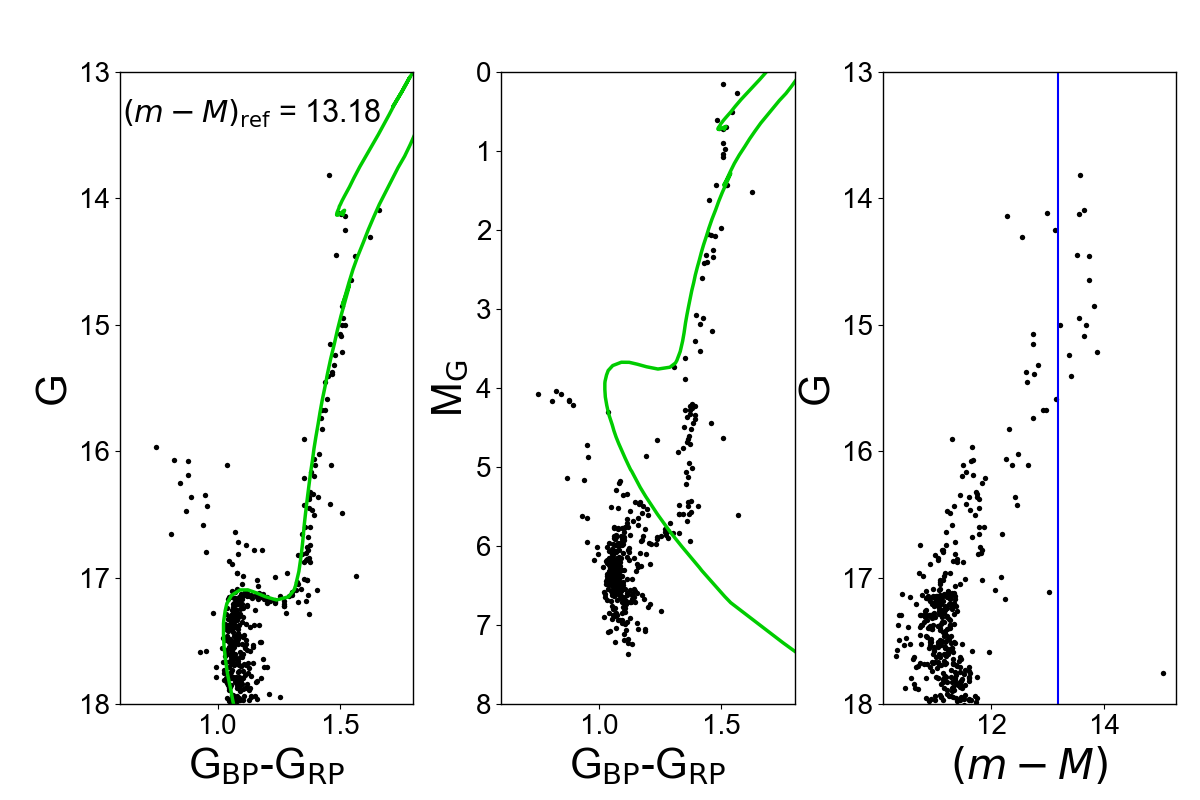}
\end{center}
\caption{Colour-magnitude diagram  of NGC~6791  (left panel),  \fieldName{mg_gspphot} vs \bprp(central panel), and G vs distance modulus (m-M) derived from \fieldName{distance_gspphot} (right panel). The blue line in the right panel shows the literature value, and the green lines in the left and central panels show the PARSEC isochrone, which has the same parameters as the cluster.}
\label{fig:cu8par_qa_nss_MG1039}
\end{figure}

We recommend using the deviation between the GSP-Phot distances and the parallax\footnote{ \texttt{(parallax - 1000/distance\_gspphot)/parallax\_error}} to filter GSP-Phot outliers. For a number of usages, it may be preferable to use the parallax to estimate the distance and absolute magnitude (see \cite{DR2-DPACP-38}) over the GSP-Phot estimates.

\subsection{Stellar evolution parameters}

The stellar evolution parameters radius, mass, age, evolution stage, and gravitational redshift are provided by the FLAME module \citep{DR3-DPACP-157}. They are derived either from GSP-Phot parameters (fields named \fieldName{\_flame} in the table \tableName{astrophysical\_parameters}) or from GSP-Spec parameters (fields named \fieldName{\_flame\_spec} in \tableName{astrophysical\_parameters\_supp}). For the mass, age, and evolutionary stage, they use solar metallicity evolution models. The estimates of these parameters for non-solar metallicity stars should therefore be used with caution. 

Figure~\ref{fig:jsdcradius} shows the comparison between the FLAME radius and the radius from the JSDC stellar diameter catalogue \citep[][v2, selecting stars with $\chi^2<2$]{2017yCat.2346....0B} for stars with relative parallax uncertainties smaller than 10\%. The parallax is used to transform the JSDC angular diameter into radius. The radius derived by FLAME using GSP-Spec Teff, \fieldName{radius\_flame\_spec}, is overestimated for blue main-sequence stars and for red giants, but it is underestimated for very red giants (\bpminrp$>$2.2). The radius derived by FLAME using GSP-Phot Teff, \fieldName{radius\_flame}, has the same properties, but fewer outliers than \fieldName{radius\_gspphot} provided directly by GSP-Phot because FLAME directly uses the parallax to derive the luminosity for this sample with a good parallax signal-to-noise ratio (see \fieldName{flags_flame}). We therefore recommend using \fieldName{radius\_flame} for a radius estimation. 

\begin{figure}\begin{center}
\includegraphics[width=0.49\columnwidth]{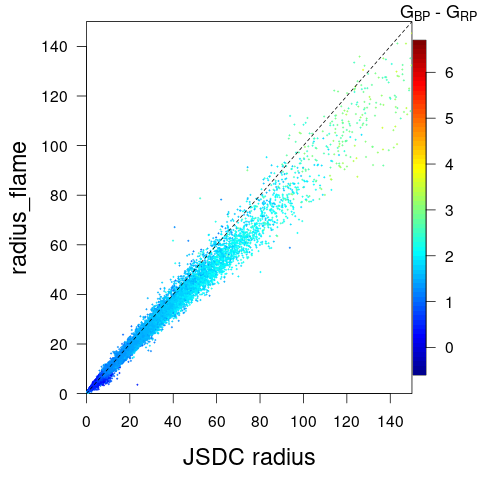}
\includegraphics[width=0.49\columnwidth]{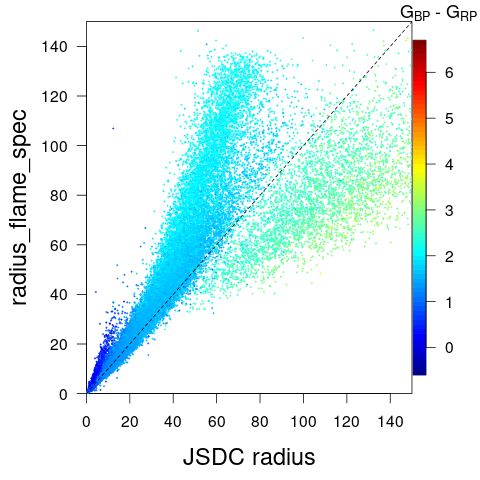}
\caption{Comparison between JSDC radius and the FLAME radii based on GSP-Phot (left) or GSP-Spec (right), colour-coded with the {\bprp} colour.}
\label{fig:jsdcradius}
\end{center}\end{figure}

Masses from FLAME compare well with asteroseismic estimates for dwarfs and subgiants \citep[using][]{2017ApJS..233...23S,2021ApJ...915...19G}, but strong outliers are seen for giants \citep[using][]{2018ApJS..236...42Y}. 
Comparison with the GUMS model confirms the presence of a high-mass tail in the FLAME data that is not predicted by the model. This tail is present in all Galactic directions, even at high latitudes. It is associated with an excess of young stars. These young ($<2$Gyr) and massive ($>2$\Msun) stars are on the giant branch. 
We therefore recommend using the FLAME masses with \fieldName{flags_flame[_spec]} first character 1 (giant flag) only within the 1-2\Msun range and with caution, and taking their large uncertainties into account.

The overestimation of the GSP-Phot extinction for low-mass stars ($M_G\gtrsim7$, Fig.~\ref{fig:gsphota0_hrd}) also has an impact on the FLAME parameters. The impact on masses $\mathcal{M}\lesssim0.7$ is illustrated in Appendix D of \cite{DR3-DPACP-100}. It has an impact on the luminosity similar to what is shown in Fig.~\ref{fig:mgbias}. 

Strong outliers in the luminosity of giants are visible in the APOGEE red clump sample. A few mismatches of the evolutionary stage may occur also for giants that are confused with dwarfs with high extinction. They can be spotted in an HR diagram using an independent extinction estimate. 

The gravitational redshift determined by FLAME is compared to the one used in GALAH DR3 \citep{2021MNRAS.508.4202Z} in Fig.~\ref{fig:galahgravred} for sources with a gravitational redshift error from FLAME lower than 1\kms. The gravitational redshift based on GSP-Spec (\fieldName{gravredshift\_flame\_spec}) has fewer outliers than the redshift based on GSP-Phot (\fieldName{gravredshift\_flame}), but it has a small bias of 0.05 \kms , corresponding to the bias in \logg\ discussed above. 
\begin{figure}\begin{center}
\includegraphics[width=0.49\columnwidth]{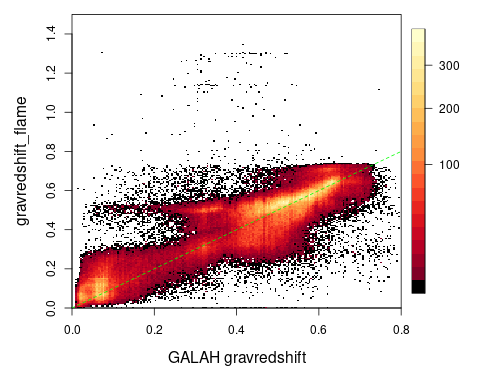}
\includegraphics[width=0.49\columnwidth]{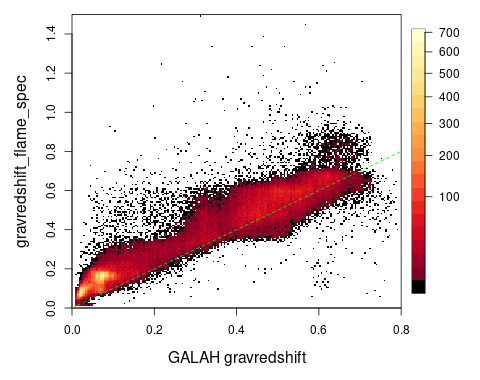}
\caption{Density plot of the comparison between the GALAH gravitational redshift and the FLAME redshifts based on GSP-Phot (left) or GSP-Spec (right).}
\label{fig:galahgravred}
\end{center}\end{figure}

%+++++++++++++++++++++++++++++++++++++++++++++++++++++++++++++++++++++++++
\section{QSO and galaxies}\label{sec:extragal}
%+++++++++++++++++++++++++++++++++++++++++++++++++++++++++++++++++++++++++

\gdrthree includes two tables of extragalactic candidate sources, one for quasars and one for galaxies, called \tableName{qso_candidates} and \tableName{galaxy_candidates} (QSO and galaxy tables, for simplicity). These tables contain two main types of added value columns: on the one hand, we can use the different labels that are provided to tune the purity-to-completeness ratio of the sample, and on the other hand, each table also contains physical properties of the objects such as redshift, size, or variability. The astrophysical parameters \citep{DR3-DPACP-157} associated with these tables, that is, classification and redshifts, are described in \cite{DR3-DPACP-158}, the surface brightness profiles are described in \cite{DR3-DPACP-153}, the variability is presented in \cite{DR3-DPACP-165} and in \cite{DR3-DPACP-167} for AGNs. Moreover, a global analysis of these tables is presented in \cite{DR3-DPACP-101}.
It is worth noting that these tables have been constructed with the aim of completeness, and as we show below, this means that their default purity is rather low. However, it is possible to obtain a high-purity subsample \citep{DR3-DPACP-101}, as discussed below.

\subsection{Purity}
% contamination 
The different labels that are included within the QSO and galaxy tables can be used to create a subsample with different properties (see section 8 of \citep{DR3-DPACP-101} for a selection leading to 94\% and 95\% purity in the QSO and galaxy tables, respectively). The QSO and galaxy tables contain three common labels: \fieldName{vari_best_class_name} = "AGN"/"GALAXY" (classification according to the stellar variability patterns \citep{DR3-DPACP-165}), \fieldName{classlabel_dsc} = "quasar"/"galaxy" (from the discrete source classifier \citep{DR3-DPACP-157}), 
\fieldName{classlabel_dsc_joint} = "quasar"/"galaxy" (similar to the previous, but more restrictive since it requires DSC-Specmod and DSC-Allosmod to agree, both with a score higher than 50\%) and \fieldName{classlabel_oa}\footnote{OA distinguishes extragalactic sources by bins of redshift, so we select all quasars/galaxies at all redshifts to create this label.} (assigned by the self-organising map (SOM) \citep{DR3-DPACP-157}). It is worth noting that the results of the SOM were not used to construct these tables. In other words, this label was attached to the sources that were previously selected as candidates by other means.
\noindent In addition to these shared class labels, we can also identify unique labels for each table. In the QSO table we have access to the \fieldName{astrometric_selection_flag} (ASF), which allows us to select only sources with a high probability of being quasars based on their astrometry. We can also use the \fieldName{source_selection_flags} to isolate the QSOs in the \tableName{qso_catalogue_name} table (bit 3 set to 1), hereafter called QuasarObject list, which effectively corresponds to the sources that are found in well-known QSO catalogues that had enough raw data to be processed successfully by the QSO pipeline. Finally, in the galaxy table, we can select the sources whose morphology was fit reliably. We refer to this subset as EO solution.

In Fig.~\ref{fig:extragal_astrometric_prop} we show the astrometric properties, normalised by their formal uncertainties, of the different subsamples described above. Because extragalactic sources are so far away from the Sun, the astrometry of these objects might be expected to be dominated by observational errors. This is what we indeed observe for some of the subsamples, for example those in panels a and c, and also in panel e in the QSOs and panel b in the galaxies. However, the other subsamples show clear deviations from the expected standard normal probability distribution. These deviations in the DSC and OA subsamples are due mostly to the astrometric signal of the Magellanic clouds and the Galactic disc (Fig.~\ref{fig:extragal_sky}). We note, however, that while $\sim 94$\% of the sources in the QSO table have a 5p or 6p astrometric solution, the galaxy table is mostly dominated by 2p sources ( $\sim 71$\%). Therefore, the conclusions we draw concern only a portion of the galaxy table. In either case, it is clear that some subsamples, namely \fieldName{vari_best_class_name} and \fieldName{classlabel\_dsc\_joint}, are purer than others (\fieldName{classlabel\_dsc} and \fieldName{classlabel\_oa}) that were built for completeness. 

\begin{figure*}\begin{center}
\includegraphics[width=0.95\linewidth]{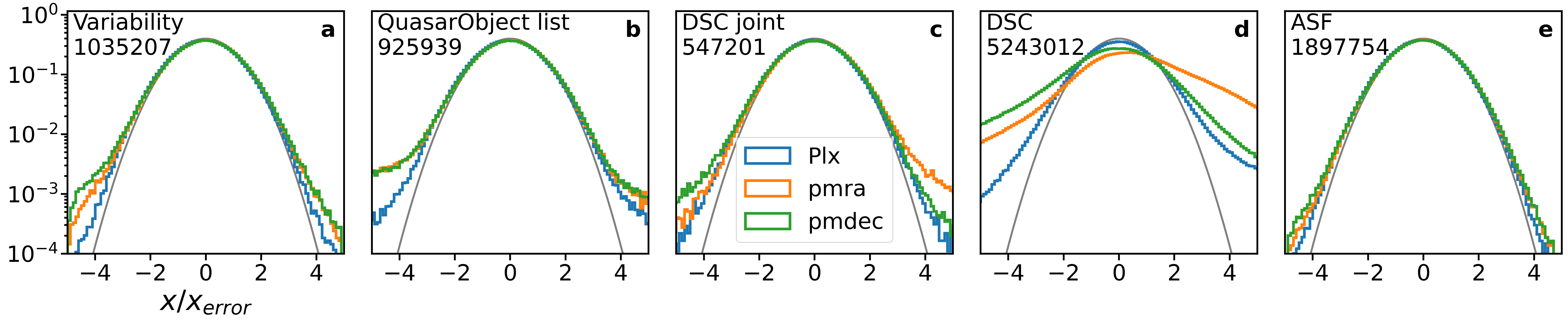}\\
\includegraphics[width=0.95\linewidth]{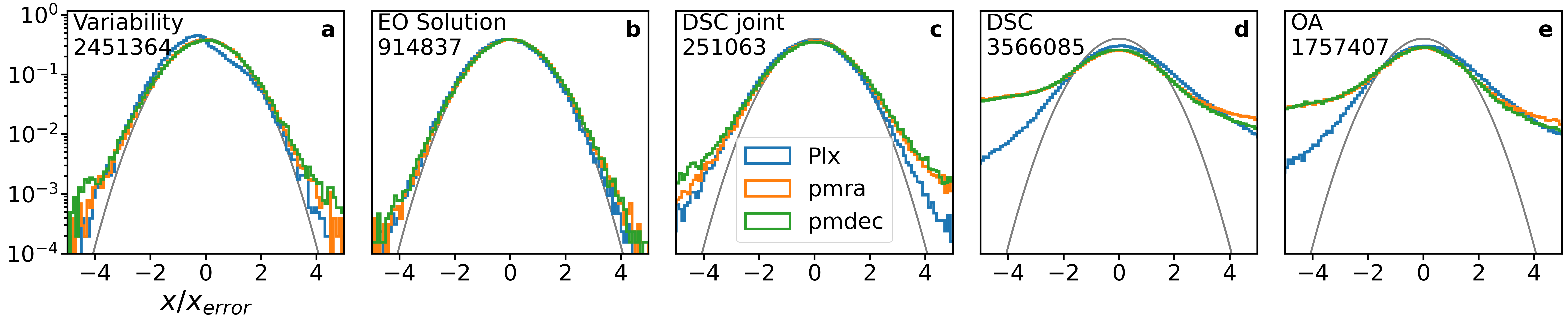}
\caption{Astrometric properties of the different subsamples contained within the QSO (top) and galaxy (bottom) candidate tables. Each panel contains the distribution of parallaxes and proper motions, normalised to their errors. The grey line corresponds to a normal distribution.}
\label{fig:extragal_astrometric_prop}
\end{center}\end{figure*}
 
The sky distribution of the sources in these subsamples is presented in Fig.~\ref{fig:extragal_sky}. These plots are difficult to relate directly to the purity as some modules remove the LMC and SMC and the disk plane by force or using criteria that depend on the density, while some others, such as DSC, do not. Moreover, a constant misclassification rate over the sky leads to a higher density of misclassified objects where the objects density is higher. 
However, in the LMC and SMC areas, more than 10\% of the sources are in the \tableName{qso_candidates} table, so here the classification does not work well.

\begin{figure}\begin{center}
\includegraphics[width=0.5\columnwidth]{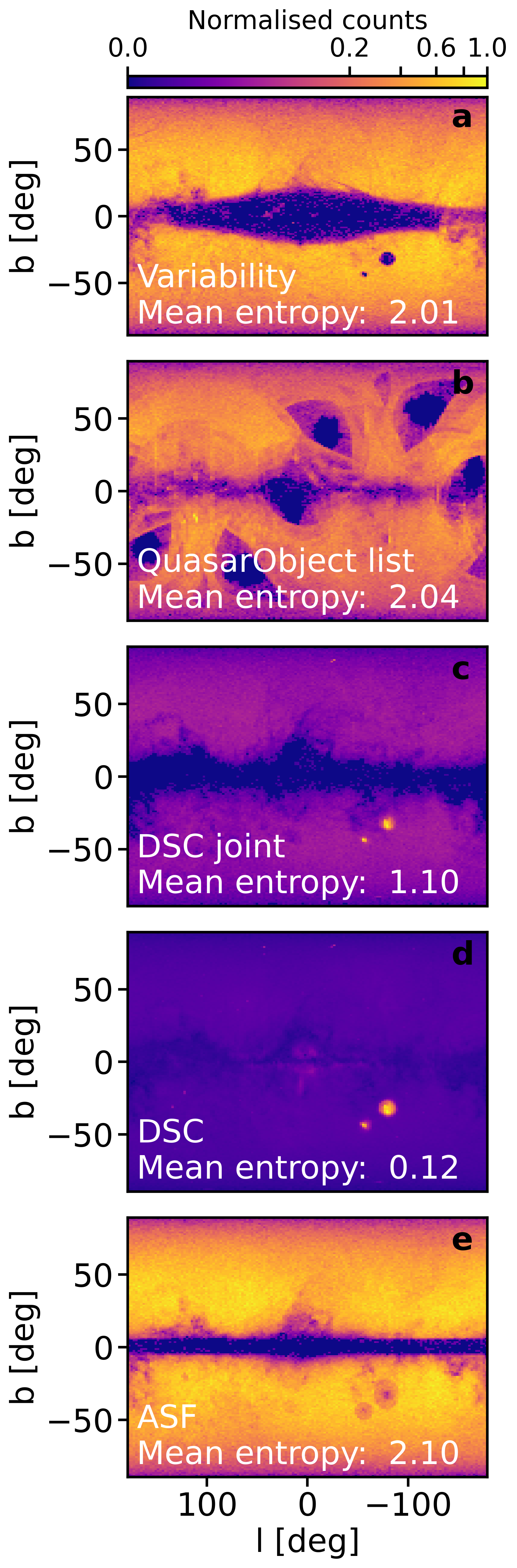}\includegraphics[width=0.5\columnwidth]{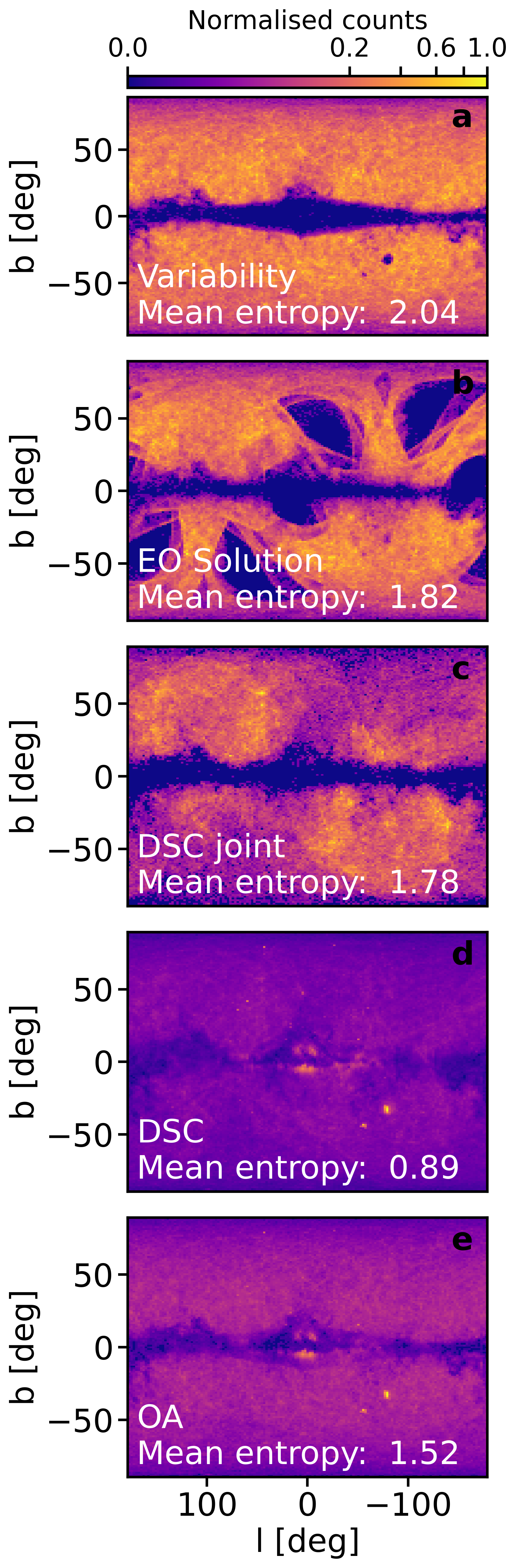}
\caption{Sky distribution of the different subsamples contained within the QSO (left) and galaxy (right) candidate tables.}
\label{fig:extragal_sky}
\end{center}\end{figure}

% contamination on the HRD
A good idea of the main stellar types of the stellar contaminants in the QSO and galaxy tables can be obtained by positioning those with a relative parallax uncertainty lower than 20\% in an HR diagram in Fig.~\ref{fig:extragal_hrd}. It shows that the QSO candidate stellar contaminants are mainly white dwarfs and stars with \bprp$\sim$0.4, while galaxy candidate stellar contaminants are mainly stars with \bprp$\sim$1.4 or 0.8. Fig.~\ref{fig:extragal_hrd} also shows that the criteria for the purer samples proposed in \cite{DR3-DPACP-101} are efficient, but still retain a few contaminants. 
%The EO contaminants are mostly young stars on nebulae and binaries.
We use here \fieldName{host_galaxy_detected='true'} instead of $\fieldName{host_galaxy_flag}<6$ as the latter leads to eight times more contaminants in our sample. These are due to the EO input catalogue, however, the host galaxy has not been detected for them.
By construction, the entire QSO sample of Fig.~\ref{fig:extragal_hrd} has \fieldName{astrometric_selection_flag='false'}.
% \footnote{For QSOs: \texttt{\fieldName{gaia_crf_source}='true' OR
% \fieldName{host_galaxy_flag}<6 OR
% \fieldName{classlabel_dsc_joint}='quasar' OR
% \fieldName{vari_best_class_name}='AGN'
% }\\ 
% For Galaxies: \texttt{\fieldName{radius_sersic} IS NOT NULL OR
% \fieldName{classlabel_dsc_joint}='galaxy' OR
% \fieldName{vari_best_class_name}='GALAXY'
% }
% } 

\begin{figure}\begin{center}
\includegraphics[width=0.8\columnwidth]{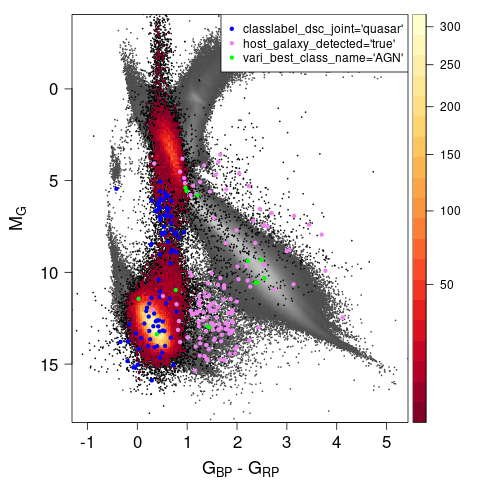}
\includegraphics[width=0.8\columnwidth]{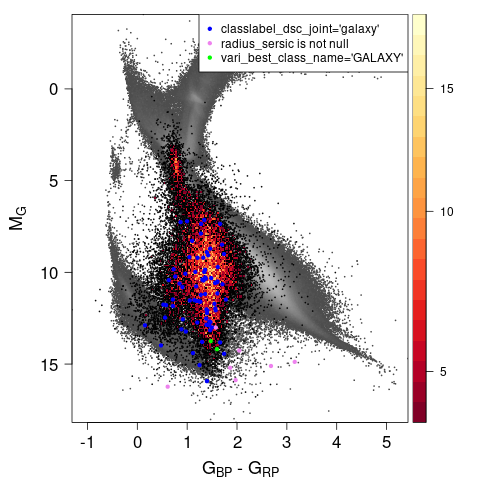}
\caption{{\gdrthree} low-extinction HR diagram (grey scale). The position of sources in the QSO (top) and galaxy (bottom) candidate tables with $\fieldName{parallax_over_error}>5$ is overplotted with a red scaling with the square root of the number of sources. Colour points correspond to the stricter selection of candidates proposed in \cite{DR3-DPACP-101}.}
\label{fig:extragal_hrd}
\end{center}\end{figure}

% EO 
\subsection{Morphological parameters}

We compared the extended object morphological parameters of the galaxy table \citep{DR3-DPACP-153} with the GAMA \citep{2012MNRAS.421.1007K} and Dark Energy Survey \citep[DES][]{Tarsitano:2018} S\'ersic profiles and with the SDSS DR16 \citep{2020ApJS..249....3A} de Vaucouleurs profiles. \autoref{fig:sersic_des_comparison} illustrates the comparison with the DES S\'ersic profiles. Saturation of the effective radius \fieldName{radius\_sersic} and \fieldName{radius\_de\_vaucouleurs} at 8000~mas and of the S\'ersic $n$-index at 8 is visible. It corresponds to the boundaries of the algorithm. Compared to DES, the \gdrthree index seems to be spread out more or less uniformly, with DES preferring $n=4$. Essentially, the comparison with external catalogues of galaxy profiles shows an overestimation of the S\'ersic index and an underestimation of the ellipticity. Both are a consequence of the fact that \gaia\ observes a smaller area around the galaxies than external catalogues, which prefer central measurements and are biased towards bulges \citep{DR3-DPACP-153}.

\begin{figure}\begin{center}
\includegraphics[width=0.95\columnwidth]{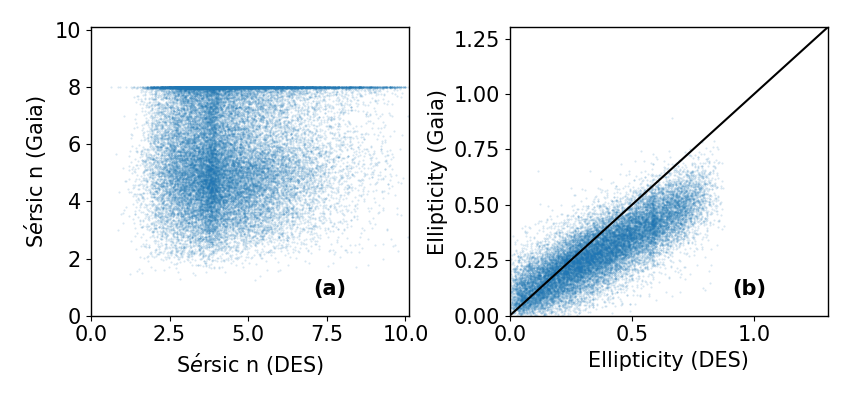}
\caption{Comparison of S\'ersic index (panel a), and ellipticity (panel b) from DES with the {\gaia} measurements.}
\label{fig:sersic_des_comparison}
\end{center}\end{figure}

The morphological parameters are accompanied by their formal uncertainties. Since these are estimated from the variance resulting from the search for a minimum in the residuals between model and observations, the provided uncertainties reflect the quality of the convergence rather than the precision of the estimation. In consequence, a fraction of sources may appear to have extremely small uncertainties while in reality, this is just the byproduct of a correlation with the  convergence velocity.

\subsection{Redshifts}

\begin{figure}\begin{center}
\includegraphics[width=0.95\columnwidth]{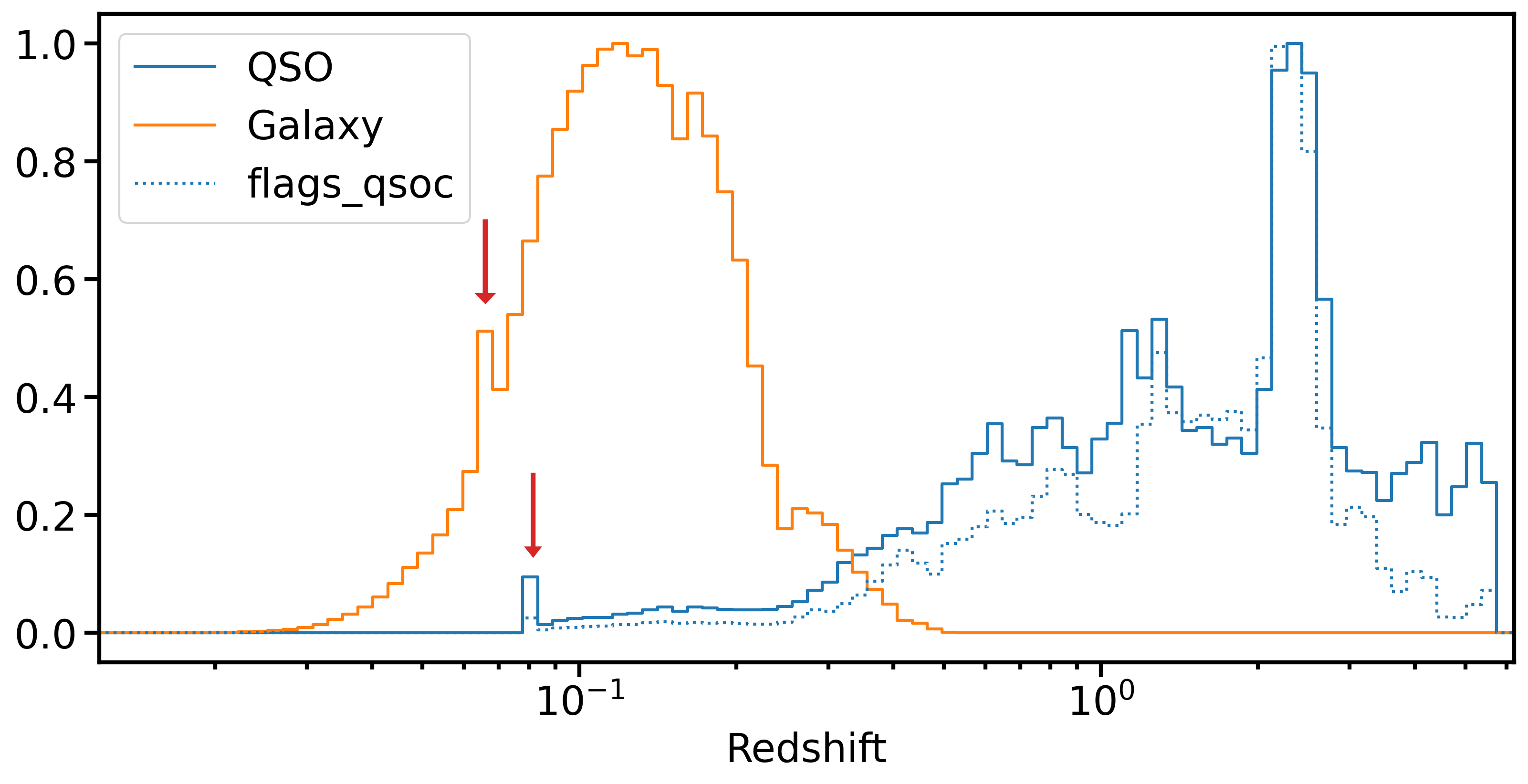}
\caption{Redshift distribution of the QSO (blue) and galaxy (orange) candidate sources. The dotted blue line corresponds to the quasars that were selected with the recommended QSOC redshift flags (\fieldName{flags_qsoc}=0 or \fieldName{flags_qsoc}=16). The two peaks marked by the red arrows are discussed in the text. }
\label{fig:extragal_redshift}
\end{center}\end{figure}

% UGC redshifts 
The provided galaxy redshift upper and lower values do not correspond to confidence intervals, but to prediction limits based on machine-learning. Still, the comparison with external catalogues indicates that $(\fieldName{redshift_ugc_upper}-\fieldName{redshift_ugc_lower})/2$ gives a good estimate of the uncertainty. A redshift peak at about 0.07 (red arrow in the orange histogram in Fig.~\ref{fig:extragal_redshift}) is found, which corresponds either to very bright galaxies or to stellar contaminants with convergence issues. The redshift range 0.070-0.071 should therefore be ignored \citep{DR3-DPACP-158}. A global overestimation of the redshifts for bright sources ($G<19$) is also observed. 

% QSO redshifts 
The QSO redshifts are log-normally distributed. To compare them to the literature, we therefore used $Z=\log(\fieldName{redshift_qsoc}+1),$ which is normally distributed with a standard deviation of $\sigma=(\log(\fieldName{redhift_qsoc_upper}+1)-\log(\fieldName{redhift_qsoc_lower}+1)) / 2$ \citep{DR3-DPACP-158}.
The comparison with LQAC5 \citep{2019A&A...624A.145S} presents 33\% of outliers at 5$\sigma$, which reduces to 8\% when the flag \fieldName{flags_qsoc}=0 or \fieldName{flags_qsoc}=16 is used. 
This is due to the degeneracies between spectral lines and redshift in the \xp spectra \citep[see][]{DR3-DPACP-158,DR3-DPACP-101}. A peak, this time at about 0.08, is also visible in the redshift distribution of the QSO (red arrow in the blue histogram in Fig.~\ref{fig:extragal_redshift}). The reason for this peak is that the MgII emission line is misclassified as H$\beta$, a characteristic emission line of this specific redshift range \citep{DR3-DPACP-158}. However, only a small number of sources contributes to this peak, and most of them have a non-zero \fieldName{flags\_qsoc}.

%+++++++++++++++++++++++++++++++++++++++++++++++++++++++++++++++++++++++++
\section{Non-single stars}\label{sec:nss}
%+++++++++++++++++++++++++++++++++++++++++++++++++++++++++++++++++++++++++

\gdrthree provides four tables for non-single stars (NSSs). The table \tableName{nss_two_body_orbit} contains orbital two-body models, covering astrometric \citep{DR3-DPACP-163,DR3-DPACP-176}, spectroscopic \citep{DR3-DPACP-178, DR3-DPACP-161}, and eclipsing \citep{DR3-DPACP-179} binaries as well combinations of these. The model that is used is indicated in the field \fieldName{nss_solution_type}, and the parameters that are solved for a given solution are described in the \href{https://gea.esac.esa.int/archive/documentation/GDR3/Gaia\_archive/chap\_datamodel/sec\_dm\_main\_tables/ssec\_dm\_nss\_two\_body\_orbit.html#nss\_two_body\_orbit-bit\_index}{\fieldName{bit_index}} field. The tables \tableName{nss_acceleration_astro} and \tableName{nss_non_linear_spectro} contain astrometric \citep{DR3-DPACP-163} and spectroscopic \citep{DR3-DPACP-178} acceleration solutions, and \tableName{nss_vim_fl} contains variability-induced mover (VIM) solutions \citep{DR3-DPACP-163}.  \cite{DR3-DPACP-100} also present the overall content of these non-single star tables.

\begin{figure}\begin{center}
\includegraphics[width=0.8\columnwidth]{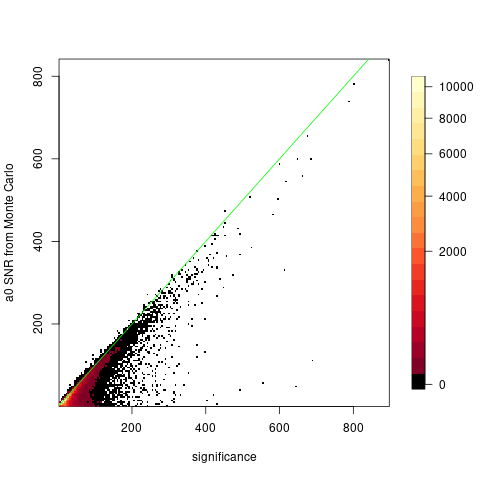}
\caption{Density plot of the signal-to-noise ratio of the semi-major axis of the photocentre orbit ($a_0$) derived from a Monte Carlo method as a function of the value provided in the field \fieldName{significance}.}\label{fig:nsssignificance}
\end{center}\end{figure}

\subsection{Astrometric orbital elements}
% TI to Campbell infos (C9VALSCI-239, C9VALSCI-178)
The orbital solutions for the astrometric binaries are presented using
what is called Thiele-Innes coefficients. They express the
orbital motion of the photocentre on the sky with a linear formulation. These coefficients replace the more
usual Campbell elements $a_0, i, \omega, \text{and }\Omega$, which are semi-major axis,
inclination, longitude of periastron, and position angle of the ascending node, respectively.
The relations between the two parameter sets are described in \cite{DR3-DPACP-163}.

In the transformation from Thiele-Innes to Campbell coefficients, it may be useful to use Monte Carlo simulations that take the correlation matrix into account instead of using local linear approximation formulas. In 87\% of the NSS sample, Gaussian errors in Thiele-Innes coefficients are transformed into asymmetric distributions for at least one of the Campbell elements. Mostly in the case of very low eccentricities, however, a number of sources shows a \fieldName{significance} parameter that disagrees with the signal-to-noise ratio that can be derived from Monte Carlo simulations (Fig.~\ref{fig:nsssignificance}). This seems to be due to an overestimation of the Thiele-Innes coefficient errors. Despite this, the local linear approximation formulas work well in deriving the error on the $a_0$ parameter even with a very strong overestimation of the Thiele-Innes coefficient errors. As Orbital solutions are filtered to have \fieldName{significance}$>5$, using a Gaussian error model for $a_0$ is reasonable while OrbitalTargetedSearch solutions need to be filtered. Due to an issue with the significance of AstroSpectroSB1 (see the on-line documentation), it needs to be verified that the signal-to-noise ratio is higher than 5.
The issue is also present for the spectroscopic part of the AstroSpectroSB1 solutions, for which local linear approximation errors on $a_1$ can be used as soon as the resulting signal-to-noise ratio is confirmed to be higher than 5. The local linear approximation formulas for the Thiele-Innes coefficients can be found in the appendix of \cite{DR3-DPACP-163}. Overall, to handle the Thiele-Innes coefficients, usual Monte Carlo techniques such as MCMC should not be used. Codes using automatic differentiation such as ADMB \citep{ADMB} and TMB \citep{TMB} have been tested to work fine for signal-to-noise ratios higher than 5.

The covariance matrix for very low eccentricity solutions may be problematic. In these cases, the eccentricity and periastron time should be set to zero. For AstroSpectroSB1 with eccentricity and argument of periastron fixed to zero (\fieldName{bit\_index}=65435), \fieldName{c\_thiele\_innes} is fixed to the non-circular value instead of zero.
The statistical properties of the distribution of the orbital elements are discussed in the appendix of \cite{DR3-DPACP-100}.

% Comparison with external catalogues
\subsection{External comparisons}
The comparison with external catalogues\footnote{\cite{2000A&AS..145..215P,2005A&A...442..365J,2019A&A...632A..31G,2018MNRAS.474..731K,2016MNRAS.458.3272K,2020MNRAS.496.1355H,2014PASA...31...24E}; SB9 \citep{2004A&A...424..727P}; APOGEE DR16 apjoker orbits \citep{2017ApJ...837...20P}; APOGEE SB2 \citep{2021AJ....162..184K}; Orb6 (http://www.astro.gsu.edu/wds/orb6.html)} shows that the orbital parameters agree well with literature values when the periods are consistent. It also confirms that the \fieldName{center_of_mass_velocity} agrees better with literature binary values than with the \fieldName{gaia_source.radial_velocity}. The strongest disagreements with external catalogues on the radial velocity semi-amplitude of the primary are for stars that are known to be SB2 (double-line spectroscopic binary), but are treated as SB1 (single-line) by NSS \citep{DR3-DPACP-178}.

The comparison of the literature orbits with astrometric acceleration solutions in the \fieldName{nss\_acceleration\_astro} table indicates that a significant fraction might have had an orbital solution and that some Acceleration7 could have been Acceleration9. This is intrinsic to the decision chain explained in \cite{DR3-DPACP-163}. The acceleration values disagree with the expectations from the known orbits and the \gaia\ observation times. The acceleration values should therefore be used with caution.

% NSS versus CU3
The NSS parallaxes show a median difference with the \tableName{gaia_source} parallaxes that is smaller than a few \muas. The HR diagram derived using NSS parallaxes (orbital and acceleration) is slightly sharper than the diagram derived with the \tableName{gaia_source} parallaxes, which indicates that the parallaxes are slightly more precise. The (statistical) improvement of the solutions does not guarantee that the accelerations are all physical, however. When they are compared to the long-term proper motion provided in the Hipparcos-Gaia catalogue of accelerations \citep{2022A&A...657A...7K,DR3-DPACP-100}, the NSS proper motions improve versus the \tableName{gaia_source} proper motions for the orbital solutions (moving from 21\% of 5$\sigma$ outliers to 9\%), but not the acceleration solutions (which have a much higher median signal-to-noise ratio of the proper motion anomaly than the orbital solutions) for which the comparison is slightly worse (moving from 80\% outliers to 85\%). This highlights that proper motion and accelerations may both have absorbed the orbital motion.

% Eclipsing binaries
The temperature ratio of eclipsing binaries corresponds well to the ratio derived by \cite{2014PASA...31...24E}, except for sources with a low \fieldName{g\_luminosity\_ratio}. The correspondence with the MSC temperature ratio is poor, but these ratios are to be used with caution (see Section~\ref{sec:APatmo}). The uncertainties on the eclipsing binary inclinations are suspiciously small. 

\subsection{Spurious solutions and error rescaling}
% C9VALSCI-114
To achieve the required radial velocity precision, the precise position of the spectra at the epoch on the focal plane needs to be known. For this purpose, the expected astrometric position as given by the predicted standard astrometric motion is used, rather than the measured astrometric position at the epoch, which would not be precise enough. However, if the astrometric motion is perturbed, or if the astrometric solution is not correct, then the computed epoch $RV$ will absorb this astrometric perturbation. This means that the epoch radial velocities could increase by up to $\approx 0.146 \times $\fieldName{astrometric\_excess\_noise} (km/s) in the case of binaries. In the best case, this would add an unmodelled  additional dispersion and possibly a small trend in the worst case. This may lead to spurious short period and large \fieldName{ruwe} SB1 solutions as well as to spurious solutions around the precession period (62.97 days), see \cite{DR3-DPACP-100}. 

The absence of a \fieldName{gaia_source.radial_velocity} value for an SB1 solution should warn the user: the source might have been considered peculiar, potentially SB2, too hot, too cool, with emission lines, or contaminated by a nearby star. Radial velocity variations can also be due to stellar pulsations instead of an orbital motion \citep{DR3-DPACP-100}, so that the variability information should also be confirmed for suspicious solutions.

% error rescaling
While the errors were rescaled according to the goodness of fit for the astrometric solutions (Orbital, AstroSpectroSB1, VIM,  and acceleration), this is not the case for the others. Because the mean goodness-of-fit distribution of SB2 and eclipsing solutions is quite large, we recommend rescaling the formal uncertainties for these solutions. The \fieldName{goodness_of_fit} provided for SB2 solutions can deviate by up to 1.6 from the one that can be recomputed using \fieldName{obj_func}.

%+++++++++++++++++++++++++++++++++++++++++++++++++++++++++++++++++++++++++
\section{Variability}\label{sec:vari}
%+++++++++++++++++++++++++++++++++++++++++++++++++++++++++++++++++++++++++

{\gdrthree} provides variability information for about 11.8 million sources, including 10.5 million variable sources of about 30 types of variability  \citep{DR3-DPACP-162} and 1.3 million sources (variable or not) in the Gaia Andromeda Photometric Survey (GAPS) \citep{DR3-DPACP-142}. Time-series photometry is released for all these 11.8 million sources in the \tableName{epoch_photometry} datalink table as well as their statistical parameters, and links to their potential other variability table are listed in the \tableName{vari_summary} table. The variability associated with galaxies provided in the \tableName{galaxy_candidates} table are mostly artefacts due to their extension \citep{DR3-DPACP-164} and therefore are not in the \tableName{vari_summary} or \tableName{epoch_photometry} tables.
Here, we present a brief overview of some issues we found during the scientific validation,  while for further details, we suggest the readers to consult the on-line documentation\textsuperscript{\ref{onlinedoc}} and papers\textsuperscript{\ref{dr3papers}}. 

A number of sources show more than one type of variability. While most overlaps between different classes can be explained scientifically, some stars have contradicting classifications. For example, 3159 sources are classified as both long-period and short-timescale variables. Detailed analyses of the final classification for these sources are provided in \citet{DR3-DPACP-171}.

 Intensity-averaged magnitudes in the BP (\fieldName{int\_average\_bp}) and RP (\fieldName{int\_average\_rp}) bands for four and two RR Lyrae stars, respectively, have unreliable negative values reaching $BP = -88 \pm 22$~mag. These six sources are faint RR Lyrae variables ($G \sim 18.5-19$~mag) for which the specific objects study pipeline for Cepheids and RR Lyrae stars (SOS Cep\&RRL; \citealt{DR3-DPACP-168}) failed to fit data points with the model line. The values that were provided were accordingly unreliable. Instead, other parameters that were calculated for these stars, such as intensity-averaged $G$ magnitudes and pulsation periods, are correct. It was therefore decided to include these sources in the DR3 sample of RR Lyrae stars despite the incorrect BP and RP intensity-averaged magnitude estimates.

 For 286 RR Lyrae stars, absorption in the $G$ passband (\fieldName{g\_absorption}) reaches unreliably high values from 10 to 3367~mag. This is
 likely caused by the imprecise estimation of the {\grp} magnitudes for the faint sources (see \citealt{DR3-DPACP-168} for further details).

%+++++++++++++++++++++++++++++++++++++++++++++++++++++++++++++++++++++++++
\section{Solar System objects}\label{sec:sso}
%+++++++++++++++++++++++++++++++++++++++++++++++++++++++++++++++++++++++++

% \begin{itemize}
%   - Accuracy of the orbits
% \end{itemize}

\gdrthree provides information for 158\,152 Solar System objects (SSO) with
more than 20 million observations (epoch astrometry). A
large data-set of ultra-accurate observations like this is made available in a
single day for the first time. 
%\sout{Here, we present a brief overview of some characteristics
%of the published asteroid population, including issues found during
%the scientific validation. For additional details and explanation, we
%suggest the readers to consult the relative documentation and papers}
%(see~\cite{GaiaDR3_CU4documentation,DR3-DPACP-150,GaiaDR3_CU9documentation}).

% Data sample
%\subsection{The data sample}
%The data sample provided in \gdrthree contains $158\,152$ objects,
%which is ten times larger than DR2 and more than 20 million
%observations (epoch astrometry single positions). 
The sample contains $156\,801$ known numbered minor planets, $1\,320$ unmatched moving
objects and $31$ natural satellites of planets. The source selection is described in~\cite{DR3-DPACP-150}.

All the main categories of Solar System bodies are present among the
known numbered minor planets: $447$ Near-Earth asteroids (NEAs),
$154\,771$ main-belt asteroids (MBAs), and a total of $1\,551$ Jupiter
Trojans, Centaurs, and more distant
objects. Figure~\ref{fig:asteroid_population} shows the different
categories in the semi-major axis and eccentricity plane.

\begin{figure}\begin{center}
\includegraphics[width=0.95\columnwidth]{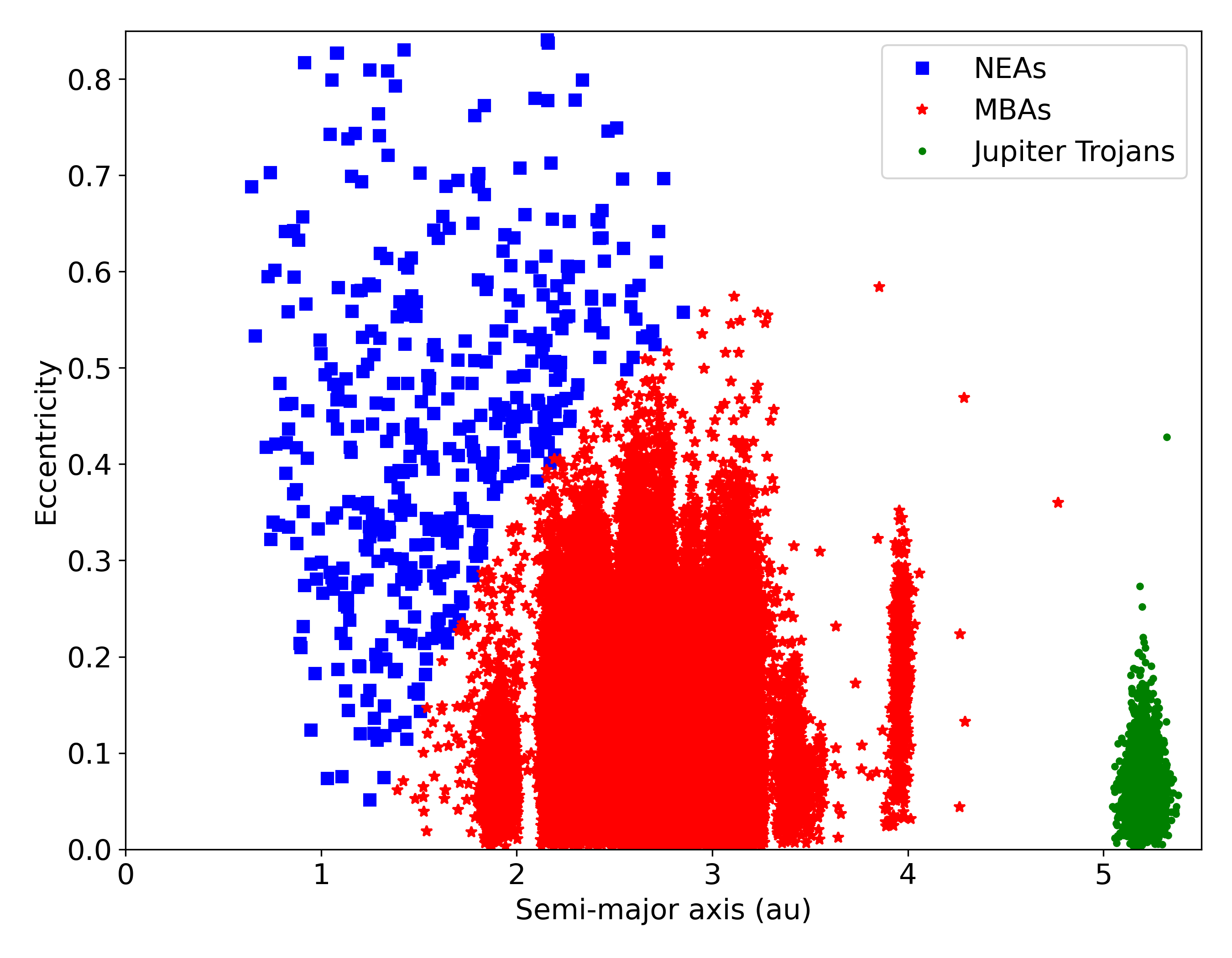}
\caption{Asteroid population in \gdrthree in the $(a,e)$ plane, where
  $a$ is the semi-major axis in au and $e$ is the eccentricity of the
  minor planets. The legend shows the different categories: blue
  squares for NEAs, red stars for MBAs, and green dots for Jupiter
  Trojans. For sake of clarity, the plot does not show Centaurs and
  more distant objects.}
\label{fig:asteroid_population}
\end{center}\end{figure}

The table \tableName{sso\_source} in the \gaia\ archive contains the
number of observations for each source. We would like to point out
that the count of the number of observations is incorrect for four
sources. The explanation for how to obtain the correct
number of observations is provided in the
\href{https://gea.esac.esa.int/archive/documentation/GDR3/Catalogue\_consolidation/chap\_cu9val/}{on-line
  documentation}.

%Unknown objects
\subsection{Unmatched sources and natural satellites of planets}

A small subsample of the data consists of $1\,320$ objects that were considered unknown at the time of processing. We refer to them as
unmatched sources.~\cite{DR3-DPACP-150} performed a search to
identify how many unmatched sources can now be identified (February
2022), and they found an identification for $712$ sources. We cannot
exclude either that some of the still-unmatched sources will be
identified or linked to known objects when the observations are
sent to the Minor Planet Center\footnote{The Minor Planet Center is
the single worldwide location for receipt and distribution of
positional measurements of minor planets, comets, and outer irregular
natural satellites of major
planets \url{https://minorplanetcenter.net/}.} All the sources will
still appear as unmatched in the \tableName{sso\_source} and
  \tableName{sso\_observations} tables in the \gaia\ archive. 

The sample also contains natural satellites of
planets for the first time. For a complete description of the process of selection, we
refer to~\cite{DR3-DPACP-150} (Sect. 3.1).

%Orbit fit 
\subsection{Orbit determination process}

We used an orbit determination process to assess the quality of the
data. This process is similar to the process that was carried out to validate
\gdrtwo~\citep{DR2-DPACP-32,2018A&A...616A..17A}.

We selected \gaia\ observations only for every known numbered minor
planet in \gdrthree. We used a modified version of the OrbFit
software\footnote{\url{http://adams.dm.unipi.it/orbfit/}} to fit the
orbits to \gaia\ observations alone. It is important to note that this
software is completely independent from everything that runs in the
\gaia\ data processing, and we improved it to fully exploit the
accuracy of \gaia\ observations.

The results of the orbital fit can be summarised as an orbit (if the fit converges), post-fit residuals in the $(\alpha \cos(\delta), \delta)$ space and in the $(AL,AC)$ space, and rejection of incorrect quality or mistakenly linked observations.

The modified version of the OrbFit software makes use of a non-linear
weighted least-squares algorithm to fit the orbits. The weight matrix
for \gaia\ is the quadratic sum of the systematic and random matrices,
available in the \tableName{sso\_observation} table.

We also corrected the observations for the light bending. This is a
different approach than was applied in the validation of
\gdrtwo. 

%Orbit determination results
\subsubsection{Orbit determination results: Orbit failures}

The orbit fit procedure worked for almost all the known sources. It
only failed for $198$ objects. The reasons for this vary: the time spanned by the observations was too short, too few observations
were  available,  or a combination of the two, as shown in
Fig.~\ref{fig:failures}.

\begin{figure}\begin{center}
    \includegraphics[width=0.95\columnwidth]{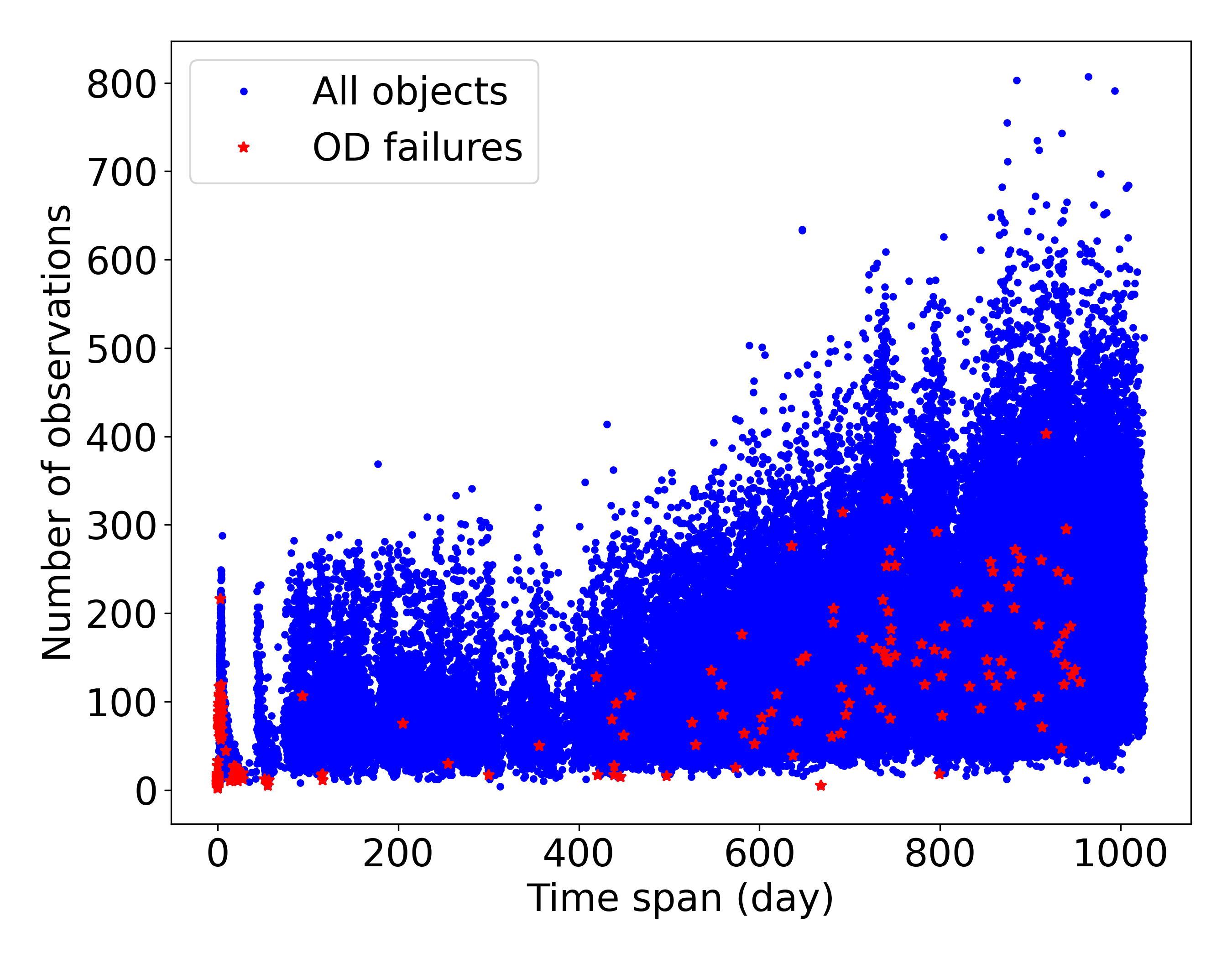}
    \caption{Time span by the observations in \gdrthree (in days)
      vs the number of observations for each known source. The red
      stars represent the objects for which the orbit determination
      process did not converge.}
    \label{fig:failures}
\end{center}\end{figure}

The quality of the observations is not affected by the non-convergence
of the orbit. They were therefore all accepted and are
available in the \tableName{sso\_observations} table.

\subsubsection{Orbit-determination results: Post-fit residuals}

The orbit-determination process is based on finding and removing bad-quality observations, so that they do not affect the goodness of
fit. The orbit-determination software we used to validate the
data rejects observations with $\chi^2>25$ ($5\sigma$). This can
happen because the quality of the data is not as expected, because the
weights that are used are too low, or because the observations do not belong to
the object. The latter case is called mistaken linkage or incorrect
identification. We decided to remove from the data only the
observations for which the absolute value of the along-scan post-fit
residuals was higher than $250$ mas and for which the absolute value of the
across-scan post-fit residuals was higher than $2\,500$ mas. In this
way, we cleaned the database from possible contaminants. At the
same time, we wished to keep the largest possible number of
observations so that the community could search for interesting features
(e.g. the presence of satellites). As a consequence, the sample can still contain some contaminants. For example,
  we may only have removed part of a transit, but we decided to adopt
  a unique approach that is valid for all the observations.
 Some observations can also be rejected during the orbit-determination
  process, but this does not affect the overall quality of the data.

After removing the bad-quality observations, we analysed the post-fit
residuals in the along-scan and across-scan
directions. Figure~\ref{fig:postfit_res_histo} shows the histogram of the post-fit
residuals along-scan ($\Delta AL$) and across-scan ($\Delta AC$). These residuals are obtained as
a rotation of the residuals in $\alpha \cos(\delta)$ and $\delta$,
where the rotation angle is the position angle as given in the
\tableName{sso_observations} table. The whole procedure has been
described in~\cite{DR2-DPACP-32}.

\begin{figure}\begin{center}
    \includegraphics[width=0.45\columnwidth]{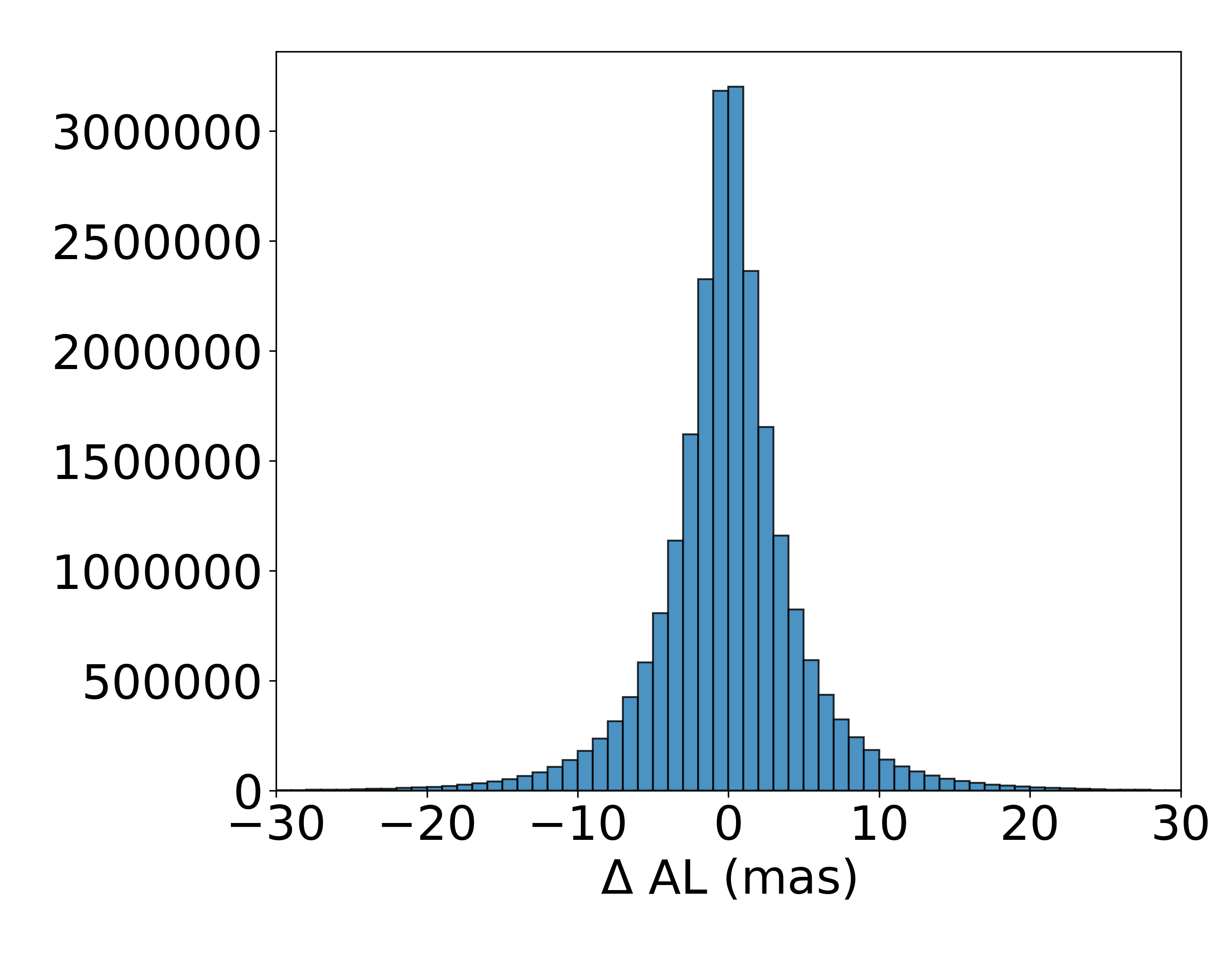}
    \includegraphics[width=0.45\columnwidth]{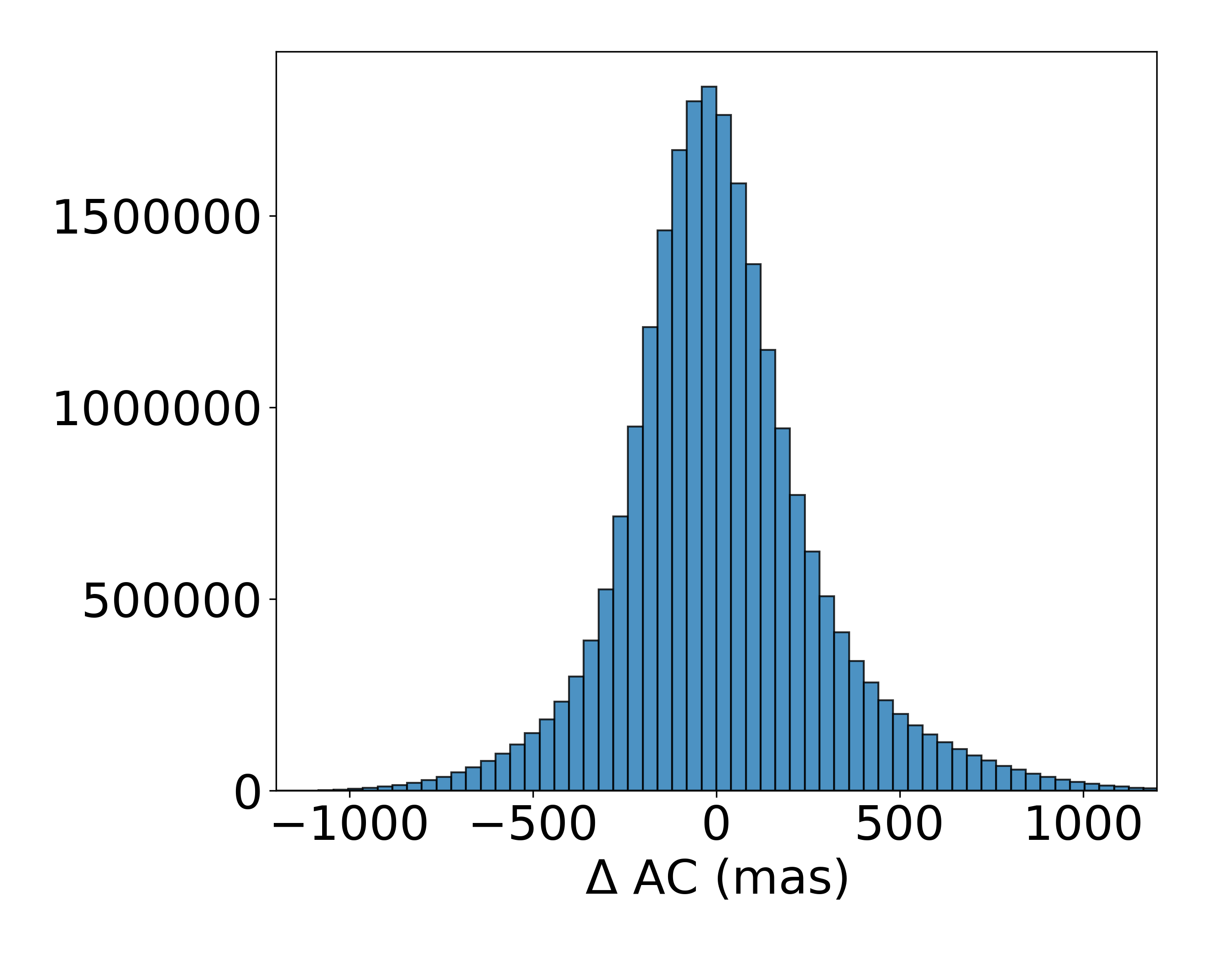}
    \caption{Histogram of post-fit residuals of the selected
      observations in the left: along-scan, right: across-scan direction.}
    \label{fig:postfit_res_histo}
\end{center}\end{figure}

The mean of the post-fit along-scan residuals is $0.03$ mas, and the
standard deviation is slightly larger than $5$ mas. This is exactly
what we expected as a result of the orbit-determination fit (we recall
that we discarded observations at $5$ $\sigma$ level). Post-fit
residuals in the across-scan direction are expected to be far larger than
the corresponding along-scan residuals as a result of the geometry of the
spacecraft observations~\citep{DR2-DPACP-32}, as the histogram in
fig.~\ref{fig:postfit_res_histo} shows. The mean in this case is close
to $13$ mas, which shows that the across-scan observations still
contain a small bias. The standard deviation is larger than $200$ mas. This is
close to what we expected.

\begin{figure}\begin{center}
    \includegraphics[width=0.45\columnwidth]{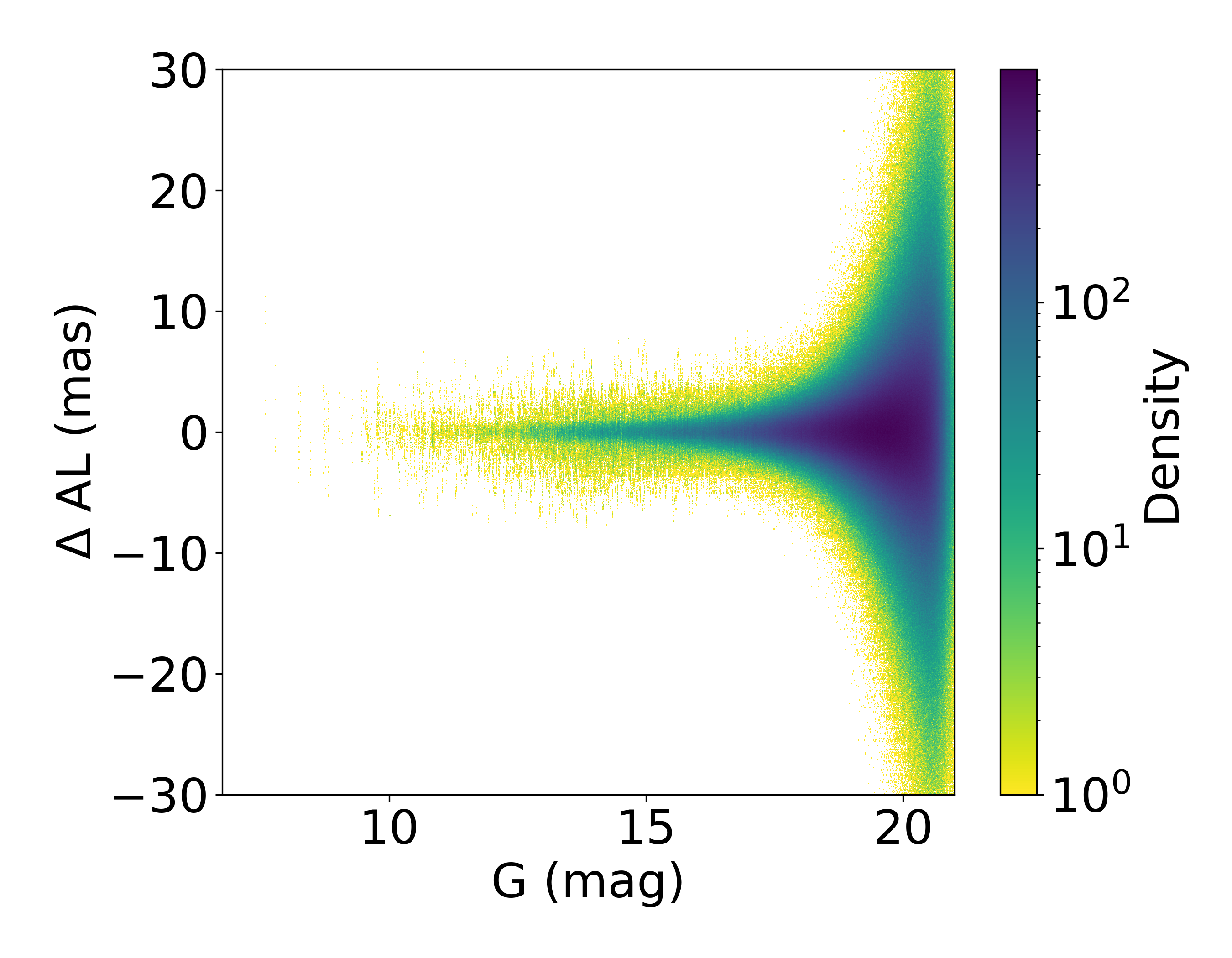}
    \includegraphics[width=0.45\columnwidth]{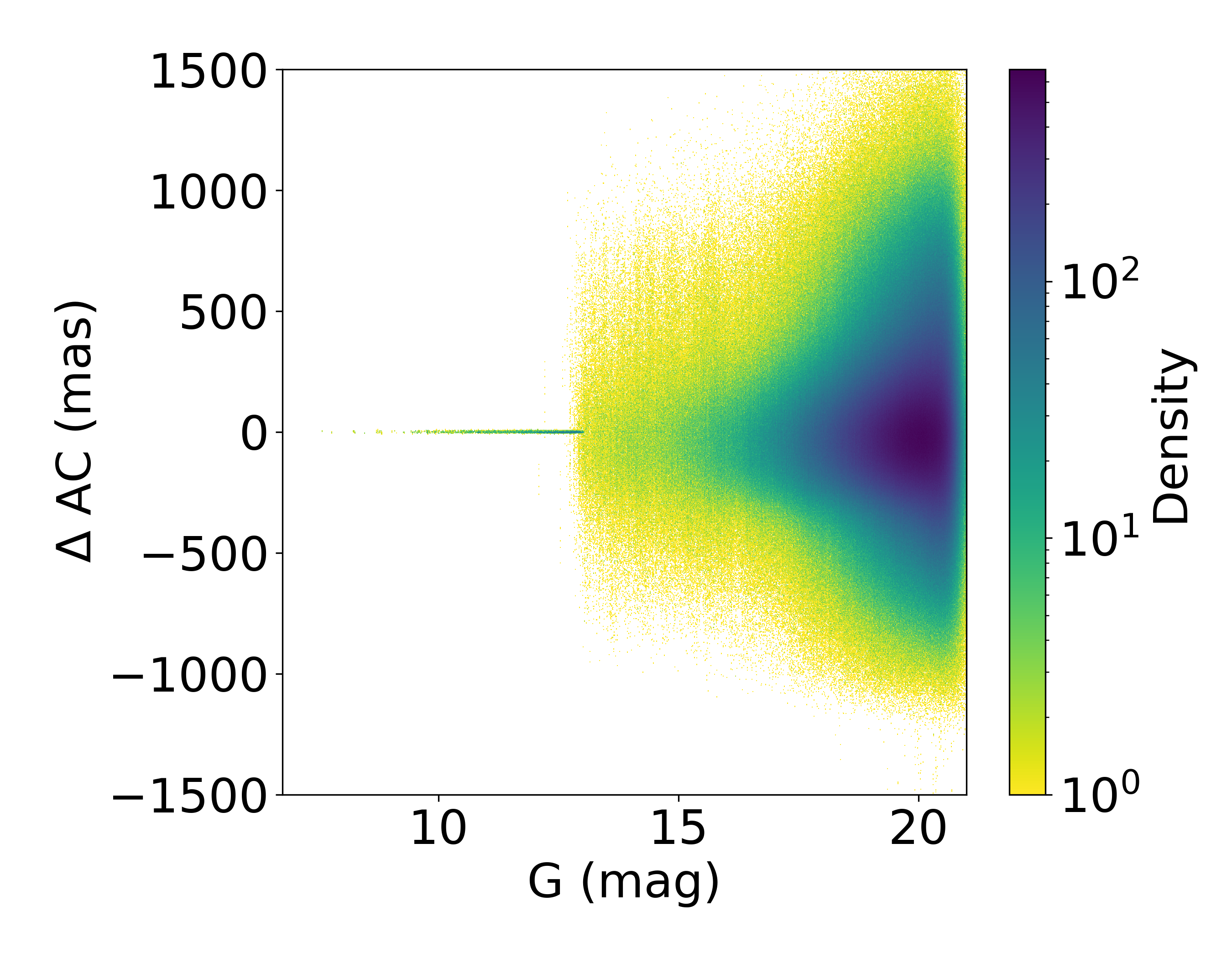}
    \caption{Density plot of the post-fit residuals as a
      function of the $G$ magnitude left: along-scan, right: across-scan}
    \label{fig:postfit_res_mag}
\end{center}\end{figure}

We now examine the $(\Delta AL, \Delta AC)$ post-fit residuals as a
function of the $G$ magnitude (Fig.~\ref{fig:postfit_res_mag}).  For
very bright sources ($G<13$ mag), a full two-dimensional window is
transmitted, which means that across-scan information is available,
corresponding to what we show in Fig~\ref{fig:postfit_res_mag}, where
across-scan residuals are at the milliarcsecond level when $G<13$
mag. Figure~\ref{fig:postfit_res_mag} shows the increase in along-scan residuals when the source is fainter ($G>19$ mag). They  almost reach
the detectability limit, but usually remain very small
(inside the $[-10,10]$ mas interval) for all the other sources.

Additional information about residuals and a comparison with \gdrtwo are
available in \cite{DR3-DPACP-150}, even though the authors used
a different set of residuals that they obtained as a result of the internal
process of the observations and not from the validation. It has been
proved in the same paper that these residuals and the corresponding
orbit can be considered equivalent to those that were computed during the
validation process.

%Orbit accuracy
\subsection{Orbit accuracy: Comparison with known catalogues}

The post-fit accuracy of the semi-major axis $(\sigma_{a})$ is a good
estimator of the orbit quality. We compared the post-fit
$\sigma_a$ obtained using \gaia\ observations alone with that
available from the JPL Small Body Database\footnote{\url{https://ssd.jpl.nasa.gov/}}
, which makes use of all the available observations; see
Fig.~\ref{fig:sigmaa_gaia_all}. The black line in the figure is the
bisector of the first quadrant: the orbits of the objects below the
line have a better uncertainty using \gaia\ observation alone. It is
clear that \gdrthree alone is still not enough to reach the final
accuracy expected for \gaia\ \citep[][Fig. 32]{DR2-DPACP-32}, but the
number of orbits for which the accuracy is now better using
\gaia\ alone has largely increased from \gdrtwo.

\begin{figure}\begin{center}
    \includegraphics[width=0.8\columnwidth]{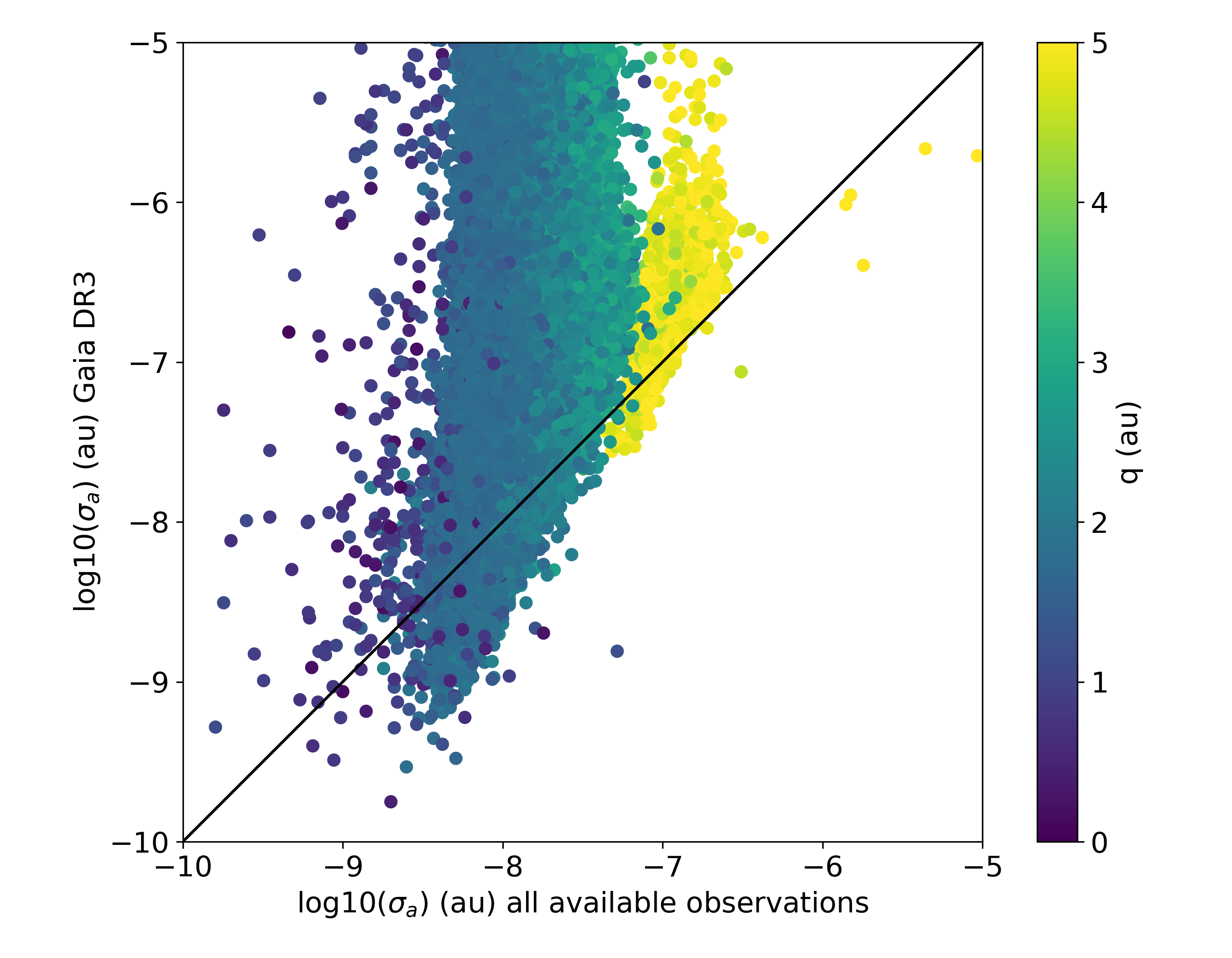}
    \caption{Quality of the orbit determination measured by the
      post-fit uncertainty of the semi-major axis for the whole sample
      of objects contained in \gdrthree with respect to the current
      measurements from the JPL Small Body Database. The black line is
      the bisector of the first quadrant.}
    \label{fig:sigmaa_gaia_all}
\end{center}\end{figure}

%+++++++++++++++++++++++++++++++++++++++++++++++++++++++++++++++++++++++++++
\section{Conclusions}\label{sec:conclu}
%+++++++++++++++++++++++++++++++++++++++++++++++++++++++++++++++++++++++++++

%Users should have an aspirin at hand. 
The third data release of \gaia, DR3, provides a very large amount of new data. This complex and diverse dataset has a number of caveats that the users should be aware of. In this paper we summarised the main issues we found during the transversal validation, and we provided links to the relevant papers or documentation and recommendations. In particular, we highlighted that flags provided with the data products should be used whenever available (e.g. \fieldName{flags_gspspec}, \fieldName{flags_flame}, \fieldName{dibqf_gspspec}, and \fieldName{flags_qsoc}). We warned about the error underestimation of the XP coefficients, GSP-Phot parameters, GSP-Spec ANN, spectroscopic and eclipsing binary solutions with a poor goodness-of-fit, and we 
provided a correction formula for the radial velocity error estimates. The DSC white dwarf and binary star classifications should not be used. A number of parameters were highlighted as to be used with caution (MSC parameters, \fieldName{mh_gspphot}, \fieldName{distance_gspphot}, \fieldName{mg_gspphot}, FLAME mass and ages for giants, radial velocities for $(G_{\rm RVS} - G) < -3$, and astrometric binary acceleration values). The corrections proposed in \cite{DR3-DPACP-186} should be applied to the GSP-Spec parameters. Some systematics such as those presented for \fieldName{vbroad}, the XP spectra dip, or extinction overestimation are to be taken into account according to the science case. Filters need to be applied to the QSO and galaxy candidates to have purer samples \citep{DR3-DPACP-101}. Monte Carlo techniques should not be used with the Thiele-Innes astrometric binary orbital parameters.

This paper focused on limitations in the data released in {\gdrthree}. We encourage a study of the other papers that accompany this data release for an overview of the high quality of the {\gaia} products and a glance at the wonderful science outcomes that can be expected from this wealth of data. We hope that this paper will help users to find their way through the data so that they can make the best of it.

%
%________________________________________________________________
%
%
\begin{acknowledgements}
This work presents results from the European Space Agency (ESA) space mission \gaia. \gaia\ data are being processed by the \gaia\ Data Processing and Analysis Consortium (DPAC). Funding for the DPAC is provided by national institutions, in particular the institutions participating in the \gaia\ MultiLateral Agreement (MLA). The \gaia\ mission website is \url{https://www.cosmos.esa.int/gaia}. The \gaia\ archive website is \url{https://archives.esac.esa.int/gaia}.

This work has made an extensive use of Aladin and the SIMBAD, VizieR
databases operated at the Centre de Donn\'ees Astronomiques (Strasbourg) in
France and of the software TOPCAT \citep{2005ASPC..347...29T}.

This work has been supported by the Agence Nationale de la Recherche (ANR project SEGAL ANR-19-CE31-0017). It has also received funding from the project ANR-18-CE31-0006 and from the European Research Council (ERC grant agreement No. 834148). ZKR acknowledges funding from the Netherlands Research School for Astronomy (NOVA).
This work was partially funded by the Spanish MICIN/AEI/10.13039/501100011033 and by "ERDF A way of making Europe" by the “European Union” through grant RTI2018-095076-B-C21, and the Institute of Cosmos Sciences University of Barcelona (ICCUB, Unidad de Excelencia ’Mar\'{\i}a de Maeztu’) through grant CEX2019-000918-M.

\end{acknowledgements}

\bibliographystyle{aa} % style aa.bst
\bibliography{biblio} % your references refs.bib

\appendix

%+++++++++++++++++++++++++++++++++++++++++++++++++++++++++++++++++++++++++
\section{Wide binaries}\label{sec:widebinaries}
%+++++++++++++++++++++++++++++++++++++++++++++++++++++++++++++++++++++++++

We used the \cite{2021MNRAS.506.2269E} catalogue of wide binaries, further limited to a chance alignment probability $\mathrm{R\_chance\_align}<0.1$, to test the radial velocity error underestimation in Sec.~\ref{sec:rverrors}. Here we test the astrophysical parameter compatibility of the two components. A summary of the results is listed in Table~\ref{tab:widebinaries}. The percentage of systems consistent within 1$\sigma$ was computed after removing the 5$\sigma$ outliers. If this percentage is significantly lower (higher) than $\text{about }$68, the errors are underestimated (overestimated).

\begin{table}
\footnotesize
\caption{\footnotesize\sl Comparison of the astrophysical parameters derived for the two components of a wide binary \citep{2021MNRAS.506.2269E}. }
\label{tab:widebinaries}
\begin{center}
\begin{tabular}{lcc} 
{\bf Parameters} & {\bf \% of 5$\sigma$ outliers} & {\bf \% within 1$\sigma$} \\
age\_flame & \textcolor{black}{3} & \textcolor{black}{65} \\
age\_flame\_spec & \textcolor{black}{0} & \textcolor{black}{66} \\
\hline
mh\_gspspec & \textcolor{black}{7} & \textcolor{black}{47} \\
alphafe\_gspspec & \textcolor{black}{5} & \textcolor{black}{49} \\
fem\_gspspec & \textcolor{black}{0} & \textcolor{black}{90} \\
sife\_gspspec & \textcolor{black}{0} & \textcolor{black}{73} \\
cafe\_gspspec & \textcolor{black}{2} & \textcolor{black}{58} \\
tife\_gspspec & \textcolor{black}{0} & \textcolor{black}{83} \\
mgfe\_gspspec & \textcolor{black}{0} & \textcolor{black}{70} \\
feiim\_gspspec & \textcolor{black}{0} & \textcolor{black}{64} \\
sfe\_gspspec & \textcolor{black}{0} & \textcolor{black}{79} \\
nfe\_gspspec & \textcolor{black}{0} & \textcolor{black}{67} \\
crfe\_gspspec & \textcolor{black}{0} & \textcolor{black}{44} \\
nife\_gspspec & \textcolor{black}{0} & \textcolor{black}{73} \\
\hline
mh\_gspspec\_ann & \textcolor{black}{20} & \textcolor{black}{32} \\
alphafe\_gspspec\_ann & \textcolor{black}{8} & \textcolor{black}{36} \\
\hline
mh\_gspphot & \textcolor{black}{45} &  \textcolor{black}{25} \\
azero\_gspphot & \textcolor{black}{52} & \textcolor{black}{26} \\
distance\_gspphot & \textcolor{black}{22} & \textcolor{black}{35} \\
mh\_gspphot\_marcs & \textcolor{black}{41} & \textcolor{black}{26} \\
azero\_gspphot\_marcs & \textcolor{black}{52} & \textcolor{black}{26} \\
distance\_gspphot\_marcs & \textcolor{black}{23} & \textcolor{black}{35} \\
mh\_gspphot\_phoenix & \textcolor{black}{46} & \textcolor{black}{29} \\
azero\_gspphot\_phoenix & \textcolor{black}{49} & \textcolor{black}{28} \\
distance\_gspphot\_phoenix & \textcolor{black}{25} & \textcolor{black}{34} \\
mh\_gspphot\_ob & \textcolor{black}{40} & \textcolor{black}{75} \\
azero\_gspphot\_ob & \textcolor{black}{72} & \textcolor{black}{25} \\
distance\_gspphot\_ob & \textcolor{black}{51} & \textcolor{black}{34} \\
mh\_gspphot\_a & \textcolor{black}{51} & \textcolor{black}{47} \\
azero\_gspphot\_a & \textcolor{black}{86} & \textcolor{black}{25} \\
distance\_gspphot\_a & \textcolor{black}{58} & \textcolor{black}{30} \\
\hline
mh\_msc & \textcolor{black}{3} & \textcolor{black}{77} \\
azero\_msc & \textcolor{black}{1} & \textcolor{black}{88} \\
distance\_msc & \textcolor{black}{4} & \textcolor{black}{68} \\
\hline
azero\_esphs & \textcolor{black}{18} & \textcolor{black}{36} \\
\hline
dib\_gspspec\_lambda & \textcolor{black}{34} & \textcolor{black}{61} \\
dibew\_gspspec & \textcolor{black}{7} & \textcolor{black}{59} \\
dibp2\_gspspec  &  \textcolor{black}{7} & 65 \\
\end{tabular}
\end{center}
\end{table}

\end{document}